\documentclass[journal,comsoc]{IEEEtran}

\ifCLASSINFOpdf
\usepackage[pdftex]{graphicx}
\usepackage{graphicx}
\usepackage{latexsym}
\usepackage{amssymb}
\usepackage{amsmath} 
\usepackage{enumerate}
\usepackage{mathtools}
\usepackage{hyperref}
\usepackage{multirow}
\usepackage{makecell}
\usepackage{longtable} % for 'longtable' environment
\usepackage{pdflscape,array,booktabs}
\usepackage[table]{xcolor}
\usepackage{footnote}

\usepackage[colorinlistoftodos]{todonotes}

\hypersetup{
    colorlinks=true,
    linkcolor=black,
    filecolor=black,      
    urlcolor=black,
    citecolor=black,        % color of links to bibliography
}

\usepackage[switch,pagewise]{lineno}
\usepackage{rotating}

\usepackage{dcolumn}

\renewcommand{\tabcolsep}{3.5pt}

\newenvironment{grey-color}{\par\color{gray}}{\par}

\definecolor{brightmaroon}{rgb}{0.76, 0.13, 0.28}

\else
\fi

\usepackage{color}
\usepackage{showexpl}
\usepackage{makeidx}
\usepackage{array}
\usepackage{bbding}
\usepackage{amssymb}

\usepackage{cite}

\graphicspath {{Figures/}}

\usepackage{adjustbox}
\usepackage{array}

\newcolumntype{R}[2]{%
	>{\adjustbox{angle=#1,lap=\width-(#2)}\bgroup}%
	l%
	<{\egroup}%
}
\makeindex
\begin{document}

\twocolumn

%\title{Software-Defined WLANs: State-of-the-Art\\ and Research Challenges}
%\title{{\huge A Review of Software-Defined WLANs: Architecture, Association Control, and Channel Assignment}}
\title{{A Review of Software-Defined WLANs: Architectures and Central Control Mechanisms}}
\author{Behnam~Dezfouli, \textit{Member, IEEE}, Vahid~Esmaeelzadeh, \textit{Student Member, IEEE},\\ Jaykumar Sheth, \textit{Student Member, IEEE}, and Marjan Radi, \textit{Member, IEEE}
\thanks{ 
% Please, write here acknowledgment for financial support if desired.
%Cognitive work was supported in part by the Spanish Council of Research under Grant  BS123456 (sponsor and financial support acknowledgment goes here). Paper titles should be written in uppercase and lowercase letters, not all uppercase. Avoid writing long formulas with subscripts in the title; short formulas that identify the elements are fine (e.g., "Nd-Fe-B"). Full names of authors are preferred in the author field. 
}
\thanks{Behnam Dezfouli, Vahid Esmaeelzadeh and Jaykumar Sheth are with the Internet of Things Research Lab, Department of Computer Engineering, Santa Clara University, Santa Clara, CA, 95053 USA.
E-mail: \{bdezfouli, vesmaeelzadeh, jsheth\}@scu.edu.
}% <-this % stops a space
\thanks{Marjan Radi is with Western Digital Corporation, Milpitas, CA, 95035 USA. E-mail: {marjan.radi}@wdc.edu.
}% <-this % stops a space
}

%\markboth{}%
%{B. Dezfouli \MakeLowercase{\textit{et al.}}: A Review of Software-Defined WLANs: Architectures and Central Control Mechanisms}

\maketitle

% The Abstract
\begin{abstract}
The significant growth in the number of WiFi-enabled devices as well as the increase in the traffic conveyed through wireless local area networks (WLANs) necessitate the adoption of new network control mechanisms. 
Specifically, dense deployment of access points, client mobility, and emerging QoS demands bring about challenges that cannot be effectively addressed by distributed mechanisms.
Recent studies show that software-defined WLANs (SDWLANs) simplify network control, improve QoS provisioning, and lower the deployment cost of new network control mechanisms. 
In this paper, we present an overview of SDWLAN architectures and provide a qualitative comparison in terms of features such as programmability and virtualization.
In addition, we classify and investigate the two important classes of centralized network control mechanisms: (i) association control (AsC) and (ii) channel assignment (ChA).
We study the basic ideas employed by these mechanisms, and in particular, we focus on the metrics utilized and the problem formulation techniques proposed.
We present a comparison of these mechanisms and identify open research problems.
\end{abstract}
% The Keywords 

\IEEEpeerreviewmaketitle

\def\blue!50!blue{blue!50!black}
\def\RevisionColor{blue!50!black}
\def\RevisionColorTwo{black!50!black}

\section{Introduction}
\label{intro}
\IEEEPARstart{W}{ireless} local area networks (WLANs) have been deployed broadly in various types of environments such as enterprises, universities, airports, shopping malls, and homes. 
The IEEE 802.11 standard, commonly known as WiFi, is becoming more popular and ubiquitous. 
In particular, the traffic demand of WiFi networks is increasing due to the emergence of high-definition video streaming, Internet of Things (IoT), cellular data offloading, online gaming, and virtual reality \cite{Tozlu2012,Baird2017,NextGen-WiFi-Survey-2016,Offloading-survey-2016,WifiAlliance1,Ericsson}.
%removed citations \cite{NGwlans-survey-2016}

\begin{figure}[!t]
	\centering
{\includegraphics[width=\linewidth]{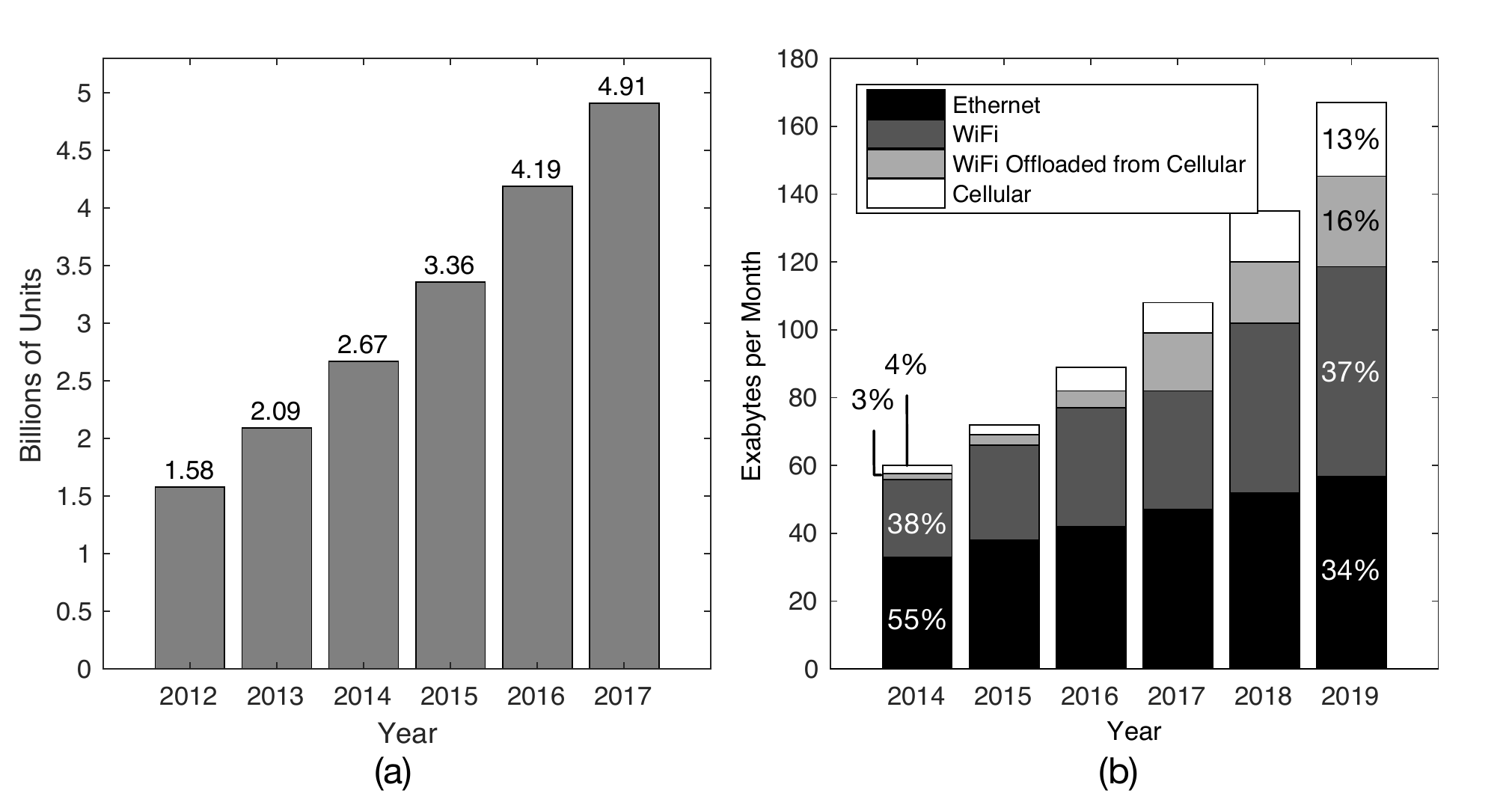}}
	\caption{(a) Increase in the number of WiFi-enabled devices shipped during 2012 to 2017. (b) Traffic growth of WiFi, cellular and Ethernet networks during 2014 to 2019 \cite{Cisco-statistics}.}
		\label{wifi-statistics}
\end{figure}

Statistics and forecasts show a continuous growth in the number of WiFi-enabled devices shipped as well as the amount of traffic conveyed through WLANs. 
Figure \ref{wifi-statistics}(a) shows the number of WiFi-enabled devices, including access points (APs), network interface cards (NICs), routers, switches, laptops, tablets, smartphones, and TVs that were shipped from 2012 to 2017. 
The shipment of WiFi-enabled devices was 1.58 billion units in 2012 and is estimated to be 4.91 billion units in 2017 \cite{WiFi-Devices,WifiAlliance1,Ericsson}. 
Furthermore, the percentage of traffic conveyed by WiFi networks is growing continuously, as illustrated in Figure \ref{wifi-statistics}(b). 
During 2014, 41$\%$ of mobile data traffic was exchanged through WiFi networks, including the traffic offloaded from cellular networks into WLANs \cite{Cisco-statistics,Offloading,Offloading-survey-2016}. 
This traffic is expected to reach $53\%$ by 2019. 
In addition, the scientific community has paid great attention to 802.11 networks, as demonstrated by the 165K citations that this topic has received until 2014, according to \cite{Experimenting802.11}.

%Based on the trend of WLANs usage, it is crucial to investigate and study on the practical and theoretical issues of these networks.
%
%
%
%\begin{figure}[!t]
%	\centering
%	{\includegraphics[width=0.8\linewidth]{wifitraffic.pdf}}
%	\caption{Traffic growth of WiFi, cellular and Ethernet (wired) networks during 2014 to 2019 \cite{Cisco-statistics}.}	
%		\label{wifi-traffic}
%\end{figure} 
%

The increase in the number of WiFi-enabled devices and their growing traffic demand requires a dense installation of APs to improve spatial reuse and enhance network capacity. 
However, AP densification intensifies channel contention, increases the interference level among APs and their associated clients, and exacerbates the challenges of client handoff among APs. 
Therefore, effective association and interference control are both necessary to satisfy the QoS demands of clients in terms of throughput, delay, reliability and energy efficiency \cite{largeScaleMeas,DenseAP,DenseWLAN2,APdensity,QoS-WLANs}. 
Satisfying these requirements is even more important and challenging when WLANs are used for mission-critical application such as industrial process control, factory automation, and medical monitoring.
In particular, these applications require bounded packet delivery delay and high link reliability, and in some cases energy efficiency is a critical performance metric as well \cite{tramarin2016use,REWIMO}.

%removed reference \ciet{NGwlans-survey-2016,DenseWLAN1}
%
%
% Some approaches try to improve the performance of WLANs by modifying the 802.11 standards which brings about high amount of overhead in terms of operations costs and modifying WiFi-enabled clients  
%%
%Generally, the handling and management of WLANs interference and mobility issues can be done either in distributed or centralized manner \cite{Channel-assignment-survey-2010}. 
%Traditionally, the distributed management approach was used in which the protocols are run on access points with no central entity. By increasing the size of WLANs and demand 
%However, in centralized management of WLANs, there is at least a central entity to run control and monitoring policies and algorithms with the global view of network in order to manage all components of WLAN, i.e., APs and clients \cite{Cisco}. 
%%
%Distributed management of WLANs was an appropriate approach before the evolution of needs and pervasive deployment of WLANs so that it has turned into the main access technology of almost all organizations and user communities. 
%Due to the quick growth in the size, coverage and traffic of modern WLANs,  

In order to tackle the challenges of controlling complex wired networks (such as ISPs and data centers), \textit{software-defined networking} (SDN) has been proposed to simplify and improve the design and development of network control mechanisms by decoupling the control and data plane.
Instead of running network control mechanisms distributively on switching devices, a controller configures the switches to operate according to the decisions made centrally \cite{SDN1,SDN2}. 
Consequently, in addition to simplifying the deployment of network control mechanisms as applications running on the controller, these mechanisms can benefit from the controller's global network view.
Not only for wired networks, SDN has gained popularity for the design of WLANs and cellular networks as well \cite{SDNcellular,Crowd, mSDN2,SDNsurvey}.

A \textit{software-defined WLAN} (SDWLAN) enables central monitoring and control of network operation.
% To this end, the network architecture should support the interaction between control mechanisms and data plane equipment. 
A SDWLAN can be studied from two point of views, as follows:

\begin{itemize}
    \item  \textbf{Architecture}. A SDWLAN architecture: (i) defines the various network components (e.g., controller, APs, switches, middleboxes) used to build the network as well as the topology employed to connect these components;
    (ii) implements application programming interfaces (APIs) through which the controller communicates with network devices to perform data collection and distribute control commands;
    (iii) specifies the separation level between the control plane and data plane.
    The architectural features highly affect the development flexibility and performance of network control mechanisms \cite{Primitives,DenseAP,OpenSDWN,CloudMAC}.
    For example, when medium access control (MAC) functionalities are transferred from the APs to the controller, network control applications may implement AsC mechanisms with various capabilities and performance levels, depending on the separation level and APIs provided.
    \item \textbf{Control Mechanisms.} To benefit from the features of SDWLAN architectures, and in particular, the global network view provided, various control mechanisms, such as \textit{association control} (AsC), \textit{channel assignment }(ChA), transmission power control, and CCA threshold adjustment have been proposed by the research community.
    In this paper, we particularly focus on centralized AsC and ChA mechanisms because they have been widely proposed and adopted to benefit from the features of SDWLANs.
    AsC mechanisms address client-AP association for both static and dynamic clients to achieve various objectives, such as seamless mobility, fairness among clients, load balancing among APs, and interference mitigation. %\cite{WiFiSeer,802.11n-AP-Association-2014,Migration-DAM,flow-level-DAM}.
    ChA mechanisms address the problem of channel assignment to APs in order to minimize the interference level experienced by APs and clients  \cite{Measurement-CA-WCNC-10,CAPWAP-based-CA-11,802.11ac-PCA}. %
\end{itemize}

%In addition, mechanisms such as scheduling are also proposed in the context of SDWLANs to improve network performance.

% distributed management approaches cannot easily provide and guarantee the new quality of service (QoS) requirements, and limits the ability to handle the scaling, security and reliability \cite{G4}. 
%%
%In particular, using distributed approaches increases the cost of applying new algorithms and protocols because of the need to update the new protocols over all access points in the WLAN \cite{G5}. 
%%
%However, in central management of WLANs, most algorithms can be run on a central entity (controller). In this way, most management and control decisions are performed in the controller with a global view of the whole network that leads to more optimal strategies with lower deployment cost because most updates are performed on the central entity instead of updating all access points across the WLAN \cite{G6}. 
%%
%Figure \ref{fig_ref_arch} shows the general architecture of a CCWLAN.
%%

\subsection{Contributions}
\label{contributions_sec}
\begin{figure}[!t]
	\centering
	\includegraphics[width=0.72\linewidth]{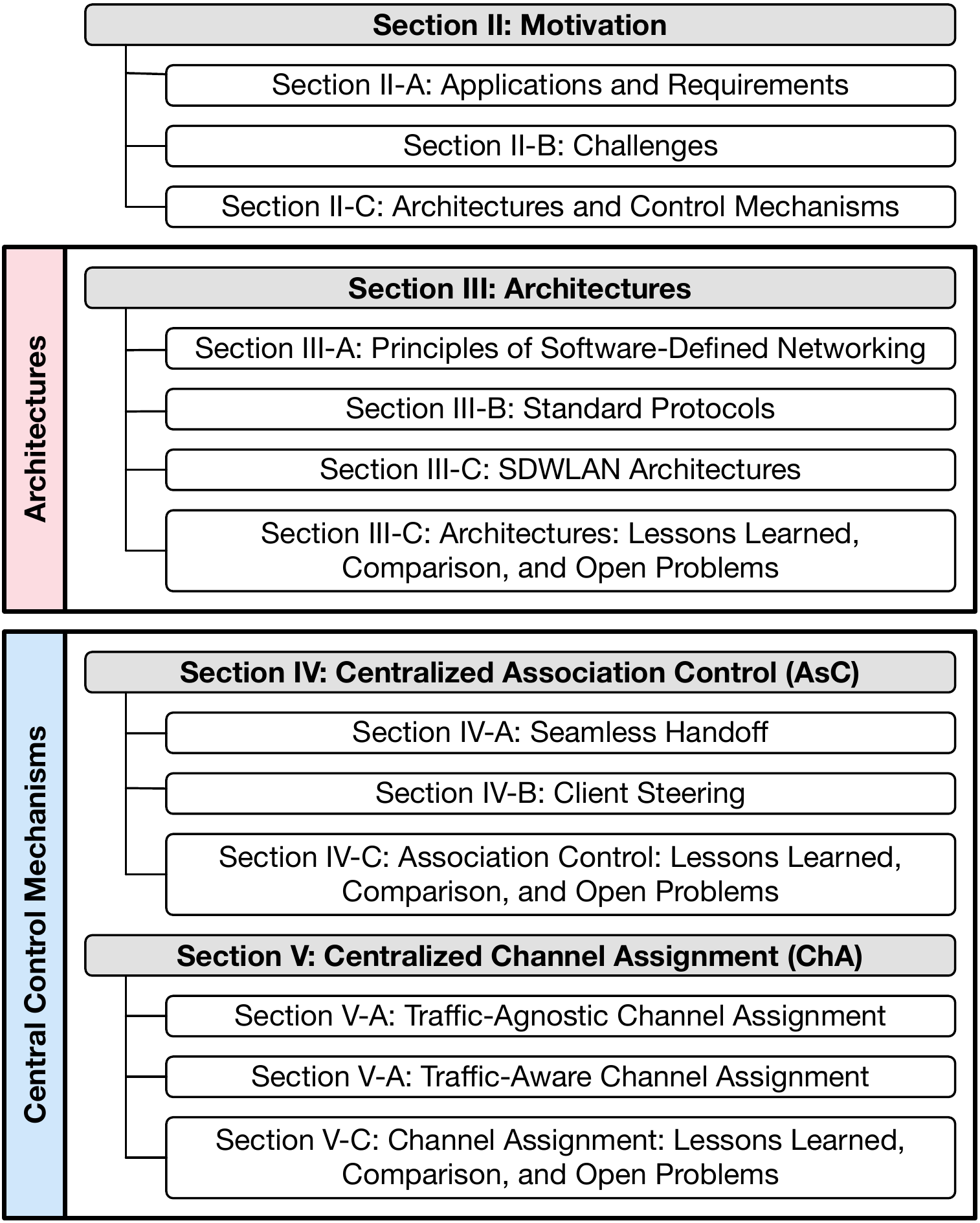}
	\caption{High-level organization of the paper.}
	\label{fig_paper_org}
\end{figure}

At a high level, this paper studies SDWLANs from two perspectives: (i) SDWLAN architectures, and (ii) the two main central network control mechanisms, AsC and ChA, that are employed to control network operation.
More precisely, the contributions of this paper are as follows:
\begin{itemize}
	\item 
	We first present an overview of software-defined networking and the standard protocols used for network monitoring, management and programming.
	Next, we review SDWLAN architectures and reveal their main objective, components, interconnection protocols, and APIs provided.
	Based on their main contribution, we classify architectures into the following categories: \textit{observability and configurability},  \textit{programmability}, \textit{virtualization}, \textit{scalability}, and \textit{traffic shaping}.
	Then, we compare these architectures and highlight future research directions considering the current and upcoming requirements of SDWLANs. 
	\item We review the AsC mechanisms proposed for SDWLANs.
	We first classify these mechanisms into two groups:  
	(i) \textit{seamless handoff}, mechanisms for reducing the overhead and delay of client handoff, 
	and (ii) \textit{client steering}, mechanisms to optimize parameters such as the airtime allocated to clients.
	Client steering approaches, in turn, are classified into two groups: 
	(i) \textit{centrally-generated hints}, where the controller relies on the global network view to generate hints for the association of clients, and (ii) \textit{centrally-made decisions}, where the controller makes the association decisions and enforces the clients to apply them.
	In addition to discussing the impact of architecture on supporting seamless handoff, we compare the reviewed mechanisms in terms of optimization scope, traffic awareness, and the ability to recognize the QoS requirements of clients. 
	By discussing the capabilities and weaknesses of AsC mechanisms, we present research directions towards designing AsC solutions for heterogeneous and dynamic SDWLANs with diverse QoS requirements. 
	We also identify how the potential features of SDWLAN architectures can be employed to design more efficient AsC mechanisms.	
	\item We review the ChA mechanisms proposed for SDWLANs.
	We classify these mechanisms into two groups, \textit{traffic-aware} and \textit{traffic-agnostic}, based on whether they incorporate APs' and clients' traffic into the decision making process.
	We specifically focus on how ChA mechanisms define and measure their interference metrics as well as the approaches employed for modeling and solving the formulated problems.
	Furthermore, we compare these mechanisms in terms of factors like client-awareness, dynamicity, and the unique features of different 802.11 standards. %(e.g., partially-overlapping channels and channel bonding). 
	We propose research directions towards designing mechanisms that address the demands of emerging applications.
	We also identify how the potentials of SDWLANs can be employed to design more efficient ChA mechanisms.

\end{itemize}
When studying AsC and ChA mechanisms, we reformulate the problems using a consistent set of notations to ease the understanding and comparison of these mechanisms.
Table \ref{acronyme-table} presents the key acronyms and notations used in this paper. 
Figure \ref{fig_paper_org} shows the high-level organization of the paper.

\begin{table}
	\centering
	\caption{Key acronyms and notations}
	\label{acronyme-table} 
	\def\arraystretch{1}
	\begin{tabular}{ m{0.32\linewidth}m{0.6\linewidth} }
		\Xhline{1\arrayrulewidth}
		\textbf{Acronym} & \textbf{Definition} \\
		\Xhline{1\arrayrulewidth}
		AsC & Association Control\\		
		AP & Access Point\\
		API & Application Programming Interface \\
		BSSID& Basic Service Set ID (AP's MAC address)\\
		ChA & Channel Assignment\\		
		CCA& Clear Channel Assessment\\
		%CHDC& Coverage Hole Detection and Correction\\
		%CUWN& Cisco Unified Wireless Network\\
		DCF & Distributed Coordination Function \\
		%FW&Firewall\\
		%LM&Local MAC\\
		ISP & Internet Service Provider \\
		LVAP&Light Virtual Access Point\\
		%MB&Middle Box\\
		MAC & Medium Access Control   \\
		NFV&Network Function Virtualization\\
		NIC&Network Interface Card\\
	    %RRM&Radio Resource Management\\
	    PHY & Physical \\
	    RSSI&Received Signal Strength Index\\	
	    SDN&Software-Defined Network\\
	    SDWLAN & Software-Defined WLAN \\
	    SINR & Signal to Interference-and-Noise Ratio \\
	    %SM&Split MAC\\
	    %TPC&Transmission Power Control\\
	    VAP&Virtual Access Point\\
	    WLAN & Wireless Local Area Network \\
	    %vMB&virtual Middle Box\\
	    %WDTX&Wireless Datapath Transmission\\
		\Xhline{1\arrayrulewidth}
		\Xhline{1\arrayrulewidth}		
		\textbf{Notation} & \textbf{Definition} \\
		\Xhline{1\arrayrulewidth}	    
		$\mathcal{C} = \{  c_{1}, ..., c_{i} \}$ & Set of clients \\
		$\mathcal{AP} =\{ AP_{1},...,AP_{i} \}$ & Set of access points \\
		$\mathcal{CH} =\{ ch_{1},...,ch_{i} \}$ & Set of channels \\		
		%$r_{AP_{i},c_{j}}$ & PHY transmission rate between $AP_{i}$ and $c_{j}$ \\			
		%$X_{AP_{i},c_{j}}$ & Association of $AP_{i}$ and $c_{j}$ \\	
				
		\Xhline{1\arrayrulewidth}
	\end{tabular}
\end{table}

\section{Motivation}
\label{motivation}
In this section, we overview the major application domains and requirements of WLANs, the challenges of wireless communication, and the importance of centralization in addressing the requirements of current and emerging applications.

\subsection{Applications and Requirements}
\label{app_req}
WLANs are being used in a variety of applications with various performance requirements.
WLANs installed in enterprises, campuses and public areas usually serve a small to medium number of stationary and mobile users.
For voice and video streams, in addition to high throughput, providing an uninterrupted service for mobile users is an important performance metric \cite{oliveira_characterizing_2016}.
For home WLANs, achieving these service requirements is more challenging due to the adjacency of multiple networks controlled individually \cite{COAP,ResFi}.

Industrial WLANs usually serve a larger number of nodes deployed for monitoring and control purposes.
In addition, while most of the nodes are static, mobility is characterized by the speed and movement scale of robot arms, rotating parts, and mobile robots \cite{li2017industrial}.
In these applications, the network infrastructure must satisfy a set of unique requirements \cite{luvisotto2017ultra,tramarin2016use,RTwifi}, as follows:
First, many of these applications require deterministic packet delivery delay.
This is particularly important when sensors and actuators communicate with controllers because the delay of the control loop directly affects control performance.
Second, a packet delivery reliability of 99\% or higher must be guaranteed to ensure the robust operation of control and monitoring systems.
In addition to sensors and actuators, industrial wireless networks also include high-rate devices such as cameras \cite{silvestre2011online}.
For example, it is important to ensure that the longer transmission duration and higher priority of video packets do not affect the timeliness of short control packets.
Similar to industrial applications, other mission-critical systems, such as medical monitoring, must satisfy a similar set of requirements \cite{REWIMO}.

Besides user devices such as smartphones, more energy-constraint devices like sensors and actuators are being used nowadays in various types of applications \cite{Tozlu2012,CYW43907,BCM4343,S1-energy-MAC,REWIMO}.
For example, wireless sensors and actuators used in industrial environments, specifically those installed in high temperature and hard-to-access areas, are usually battery-powered. 
Furthermore, a greater number of low-power IoT devices and wearables are being used for applications such as medical monitoring.
Therefore, energy is an important requirement of current and emerging applications of WLANs.

\subsection{Challenges}
\label{motiv_chall}
Wireless networks introduce a new set of network control challenges, compared to wired networks.
The broadcast nature of the wireless medium causes interference, resulting in link quality and throughput variations.
For example, an AP may under-utilize its spectrum and hardware capacity due to the high interference level caused by neighboring APs or clients \cite{DenseWLAN1,decIntf,smog}.
%Interference is caused by 802.11 equipment as well as those devices sharing the frequency band of 802.11 \cite{Dezfouli2014c} (e.g., 802.15.4 and microwave ovens). 
Interference level depends on \textit{dynamic} factors such as mobility, traffic rate, number of devices, and environmental properties (e.g., path loss, multipath effect, obstacles, etc.) \cite{Pelechrinis2011,RSSIranging,largeScaleMeas,Dezfouli2014c}.
The high path loss of wireless signals, and the inability of wireless transceivers to detect collisions, bring about new challenges such as hidden and exposed terminal problems \cite{transOpp}.

From the network control point of view, \textit{channel assignment} is a widely used mechanism to cope with interference, enhance throughput and reduce channel access delay.
Channel assignment refers to the process of assigning RF channels to APs based on various factors, such as the interference relationship between APs and clients and the traffic demand of clients.
The dynamics of wireless channel, the variability of user traffic, and variations in transmission power make channel assignment a challenging process.
This process is further complicated by the introduction of channel bonding supported by 802.11n and 802.11ac standards.
Channel bonding allows a device to combine multiple channels in order to transmit at a higher rate over a wider channel.
However, the performance benefit of this technique depends on several factors and requires collaboration among APs \cite{deek2013joint,faridi2017analysis}.
For example, using a larger bandwidth might increase contention with nearby APs and result in a longer channel access delay.

In addition to affecting interference level and the load of APs, \textit{mobility} results in connection interruption because of the delay of re-association process \cite{Odin,CloudMAC3}.
Although most of the large WLAN deployments employ a centralized user authentication scheme, the delay of association process is not completely eliminated.
In particular, when a client associates with a new AP, the AP needs to establish the data structures pertaining to the new client.
However, this process might result in communication interruption.

To increase the throughput of WLANs, extend coverage, and facilitate reliable handoff, \textit{network densification} is employed in various types of environments such as enterprises and campuses.
With this approach, the coverage range of the APs is reduced and the density of APs deployed per square meter is increased \cite{APdensity}.
Despite the potential benefits of this approach, the network dynamics caused by interference and mobility are exacerbated by densification.
For example, densification increases the number of re-associations as a client moves.
In addition, due to the limited number of channels available in the 2.4 and 5GHz bands, channel assignment becomes a more challenging process.

% As a typical enterprise WLAN includes several hundreds of APs, hardware or software updates to APs in order to address the challenges of network control would be very costly and time consuming \cite{Demand-aware-LB-Association-15,BeHop,Ethanol,DenseAP,Cai2009a}.

From the energy point of view, the 802.11 standards provide two mechanisms: power saving mode (PSM), and automatic power saving delivery (APSD).
However, the efficiency of these algorithms reduces as network density increases \cite{ting2010power,zeng2011sofa}.
For example, the concurrent wake-up of multiple clients results in collision and higher energy consumption \cite{liu2014energy}.
Furthermore, energy is wasted when a node wakes up during a bad channel condition.

% While all of the aforementioned challenges and requirements can benefit from centralized network control, it is worth noting that the interaction between these requirements further justifies the need for centralized decision making.

\subsection{Architectures and Control Mechanisms}
\label{motiv_centralization}
Central network control plays an important role to satisfy the requirements of existing and emerging applications.
First, \textit{architectures} are required to provide an infrastructure for centralized network control. 
Second, \textit{centralized control mechanisms} are necessary to make control decisions based on network-wide dynamics and requirements.
For example, utilizing centralized network control facilitates addressing the following challenges:
To offer seamless handoff, we can eliminate the need for re-association by maintaining the association information of clients at a central controller;
to enhance throughput, the controller can steer the association point of clients to balance the load of APs;
the performance of channel assignment can be improved by utilizing centralization because channel assignment is, in essence, a graph coloring problem.
In addition to client steering, seamless handoff, and channel allocation, which significantly affect throughput, delay, and energy consumption of clients, centralization enhances the adoption and performance of other techniques, such as scheduling, to further improve the aforementioned metrics \cite{zeng2011sofa,REWIMO,liu2014energy,MARS}.

% {\color{red!50!black}
% Although studies prove the higher performance of centrally-controlled wireless networks, the architectures employed by cellular networks are not applicable in WLANs.
% Cellular networks employ a dedicated control plane, which is used for control purposes, such as scheduling, QoS provisioning, power control and handoff \cite{Crowd,SDNcellular}.
% This control plane uses a significant amount of bandwidth and requires microsecond level synchronization.
% However, WLANs are meant to be simple and low cost, and such a complex control plane would not be a suitable solution \cite{FlashBack}.

% %The literature propose architectures to enable centralized and flexible control
% %\footnote{While some papers \cite{Primitives} separate network control (e.g., rate adaptation) from management (e.g., scheduling), in this paper we refer to both categories as "network control".} 
% %of WLANs in response to network dynamics and emerging applications, as reviewed in the next section.
% }

\section{Architectures}
\label{Architectures}
In this section we overview the existing SDWLAN architectures.
Before presenting these architectures, we first overview the basic concepts of software-defined networking and the standard protocols used for north and south-bound communication.
% As we will discuss later, existing architectures either rely on these standard protocols or propose proprietary protocols to address the shortcomings of standard protocols.

{\color{black!50!black}
%------------------------------------------------------------------ COLOR
\subsection{Principles of Software-Defined Networking}
\label{sdn-intro}
A software-defined network (SDN) decouples control mechanisms from data forwarding paths.
In fact, a SDN has two planes: a (i) \textit{data plane}, such as switches and APs, and a (ii) \textit{control plane}, which is a controller that runs an operating system (e.g., NOX \cite{NOX}, Floodlight \cite{Floodlight}) and mechanisms that control the network operation (e.g., AsC, ChA).
%such as association and channel assignment, routing and load balancing \cite{Ananta,feamster2014road,jain2013b4}.
%
The controller\footnote{To achieve reliability, a logical controller may represent multiple physical controllers \cite{Levin2012,SDNscalability}.} communicates with data plane equipment through a south-bound protocol (e.g., OpenFlow \cite{OpenFlow}, SNMP \cite{SNMP}), and exposes a set of north-bound APIs (e.g., REST) through which network control mechanisms are developed.
In fact, the global network view of the controller and the reprogrammability of data plane significantly simplifies the design and deployment of network control mechanisms.
Network control mechanisms are also referred to as \textit{network applications}, as they are actually applications running on the network operating system.
%

% REVISION
SDN is an enabler of \textit{network virtualization}, a.k.a., \textit{network slicing}, which refers to the abstraction, slicing and sharing of network resources \cite{wang2013network}. 
For example, an abstraction layer (e.g., FlowVisor \cite{sherwood2009flowvisor}) is used to provide applications with an isolated view of resources.
Network virtualization provides control logic isolation, as each application can see and control only the slice presented to it.
Therefore, when a SDWLAN is shared by multiple operators, network slicing enables the enforcement of multiple access policies to the users associated with various network operators.
In addition, virtual networks corresponding to various services can be established to support differentiated services and achieve higher QoS control over resources.
Virtualization also expedites the design and development of novel wireless technologies.
For example, one or multiple experiments can be run on slices of a wireless network while the production network is in operation.
Three levels of slicing are applicable to a SDWLAN: 
(i) Spectrum (a.k.a., link virtualization): requires frequency, time or space multiplexing. 
(ii) Infrastructure: the slicing of physical devices such as APs and switches. 
(iii) Network: the slicing of the network infrastructure.
%As we will see in Section \ref{Archs}, spectrum slicing is used to slice infrastructure, and infrastructure slicing is used to slice network.

Given the simplified configuration of data forwarding paths, SDN promotes the use of \textit{network function virtualization} (NFV). 
Instead of using dedicated hardware, NFV relies on the implementation of services using general computing platforms.
NFV is the virtualization of network functions such as domain name service (DNS), firewall, intrusion detection, load balancing and network address translation (NAT).
%REVISION
To this end, a network service is broken into a set of functions that are executed on general purpose hardware.
These functions can then be introduced to or moved between network devices whenever and wherever required.
As we will show later, software-defined radio (SDR) and virtual APs (VAPs) are the two most notable types of NFV in SDWLAN to offload the processing of APs into general computing platforms.

%------------------------------------------------------------------ COLOR
}

%------------------------------------------------------------------ COLOR
\subsection{Standard Protocols}
\label{Protocols}
In SDN architectures, the controller communicates with the data plane through a south-bound protocol, and provides north-bound APIs for application developers.
To avoid exposing the complexities of south-bound APIs, the north-bound APIs are usually REST-based \cite{zhou2014rest,richardson2008restful}, and network engineers use language abstractions, such as Frenetic \cite{foster2011frenetic}, Pyretic \cite{reich2013modular}, and NetCore \cite{monsanto2012compiler}, for application development \cite{SDNLang}.
For example, FloodLight \cite{Floodlight} and OpenDaylight \cite{medved2014opendaylight} both support REST north-bound APIs. 
In the rest of this section we review the widely-adopted standard south-bound protocols.

\textbf{Simple Network Management Protocol (SNMP)} \cite{case1990simple,SNMP,harrington2002rfc,stallings1998snmp}. 
This is an application layer protocol used to monitor and configure network elements.
SNMP exposes \textit{management data} in the form of variables on the \textit{managed systems}. 
The nature of the communication is fetch-store: either the manager fetches resources or stores the value of an object on the agent.
SNMP cannot directly implement actions. For example, to restart an agent, the manager cannot send a command to reboot the agent; instead, it sets the value of the reboot counter, which indicates the seconds  until next reboot.
%In addition, network elements may send push notifications to the manager.
SNMP  has  been  the  most widely used management protocol in production networks due to  its  simplistic  nature  and  ease  of  usage.
Although SNMPv3 introduces significant security enhancements, SNMP has been mostly used for monitoring, rather than configuration \cite{clemm2006network}.
Furthermore, due to the sequential execution of commands, SNMP introduces high network overhead and may result in network failure in large-scale deployments \cite{kona2002framework,silva2005defining}.

\textbf{Netconf} \cite{enns2011network}. 
%This protocol supports transaction execution, which avoids leaving a device in an unknown state if part of the configuration fails.
%This is in contrast with SNMP where instructions are executed sequentially.
Netconf uses remote procedure call (RPC) to exchange messages and apply configuration through Create, Retrieve, Update, and Delete (CRUD) operations.
In addition, it enables the manager to define what action (e.g., continue, stop, roll-back) should be taken if a command fails.
Netconf has been mostly used for configuration, and the assumption is that another protocol such as SNMP is used for monitoring purposes \cite{clemm2006network}.
YANG is a tree-structured  modelling  language used by Netconf to  describe  the  management  information of network elements.
Since YANG enables the definition of data models, Netconf has been widely used by industry; however, vendors may not support a consistent set of data models.

\textbf{OpenFlow }\cite{OpenFlow}. This protocol assumes that data plane elements are simple packet forwarding devices and their operation is determined by the forwarding rules received from the controller.
Whenever a new flow enters a switch, the forwarding table of the switch is queried to find a matching forwarding rule.
If no forwarding rule is found, either the packet is dropped or an inquiry is sent to the controller to retrieve the required \textit{action}.
OpenFlow 1.1 replaces actions with \textit{instructions}, which enables packet modification and updates actions.
The possibility of connecting to multiple controllers has been introduced in OpenFlow 1.2. 
OpenFlow 1.3 enables per-flow rate control. %this simplifies handover
%As we will see in the next section, most of the SDWLAN architecture use OpenFlow due to the flexibilities of this protocol.

\textbf{Control And Provisioning of Wireless Access Points (CAPWAP) }\cite{CAPWAP}. This protocol enables a controller to establish DTLS \cite{DTLS} connections with APs to exchange data and control messages.
Data messages encapsulate wireless frames, and control messages are used for monitoring and control purposes.
%End point devices send discover messages in search of a controller. 
%If a controller intercepts those requests, then a response is sent to each end point.
In addition to centralizing authentication and network management, CAPWAP defines two MAC operation modes \cite{shao2015ieee}: \textit{local MAC} (LM) and \textit{split-MAC} (SM). 
Using the LM mode, most of 802.11 MAC operations are implemented in APs and the role of controller is almost negligible. 
Using the SM mode, the real-time operations of 802.11 MAC are run on APs, and the rest of operations, such as the generation of beacons and probe responses, are handled by the controller.
% Unfortunately, CAPWAP does not provide any mechanism for fast handoff, virtual access points and network slicing.

\textbf{CPE WAN Management Protocol (CWMP)} \cite{TR069} (a.k.a., TR-069). Published by Broadband Forum, CWMP is a text-based protocol for communication between Auto Configuration Servers (ACS) and Customer Premise Equipment (CPE) such as modems, APs and VoIP phones.
The primary features of CWMP are secure auto-configuration, dynamic service provisioning, software/firmware image management, status and performance monitoring, and diagnostics.
One of the main capabilities of CWMP is to enable the devices behind NAT to securely discover ACSs and establish connection.
In addition, an ACS can request a session start from a CPE.
The commands (e.g., get, set, download, upload, etc.) run over a “SOAP/HTTPS” application layer protocol.
In fact, a CPE acts as a client and the ACS is the HTTP server.
Configuration and monitoring of CPEs is performed by setting and getting the device parameters, which are defined as a hierarchical structure.
CWMP is considered to be an important management protocol in M2M architectures \cite{swetina2014toward,al2015toward}.
However, this protocol does not address client mobility in WLANs.
%This protocol runs on an Auto Configuration Server (ACS). 

As OpenFlow has been primarily designed to configure the flow table of switches, an additional protocol, such as SNMP, Netconf, or a proprietary protocol, should be used to enable the control of wireless devices, as we will see in Section \ref{Archs}.
Therefore, both SNMP and Netconf are widely supported by SDN controllers (e.g., OpenDaylight, ONOS \cite{berde2014onos}).
We provide further discussion about protocols in Section \ref{arch-program}.

There are other protocols, in addition to those mentioned in this section, used for south-bound communication.
For example, although REST APIs are mostly used for north-bound interactions, REST is also employed for south-bound communication by controllers such as Ryu \cite{ryu2015framework}.
%Other protocols include Netflow/IPFix \cite{pras2009using}, CLI, CMIS/CMIP
%In addition to north-bound and south-bound protocols, west-bound protocols enable the collaboration between controllers, which is particularly important to achieve scalability.

%------------------------------------------------------------------ COLOR

\subsection{SDWLAN Architectures}
\label{Archs}
In this section we review the architectures proposed for SDWLANs.
We categorize these architectures based on the main contributed feature into five categories: \textit{observability and configurability}, \textit{programmability}, \textit{virtualization}, \textit{scalability}, and \textit{traffic shaping}. 
However, it should be noted that some of these architectures present multiple features, as we will show in Figure \ref{evolution} and Table \ref{ArchTable}.
Please note that the AsC and ChA mechanisms proposed to benefit from these architectures will be studied in Section \ref{AMmech} and \ref{CMmech}, respectively.

%% ----------------------------------------------------

%------------------------------------------------------------------ COLOR
\subsubsection{\textbf{Observability and Configurability}}
The architectures of this section provide the means for central monitoring and configuration, however, they cannot be used to develop new network control mechanisms.
% REVISION
%------------------------------------------------------------------ COLOR

\textbf{DenseAP. }
\label{DenseAP_arch}
This architecture \cite{DenseAP} addresses the challenges of densely-deployed WLANs in terms of AsC and ChA.
The two main components of this architecture are \textit{DenseAP APs} and \textit{DenseAP Controller}. 
DenseAP APs are off-the-shelf PCs running Windows operating system.
%In addition to the networking stack provided by the operating system, 
Each AP runs a \textit{DenseAP daemon}, which is a user-level service responsible for managing AP functionality.
This daemon monitors NIC and reports to DenseAP Controller periodically.
A report includes the list of associated clients, sample RSSI values, traffic pattern of clients, channel condition, and a list of new clients requesting to join the network. 
%The other responsibility of DenseAP deamon is to configure the AP based on the commands received from the DenseAP Controller.
The DenseAP Controller enables the user to tune parameters, makes control decisions, and informs the daemon to apply the configuration.
We will review the AsC mechanism of DenseAP in Section \ref{DenseAP-AM}.

\textbf{Cisco Unified Wireless Network (CUWN). }
\label{CUWNarch}
%
%
%\begin{figure}[!t]
%	\centering
%	\includegraphics[width=1\linewidth]{CAPWAP.pdf}
%	\caption{Cisco Unified Wireless Network (CUWN) \cite{Cisco} architecture. This architecture enables network monitoring and tuning of network control parameters.}
%	\label{fig_Cisco}
%\end{figure}
%
%why the control plane is extended to clients?
%
CUWN \cite{Cisco} proposes an architecture and a set of configurable services, including seamless mobility control, security and QoS provisioning.
%Figure \ref{fig_Cisco} shows this architecture \cite{Cisco}.
CUWN uses either Lightweight Access Point Protocol (LWAPP) \cite{LWAPP} or CAPWAP \cite{CAPWAP} as its south-bound protocol. 
As supported by CAPWAP, CUWN provides two MAC modes: \textit{local MAC} (LM) and \textit{split-MAC} (SM).
The Radio Resource Management (RRM) \cite{CiscoRRM} of CUWN provides several network control services.
For example, RRM establishes \textit{RF groups} and determines a leader controller for each group.
To this end, each AP periodically transmits Neighbor Discovery Protocol (NDP) messages on all channels.
NDP messages contain information such as the number of APs and clients.
All APs forward the received NDP messages to the controller, which establishes a network map, groups APs into RF Groups, and chooses a leader controller for each group. 
%Through exchanging NDP messages among controllers, different RF groups are established and the leader controllers are selected. 
We will review the AsC and ChA mechanisms of CUWN in Section \ref{CUWN_AM} and \ref{CUWN_CM}, respectively.

\subsubsection{\textbf{Programmability}}
In addition to central monitoring, these architectures expose north-bound APIs that enable the implementation of control mechanisms as applications running on a controller.
We overview these architectures as follows.

\textbf{DIRAC.}
This architecture \cite{DIRAC} proposes a proprietary south-bound protocol using two main components: a \textit{router core} (RC) running on a PC, and \textit{router agents} (RA) running on APs.

RAs receive messages from RC and convey them to the NIC's device driver.
There are three types of interactions between the RC and RAs: \textit{events}, \textit{statistics}, and \textit{actions}.
Sample events are association, re-association, and authentication, which are reported by RAs to the RC.
RAs periodically report statistics, such as packet loss rate and SNR.
The RC enforces its policy by sending actions to RAs.
For example, the RC may use \texttt{set\_retransmissions()} to limit the number of retransmissions.
The control plane of the RC has two main components: (i) \textit{management mechanisms}, which are the network applications, and (ii) a \textit{control engine}, which includes components to simplify network application development.
% (i) \textit{EventProcessor}: accepts messages from RAs and notifies interested parties, 
% (ii) \textit{StatisticsMonitor}: includes channel quality information for each client, 
% (iii) \textit{ActionProcessor}: sends a requested action to an RA, and
% (iv) \textit{RegistrationDB}: stores client-AP associations and assigns a unique ID to each client.
%The data plane forwarding engine also supports the development of various mechanisms, such as scheduling.
%DIRAC proposes a simple south-bound protocol.
%the handoff technique is mostly focused on implementation and tunneling- nothing new about decision making- also this is ols

%2008

\textbf{Trantor. }
This architecture \cite{Trantor} is an improved version of DenseAP \cite{DenseAP}, and its south-bound protocol extends the exchange of control messages with clients (in addition to APs).
These messages include packet loss estimation, the RSSI of packets received from APs, channel utilization, and neighborhood size. 
Trantor also supports "active monitoring", through which clients and APs are directed to exchange packets to measure metrics that are used by network control mechanisms.
In addition, Trantor provides a set of APIs used to control the association, channel, transmission rate and transmission power of devices.
For example, using $\texttt{associate }(AP_{i})$ a client is instructed to associate with $AP_{i}$. 
Unfortunately, no network control mechanism has been proposed to benefit from these primitives.

%%%%%% Dyson 2010
\textbf{Dyson. }
This architecture \cite{Dyson} (which is an extension of Trantor) enables both clients and APs to communicate with the controller.
%Figure \ref{fig:dyson-arch} shows this architecture.
% \begin{figure}
% 	\centering
% 	\includegraphics[width=0.9\linewidth]{Figures/Dyson-arch}
% 	\caption{Dyson \cite{Dyson} extends data plane programmability to Dyson-aware clients.}
% 	\label{fig:dyson-arch}
% \end{figure}
Clients are categorized into two groups: \textit{Dyson-aware clients} and \textit{legacy clients}. 
Only Dyson-aware clients can communicate with the controller.
The measurement APIs can be used to collect information such as: the sum of the RSSI values (total RSSI) of all received packets during a measurement window, the number of transmitted packets per PHY rate, the number of transmission failures, and the channel airtime utilization.

% \begin{itemize}
% 	\renewcommand\labelitemi{--}
% 	\item The number of received packets and total bytes 
% 	\item Sum of the RSSI values (total RSSI) of all received packets during a measurement window 
% 	%\item Three-tuple $<$\textit{source node}, number of \textit{received packets} from the source node, \textit{total RSSI} of received packets from the source node$>$
% 	\item Number of transmitted packets per PHY rate 
% 	%\item Total airtime used by packets, which is calculated as the multiplication of packet size by PHY rate
% 	\item Number of transmission failures
% 	\item Channel airtime utilization.
% \end{itemize}
%For each measurement metric, a counter is incremented for each packet received, and the average values over a measurement window are calculated through dividing the counters by the number of received packets. 

%% CUT CANDIDATE

Dyson establishes a network map based on the information collected from APs and Dyson-aware clients.
The network map provides the following components: 
(i) \textit{node locations}: reflects the location of APs (which are fixed) and clients (using \cite{WLANlocalization});
(ii) \textit{connectivity information}: a directed graph that shows from which APs/clients an AP/client can receive packets;
(iii) \textit{airtime utilization}: reflects channel utilization in the vicinity of each node;
(iv) \textit{historical measurement}: a database to store measurements.

%The controller is responsible for applying a set of policies to the current network map, making configuration decisions, and issuing commands to configure the operation of clients and APs.
% The policies are network applications that are developed by network engineers using Dyson's APIs.
Dyson's APIs include commands to configure the channel, transmission power, rate and CCA of APs and clients. 
The API set also includes commands to manage AP-client associations.
A reduced set of control APIs is used to support legacy clients. 
This set enables the controller to control the association point and operating channel of these clients. 
%In addition, APs take into account the impact of legacy clients when report their traffic level to the controller.

%\begin{itemize}
%	\renewcommand\labelitemi{--}	
%	\item \small{\texttt{SetRate(r)}}: sets the PHY rate of a client/AP to rate $r$
%	\item \small{\texttt{SetChannel(c)}}: sets the operating channel of a client/AP to channel $c$
%	\item \texttt{SetTxLevel(t)}: adjusts the transmission power of a client/AP to level $t$
%	\item \texttt{SetCCAThresh(t)}: tunes the CCA threshold of a client/AP to threshold $t$
%	\item \texttt{SetPriority(p)}: sets 802.11e priority to $p$. The 802.11e priorities are voice, video, background and best effort. %APPLIED: NOT CLEAR
%	\item \texttt{Throttle(r)}: throttles the outgoing traffic of an AP or a client at the specific rate $r$.  The Throttle command limits the rate of outgoing traffic. %APPLIED: WHAT DO YOU MEAN BY THROUTTLE- EXPLAIN BRIEFLY
%	\item \texttt{Handoff(c, ap, chan)}: hand-offs the client $c$ to access point $ap$ on channel $chan$
%	\item \texttt{AcceptClient(c)}: commands an AP to accept the client $c$ for association.
%	\item \texttt{EjectClient(c)}: commands an AP to disassociate the client $c$.	
%\end{itemize}

%In addition to the architecture proposed, \cite{Dyson} shows how the provided APIs can be used to implement mechanisms such as AsC, client-specific airtime reservation and uplink/downlink load balancing.
%For example, the authors show that Dyson is capable of throttling the traffic of clients and APs in order to adjust airtime assignment or balance uplink/downlink channel allocation.
%

%%%%% AethetFlow 2015
\textbf{$\AE$therFlow.}
\label{AEtherFlow}
This architecture \cite{AEtherFlow} simplifies data plane programmability by extending OpenFlow based on its formal specifications.
The control interfaces are categorized according to the OpenFlow specification as follows:
(i) \textit{Capabilities}: interfaces through which a controller inquires the capabilities of an AP's radio interfaces. 
These capabilities include the number of channels, transmission power levels, etc.
(ii) \textit{Configuration}: these interfaces enable the configuration of physical ports (e.g., channel, transmission power) and logical wireless ports (e.g., SSID, BSSID).
(iii) \textit{Events}: interfaces for event reporting (e.g., probe, authentication, association) to a controller. 
(iv) \textit{Statistics}: these interfaces enable a controller to query wireless related statistics.
$\AE$therFlow has been implemented in CPqD SoftSwitch \cite{ofsoftswitch}, which is installed on OpenWRT \cite{OpenWRT}.
The performance and capability of this architecture has been evaluated using a predictive AsC mechanism, which we will explain in Section \ref{AMmech}.

%%%%%% COAP 2015
\textbf{COAP. }
Coordination framework for Open APs (COAP) \cite{COAP} is a cloud-based architecture for residential deployments. 
In addition to using OpenFlow, this architecture proposes an open API to provide a cloud-based central control of home APs. 
COAP improves client QoS by running central control algorithms as well as enabling cooperation among APs.
%The proposed API set is an extension of OpenFlow API and uses the open-source Floodlight controller \cite{Floodlight}. 
A COAP AP implements three modules: (i) \textit{APConfigManager}, (ii) \textit{DiagnosticStatsReporter}, and (iii) \textit{BasicStatsReporter}. 
APConfigManager is used for AP configuration, and the other two modules provide diagnostic and basic statistics for the cloud-based COAP controller. 
The COAP controller has three modules: (i) \textit{StatsManager}, (ii) \textit{COAPManager}, and (iii) \textit{ConfigManager}, which are implemented within Floodlight. 
The controller collects the required information using the StatsManager module and performs control configurations through ConfigManager. 
The COAPManager module allows network administrators to implement new strategies and policies. 
%
%From paper OpenSDWN
%Flashback [10] proposes a control channel technique for
%WiFi networks, by allowing stations to send short control messages
%concurrently with data transmissions, without affecting throughput.
%This ensures a low overhead control plane for WiFi networks that
%is decoupled from the data plane.

%% 2016
\textbf{ResFi.}
\cite{ResFi} enables information collection and sharing in residential networks. 
\textit{ResFi Agent} is a user-space program that runs on APs and provides IP connectivity over the wired backhaul network.
This agent instructs the APs to exchange messages through which APs discover each other's IP address.
After the discovery phase, each AP establishes a secure point-to-point connection to the agents of discovered APs using the wired backbone.
Each agent can then share information with other APs or a remote controller.
For example, an AP can subscribe to the publish socket of its neighbors to receive their updates.
Using ResFi Agents on an AP does not require any kernel or driver modification as it simply relies on the interfaces provided by hostapd \cite{hostapd}.
The APIs provided enables both distributed and centralized management of residential WLANs.
%mainly evaluated based on the overhead of its security features

% ---------------------------------------------------------------------

%------------------------------------------------------------------ COLOR

\subsubsection{\textbf{Virtualization}}
\label{AP-virtualization}
In this section we review the architectures that offer virtualization to support seamless mobility, slicing, and NFV.
In particular, we discuss \textit{AP virtualization}, which refers to the mapping of multiple virtual APs to a physical AP (e.g., Odin \cite{Odin}, CloudMAC \cite{CloudMAC3}), or mapping a virtual AP to multiple physical APs (e.g., SplitAP \cite{SplitAP}, BIGAP \cite{BIGAP}).
%TODO: mention that VAP may be running concurrently on an AP, or they may be running in a round robin fashion such as WVT
%mention when each case would be more useful - for example, for mobility, slicing, etc.

%------------------------------------------------------------------ COLOR
\textbf{WVT.} Wireless Virtualization Testbed (WVT) \cite {smith2007wireless} relies on AP virtualization to run isolated experiments on a testbed.
Corresponding to each experiment, each physical AP runs a \textit{Node Agent}, where each Node Agent is controlled by a \textit{Node Handler} running on the controller. 
Node Handlers provide the user with APIs for monitoring and controlling Node Agents.
In addition to Node Agents, the APs run a \textit{Node Overseer}, which is responsible for activating the Node Agents in a round robin fashion using TDMA scheduling.
In fact, at each slot boundary, the Node Overseer records the state (e.g., SSID, channel) of the currently running Node Agent and restores the state of the next Node Agent.
Node Overseers can be configured through the \textit{Master Overseer} running on the controller.
This architecture uses an NTP time synchronization server, which is required for TDMA scheduling.

\textbf{OpenRoads}. This architecture \cite{Blueprints,yap2010openroads} adds OpenFlow to APs.
In addition, due to the limitations of OpenFlow, OpenRoads uses SNMP for configuring the APs' parameters.
Resource slicing is provided through using FlowVisor and SNMPVisor; the former slices data path, and the latter slices configuration commands through sending them to the appropriate data path element.
%Given the support for network slicing, OpenRoads has been touted as an enabler of innovation using campus networks.

%REVISION

%------------------------------------------------------------------ COLOR

\begin{figure}[!t]
	\centering
	\includegraphics[width=0.78\linewidth]{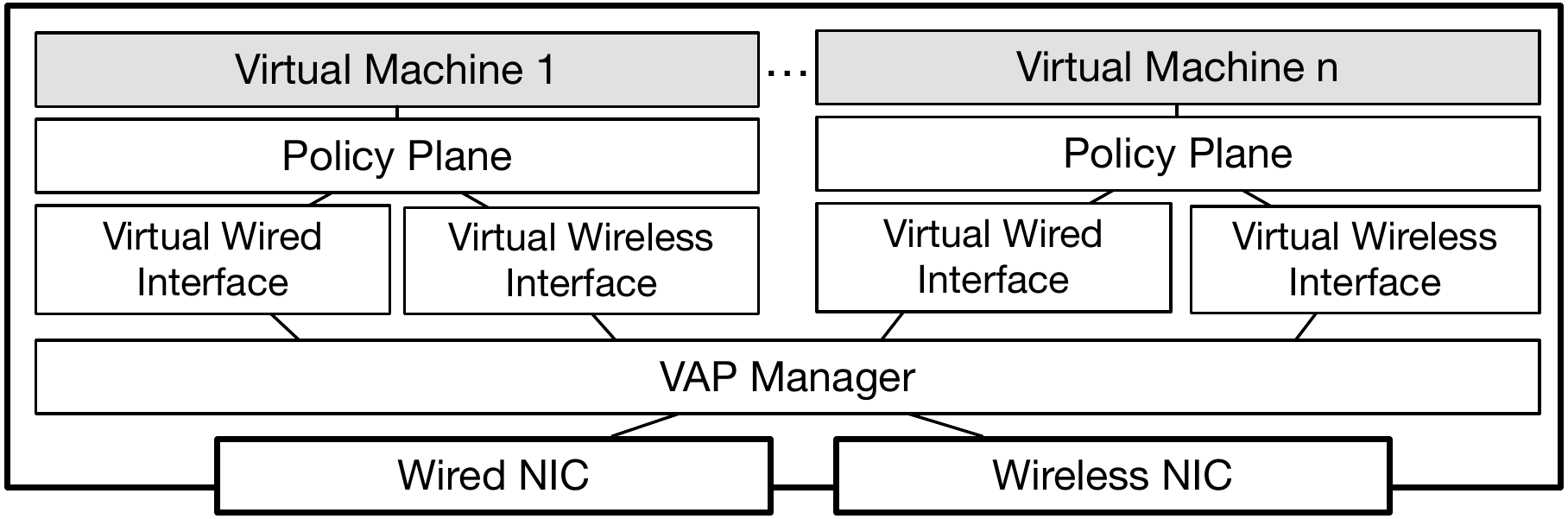}
	\caption{The architecture of an AP used by AP Aggregate \cite{nagai2011framework}. }
	\label{fig_ap_aggregate}
\end{figure}

\textbf{AP Aggregate.}
%REVISION
\label{ap_aggregate}
\cite{nagai2011framework} presents a flexible architecture that enables the aggregation of multiple virtual APs (VAPs) inside an AP, as well as slicing resources on APs.
VAP aggregation is also used for interference reduction by turning off lightly-loaded APs. 
Figure \ref{fig_ap_aggregate} shows the architecture of a physical AP.
The \textit{VAP Manager} is responsible for creating and destroying VAPs based on the commands received from the controller.
The \textit{Policy Plane} is a set of settings that are applicable to the VAP.
For example, these policies may specify the QoS and firewall settings of the VAP.
A \textit{Virtual Wireless Interface} has a SSID, MAC and IP address.
Multiple Virtual Wireless Interfaces are multiplexed into the Wireless NIC using TDMA.
Based on its collected information, a controller moves VAPs between APs.
This architecture employs layer-2 tunneling techniques when VAPs are moved between subnets.
%NOTE: the authors used VLAN and VPN to design a layer-2 flat network in multi-subnet networks.
%the authors evaluated the overhead of ap migration and the effect of l-2 flat network.

\begin{figure}[!t]
	\centering
	\includegraphics[width=0.85\linewidth]{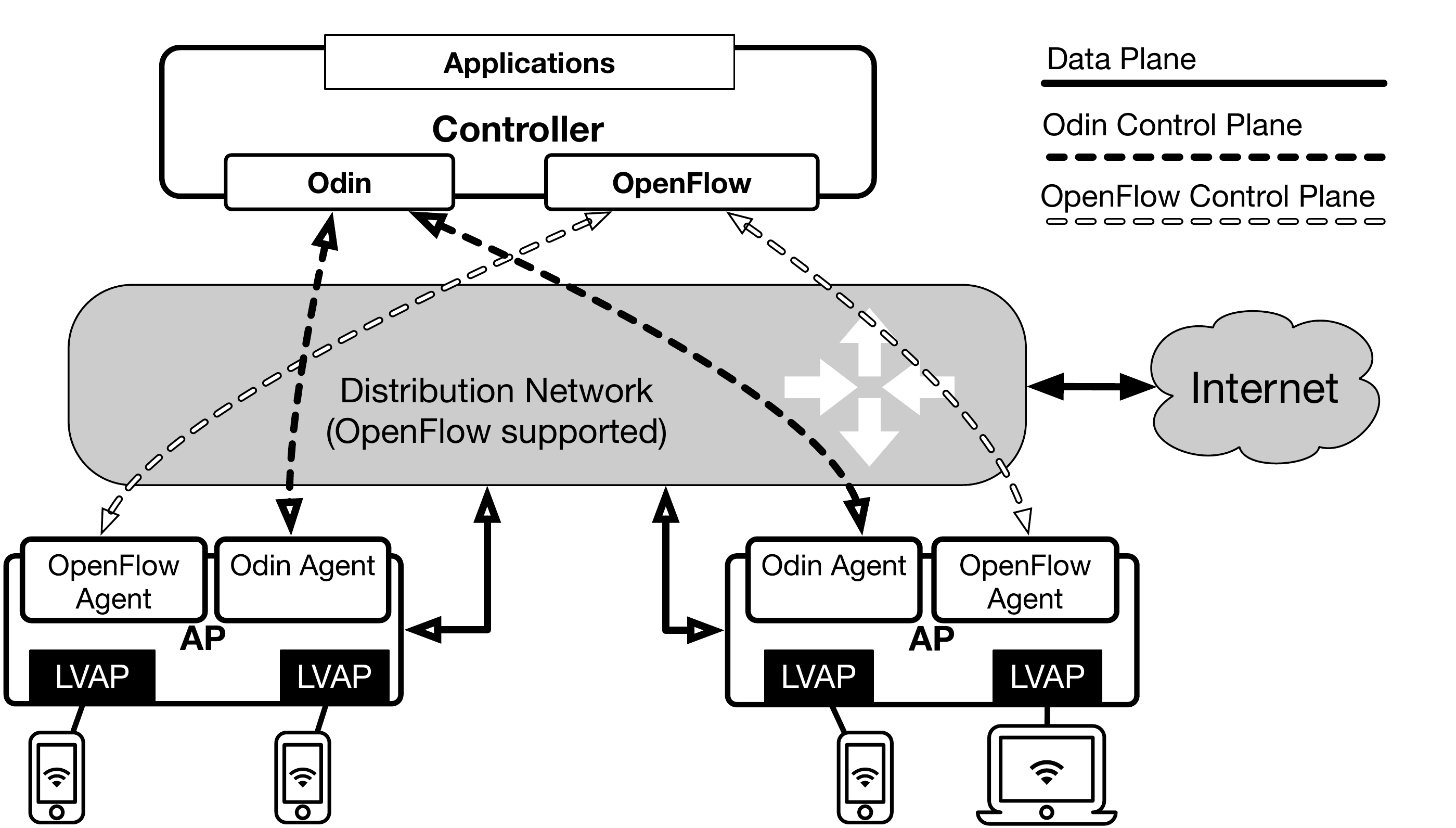}
	\caption{Odin \cite{Odin2} introduces the concept of lightweight virtual APs (LVAP) to reduce the overhead of handoffs.}
	\label{fig_Odin}
\end{figure}

%%%%%%----Odin-2012
\textbf{Odin.}
\label{Odin_arch}
Figure \ref{fig_Odin} shows the architecture of Odin \cite{Odin,Odin2,OdinThor,OdinSource}.
%\footnote{The source code of Odin is available at: \texttt{https://sdn.inet.tu-berlin.de}.}.
%
This architecture is composed of the following components: 
(i) \textit{Odin Controller}: maintains a global network view including the status of APs, clients and OpenFlow switches.
(ii) \textit{Odin Agents}: run on APs, and enable communication through a proprietary south-bound protocol. Time-critical operations (e.g., ACK transmission) are performed by APs, and non-time-critical operations (e.g., association) are handled by the controller.
This operation results in a split-MAC protocol, which is inspired by CAPWAP. 
(iii) \textit{Applications}: implemented on the controller. Applications may use the information provided by Odin Agents, OpenFlow, and SNMP.
The APIs  support the implementation of various control mechanisms such as AsC, load balancing and hidden-terminal handling.

Odin introduces the concept of \textit{light virtual AP} (LVAP) to separate clients' association states from physical APs; thereby reducing the overhead of handoffs.
An LVAP is characterized by the following four-tuple: 
\begin{itemize}
	\item client IP address,
	\item a unique virtual basic service set ID (BSSID),
	\item one or more service set ID (SSID),
	\item a set of OpenFlow rules.
\end{itemize}

For each client, an LVAP resides in an AP (hosted by Odin Agent) to represent client-AP association. 
When the controller decides to associate a client with a new AP, it can simply transfer the LVAP of that client to the new AP. 
%This process is performed without the intervention of clients, and it does not require any layer-2 or layer-3 message exchange.
From a programmer's point of view, multiple clients are connected to different ports (the LVAPs) of a physical AP. 
%More importantly, using LVAPs does not require any client-side modification.
LVAPs also simplify client authentication through adding a session key to the client's LVAP.
% Due to the limitations of OpenFlow to manage the operation of 802.11 devices \cite{Blueprints}, Odin employs the \textit{Odin protocol} to communicate with Odin Agents, and OpenFlow is used for communication with switches.
% Odin controller maintains a permanent TCP connection to Odin Agents to collect statistics and distribute configuration commands.
We will discuss about Odin's AsC mechanism in Section \ref{AMOdin}.

% Future work on the Odin 
% We are exploring further abstractions in order to support the needs of a more diverse set of network applications. 
% We also plan to explore
%           fault- tolerance and 
%           fail-over capabilities and 
%           policy management for Odin.

%%%%%%%  2010
\textbf{SplitAP.}
\cite{SplitAP} proposes an architecture to support network virtualization and manage clients' share of airtime, especially for uplink traffic.
Each AP runs a \textit{SplitAP controller}, which is responsible for VAP management and computing the uplink traffic of each slice.
By relying on the VAP concept, each physical AP broadcasts the beacons of independent virtual networks.
For example, when three different ISPs utilize an infrastructure, each AP emulates three VAPs.
This architecture requires client modification to enforce controller commands.
Specifically, each client runs a \textit{client controller} that adjusts traffic based on the commands received from the associated AP.
When a SplitAP controller realizes that the usage of a particular slice is higher than its threshold, the AP broadcasts a new maximum uplink airtime utilization that can be consumed by clients in that slice.
All the APs are connected to a shared backhaul, through which they receive channel allocation commands from a controller.

%To this end, clients are grouped as slices, and fairness of uplink airtime among different slices is achived through using LPFC and LPFC+ algorithms.

%REVISION - MOVED HERE
%%% 2013
\textbf{VAN.}
\label{VAN_arch}
Virtualized Access Network (VAN) \cite{VAN} proposes an architecture for central control of residential networks.
The architecture is composed of the following components: (i) \textit{residential APs}, (ii) a \textit{controller}, which is part of the ISP, and (iii) \textit{content provider servers} that are part of the content provider network (e.g., Netflix).
The controller provides a set of APIs through which content providers control home networks.
These APIs are suitable for different traffic classes such as video streaming and file transfer.
For example, when a content provider receives the request of a user, the content provider communicates with the controller and reserves resources for the requested data flow.

In this architecture, IP address assignment and authentication are performed centrally per user device; therefore, it is possible to transfer the association of a user device between APs.
More importantly, this architecture enables the sharing of APs' bandwidth between neighbors.
Given a flow bandwidth requirement and the neighborhood of each client, the controller associates clients to APs by employing a heuristic algorithm.

\begin{figure}[!t]
	\centering
	\includegraphics[width=1\linewidth]{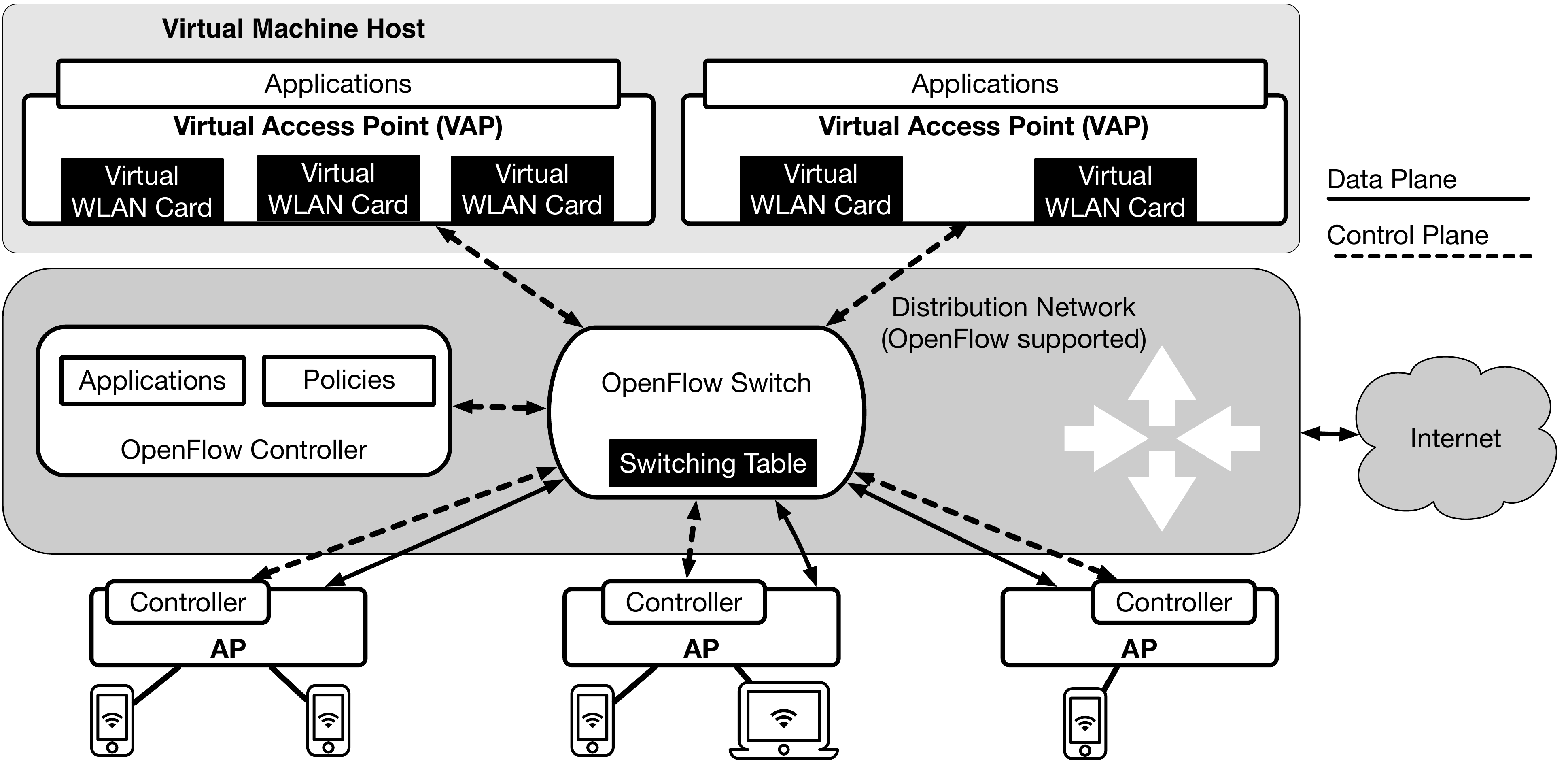}
	\caption{CloudMAC \cite{CloudMAC} reduces the complexity of APs through running VAPs on a cloud computing infrastructure.}
	\label{fig_CloudMAC}
\end{figure}

%REVISIION
%the legend size has been reduced
%%%%%%----CloudMAC-2013
\textbf{CloudMAC. }
Figure \ref{fig_CloudMAC} illustrates the CloudMAC \cite{CloudMAC,CloudMAC3} architecture. 
CloudMAC decomposes AP operations into two modules and employs OpenFlow switching tables to manage the communication between these modules.
The main components of CloudMAC are (i) APs, (ii) VAPs, (iii) an OpenFlow switch, and (iv) an OpenFlow controller. 
APs only forward MAC frames. 
All other MAC functionalities are performed by VAPs that reside in a cloud computing infrastructure. 
VAPs are operating system instances on a hypervisor such as VSphere.
Each VAP can include multiple virtual WLAN cards, where each card is implemented as a driver in a VAP; therefore, standard software tools can utilize these cards.
Virtual machines are connected to physical APs by an OpenFlow-based distribution network. 
The generation and processing of MAC frames are performed by VAPs. 
A physical AP, which is a slim AP, sends and receives raw MAC frames from clients and handles time-sensitive operations (e.g., ACK generation and re-transmission).

APs and VAPs are connected by layer-2 tunnels \cite{capsulator} and the OpenFlow switch.
The forwarding table of switches are used to decide how packets must be routed between physical APs and VAPs.
This mechanism simplifies the implementation of control mechanisms such as AsC.
For example, while Odin requires LVAP migration to handle mobility, CloudMAC simply reconfigures the switching tables to change the data forwarding path.

%From the paper: "By reconfiguring the switch table, one AP can easily be moved from one WTP to another, together with all flows passing through it. As each physical card on a WTP can be bound to multiple virtual WLAN cards (traffic can be distinguished by the BSSID and MAC addresses), CloudMAC inherently supports network virtualiza- tion."

When a control command is generated by a network application, the command is processed by VAP and a configuration packet is forwarded toward the OpenFlow switch.
The OpenFlow switch forwards the packet to the OpenFlow controller to check the legitimacy of the configuration command according to user-defined policies. 
If the command is legitimate, the OpenFlow controller forwards the packet to the AP. 
%The controller application inside the AP executes the command and returns the execution result to the OpenFlow controller and the VAP.

%Since some OpenFlow implementations (e.g., OpenVSwitch \cite{OpenvSwitch}) enable the modification of control headers, the implementation of control mechanisms (e.g., power and rate adaptation) is well supported by CloudMAC.

%An AP broadcasts the beacon messages that reduces the available network capacity for data communication. %?
%CloudMAC can enable a scenario that OpenFlow switch does not forward beacon messages from VAPs to the physical AP. 
%In this way, it is possible to have an application on OpenFlow controller to detect the probe requests of clients, and then enable the beacon messages dynamically. 
%This approach is called on-demand AP which can reduce the number of beacon messages in WLAN. 

To study the performance of CloudMAC, the round trip time (RTT) is measured between a client and a VAP. 
Due to the processing overhead of the OpenFlow switch and the delay overhead added by the tunnels, the average RTT is increased from 1.79 ms to 2.28 ms, compared to a standard WLAN.  
However, time-critical MAC frames (e.g., association response frames) are delivered fast enough to satisfy the timeliness requirements.
For large TCP packets, throughput is decreased by almost 8.5$\%$, compared to a standard WLAN. 
This performance degradation is because of the tunnels implemented in user space, which requires context switching. 
%Using kernel-space tunnels is a future work to improve performance. 
%Since a small and simple WLAN system was used to evaluate the performance of CloudMAC, it would be interesting to measure the overhead of this architecture in large-scale scenarios.

%%% LEAVE IT AFTER CLOUD MAC AS IT RELIES ON VAP CONCEPTS OF THE FIRST VERSIONS OF ODIN AND CLOUD MAD
%%%%%%% EmPower 2013

%------------------------------------------------------------------ COLOR
\textbf{EmPOWER.}
\label{EMPOWERarch}
%EmPOWER \cite{EmPOWER} is an SDN/NFV-based testbed for WLANs. 
EmPOWER \cite{EmPOWER} relies on the concept of LVAP (proposed by Odin) in order to decrease the overhead of client mobility.
The controller runs Floodlight as the operating system and FlowVisor to enable virtualization.
In this architecture, each AP is equipped with an Energino \cite{Energino} add-on, which is an open toolkit for energy monitoring.
This add-on provides REST-based APIs that enable network administrators to turn on and off the APs.

%Furthermore, it uses an add-on, named Energino \cite{Energino}, at each AP which measures the energy consumption of access point. The measurement circuit includes a current sensor that works based on the Hall effect, and a voltage sensor that works based on a voltage divider. 

\textbf{EmPOWER2.}
\label{empower2_arch}
%2015
An improved version of EmPOWER has been proposed in \cite{Primitives,EmPOWER-src}.
We refer to this architecture as EmPOWER2.
This architecture provides a full and open set of Python-based APIs by introducing four key abstractions:
(i) \textit{LVAPs}: a per-client VAP similar to Odin.
(ii) \textit{Resource pool}: each resource block is identified as $( (frequency, bandwidth), time)$. For example, an AP working on channel 36 with bandwidth 40MHz is represented as $( (36, 40), \infty)$. 
Similar notation is used to represent the 802.11 standard supported by LVAPs. This representation provides a mean for LVAP to AP mapping. 
(iii) \textit{Channel quality and interference map}: provides network programmers with a global network view in terms of the channel quality between APs and LVAPs. This view enables the programmers to allocate resources in an efficient manner.
(v) \textit{Port}: specifies the configuration of an AP-client link in terms of power, modulation and MIMO configuration.
We will study its AsC and ChA mechanisms in Section \ref{AM-InvidualOpt} and \ref{ChA_traffic_ag}, respectively. 

% the authors have presented a mechanism for monitoring uplink communications.
% They propose a passive time synchronization approach: when multiple APs receive a packet from a client, they all forward the sequence number and their timing to the controller, which performs time synchronization.
% The controller collects frame delivery statistics every 500ms to provide network applications with real-time link quality.

%------------------------------------------------------------------ COLOR
%%%%%SDWLAN 2014
\textbf{Sd-wlan.}
\label{SDWLANarch}
The main idea of Sd-wlan \cite{SDWLAN,SDWLAN2} is the implementation of most MAC functionalities at a controller, which is essentially similar to Odin \cite{Odin2}. 
%APs perform RTS/CTS exchange and acknowledgment generation.
%Other MAC operations including beaconing, probe response, client association/re-association, and client authentication/re-authentication are performed by the controller. 
To this end, an extended version of OpenFlow is proposed as the south-bound protocol. 
Using this protocol, the controller can instruct the APs whether they should send ACK packets or not.
When a client needs to change its point of association, the controller must install new rules on the new AP.
The controller also updates OpenFlow switches to direct the traffic being exchanged between the controller and APs.

%SDWLAN develops an application called \textit{virtual AP management} to provide \textit{One Big AP} illusion. 
%This application creates multiple virtual APs (VAP) in the controller and stores association-related information of all clients so that the clients will be unaware of AP handoff.  
%The authors of SDWLAN also proposed a fast handoff mechanism, which we will explain in Section \ref{SDWLANdam}. 

%%%%Three main new features of SDWLAN in comparison to Odin and CloudMAC architectures are:
%%%%\begin{itemize}
%%%%	\item Fast AP handoff: In Odin, AP handoff is performed by moving the LVAP between APs that takes some time. However, SDWLAN does the AP handoff based on the extended OpenFlow protocol that eliminates the time overhead of handoff.
%%%%	\item Central and secure key management: In Odin, LVAPs stores the keys of clients so that the mobility of clients can lead to key scattering in multiple APs. This key scattering brings about some security issues. However, SDWLAN manages all keys centrally through an en/decryption appliance.
%%%%	\item Per-client AP handoff: In CloudMAC, handoff is done by switching all the associated clients between APs and does not provide per-client AP handoff.
%%%%\end{itemize}

%%%%% BeHop 2015
\textbf{BeHop.}
%BeHop architecture \cite{BeHop} is depicted in Figure \ref{fig_BeHop}. 
%%
%\begin{figure}[!t]
%	\centering
%	\includegraphics[width=0.7\linewidth]{BeHop.pdf}
%	\caption{BeHop \cite{BeHop} architecture. Each BeHop AP creates a VAP per client.}
%	\label{fig_BeHop}
%\end{figure}
%
The main components of BeHop \cite{BeHop} are: (i) \textit{BeHop APs}, (ii) a \textit{BeHop collector}, and (iii) a \textit{BeHop controller}. 
BeHop collector is responsible for monitoring and collecting information from the network.
The controller includes network control mechanisms, in addition to processing and responding to probe, authentication and association requests.
The collector may be implemented internally (inside the controller) or externally as a separate component.
%However, the latter is proffered as decomposing collector from controller separates information plane and control plane, and eases the collection and processing of large amount of data without affecting responsiveness and stability.

A proprietary south-bound protocol has been implemented by running \textit{monitoring agents} on APs.
This protocol is used to collect and forward statistics (e.g., channel utilization, SNR, RSSI, and PHY rate) to the collector.
The controller communicates with the collector through a remote procedure call (RPC) interface to update its status about BeHop APs. 

BeHop APs create a VAP per client, and each AP maintains a \textit{client table}.
%BeHop APs compete to acquire the clients by responding to their probe requests earlier than non-BeHop APs. 
%After acquiring a client, a VAP is created per client. A VAP operates as a physical AP for clients. In other words, the clients only see the VAPs, not physical APs. 
Each entry of the client table includes client control information such as client-VAP mapping and client data rate. 
VAPs can be added to or removed from APs using the south-bound protocol implemented.
Each BeHop AP operates as an OpenFlow switch and exposes APIs for controlling channels, power, and association.
APs forward control traffic (e.g., probe, association) to the controller for further processing.
\textbf{RCWLAN}. This architecture \cite{nakauchi2012airtime} addresses the coexistence of multiple virtual networks (slices).
For each virtual network, a VAP is created on the physical APs, where each VAP has its own virtual machine and MAC queue.
The set of resources allocated to each slice is controlled by configuring the MAC parameters of the VAPs.
The extension of RCWLAN is vBS, which we explain next.

%
% 2015
\textbf{vBS.} \cite{vBS} supports service-based wireless resource reservation, which is similar to the concept of flow-based virtualization \cite{sherwood2009flowvisor}.
Specifically, by reserving resources for time-critical services (e.g., VoIP), vBS provides QoS guarantees even in the presence of background traffic.
%Figure \ref{fig:vbs} shows this architecture.
%
% \begin{figure}[t]
% 	\centering
% 	\includegraphics[width=0.95\linewidth]{Figures/vBS}
% 	\caption{vBS \cite{vBS} maps multiple physical APs to a virtual base station. %vBS1 is a common network, and vBS2 is a service-specific network. A client connected to vBS1 is handedoff to vBS2 when it initiates a service supported by vBS2.
% 	}
% 	\label{fig:vbs}
% \end{figure}
%
In contrast with SplitAP, which supports virtualization through broadcasting different ESSIDs, this architecture relies on the concept of \textit{virtual Base Station} (vBS).
A vBS is a virtual multichannel AP that uses the resources of multiple physical APs.
All of the APs share the same BSSID (MAC address) and ESSID; therefore, AP selection and handoff are completely handled by the controller.
When a network administrator registers a new service, the controller creates a vBS for that service.
When a client joins the network for the first time, the controller associates the client with an AP of the common vBS.
The common vBS provides a service for non-prioritized traffic.
When the client initiates a service, the controller performs a client handoff to the vBS of that service.
%Therefore, flows pertaining to a service are bound with the virtual network (vBS) providing that service.
The association status of the client is transferred to the new AP, and the OpenFlow switch is updated.
%When the number of clients using a service-specific vBS exceeds a threshold, then the controller assigns more APs to that vBS.

%NOTE: the handover mechanism is not for mibility. it is for handing off a client from a common AP to a dedicated AP.

%%2016
\textbf{BIGAP.}
The main goal of this architecture \cite{BIGAP} is to support seamless mobility by mapping physical APs to a single VAP.
BIGAP assumes that each AP has two wireless interfaces: one for serving as AP, and the second one for statistics collection which works in promiscuous mode to periodically overhear packets on all channels.
The collected information is used by the controller to provide a fast AsC mechanism, which will be explained in Section \ref{BIGAPhandoff}.
In order to reduce the complexities of the handoff process, BIGAP requires all of the APs to share the same BSSID.
However, to avoid collisions, BIGAP requires each AP to select a channel that is not being used by its two-hop neighborhood.
Association of a client with another AP is achieved through a channel switch command and transferring client's status to the new AP.
%Therefore, this architecture can be used with 5GHz networks where a higher number of orthogonal channels are available (25 channels). 
BIGAP provides a rich south-bound protocol for statistics collection and association control.
The AP-related APIs are implemented by modifying hostapd \cite{hostapd}, and the APIs are available as RPC by relying on ZeroRPC \cite{zerorpc} and ZeroMQ \cite{zeromq}.

\begin{figure}[!t]
	\centering
	\includegraphics[width=1\linewidth]{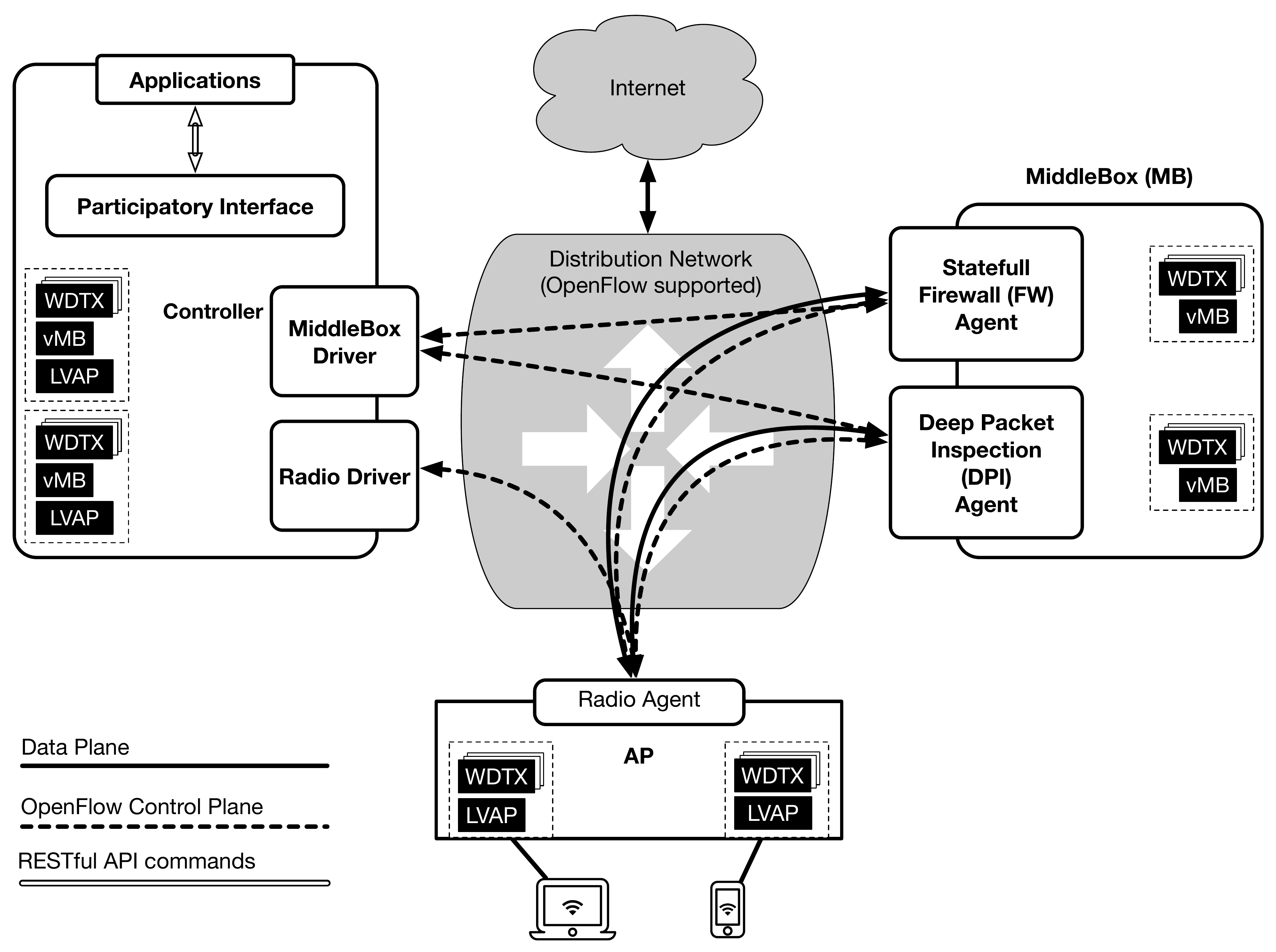}
	\caption{OpenSDWN \cite{OpenSDWN} enhances data plane programmability by introducing virtual middle boxes (vMBs).}
	\label{fig_OpenSDWN}
\end{figure}

%%%%%%% OpenSDWN 2015
\textbf{OpenSDWN.}
%This architecture \cite{OpenSDWN} enables per-flow MAC and PHY transmission settings to enhance data plane programmability.
Figure \ref{fig_OpenSDWN} illustrates this architecture \cite{OpenSDWN}. 
%This virtualization results in seamless user mobility and dynamic resource allocation.
%
%
%OpenSDWN defines per-client LVAP and vMB in order to facilitate the management of WLANs and support QoS provisioning. 
Similar to Odin, OpenSDWN provides per-client VAP; however, the new concepts of OpenSDWN are \textit{virtual middle boxes} (vMB) and \textit{wireless datapath transmission} (WDTX) rules. 
A vMB encapsulates a client's middle box state. 
Due to the small size of vMBs, their transfer across network does not introduce any significant overhead.
Each vMB includes the following information:
(i) the list of tunable parameters of the client,
(ii) the states of client's active connections,
(iii) the packet-based statistics, and
(iv) an event list that defines the behavior of the client. %(already submitted by the controller). 

The controller implements a MB driver and Radio Driver to communicate with the agents of MBs and APs, respectively. 
Through the Radio Driver, the controller manages LVAPs and collects the status of clients. 
Furthermore, the controller can get/set per-flow transmission rules through its radio interface. 
To this end, OpenFlow rules are combined with WDTX rules within APs.
WDTX rules provide per-flow service differentiation using a newly defined action that is compatible with OpenFlow protocol. 
The controller manages vMBs, gathers the statistics of vMBs, and receives the required events from different MBs. 
The latter, in particular, is supported through a publish/subscribe interface to ensure controller update when specific events occur.

A middlebox in OpenSDWN can have two types of interfaces: (i) a \textit{stateful firewall} (FW) agent and (ii) a \textit{deep packet inspection} (DPI) agent. 
The FW agent keeps track of all traffic passing through the MB in both directions and provides statistics per client and per flow. 
Also, the FW agent runs the WDTX rules defined by the controller for different flows and frames. 
The DPI agent is responsible for detecting events such as denial of service (DoS) attacks. 
The agents of a MB generate and send events that are of the interest to the controller. 
For instance, a significant change in the load of an AP may trigger sending a report to the controller. 
The controller can submit a report list to a MB in order to obtain only requested events. 

The OpenSDWN controller exposes a participatory interface that enables network applications to define flow-based and client-based priorities.
This is achieved by providing RESTful APIs through which LVAPs, WDTXs and vMBs are monitored and controlled.  

%The authors conducted experiments to reveal the effects of new features of OpenSDWN, i.e., vMB and WDTX rules, to implement WLAN control mechanisms.
%The testbed includes 25 APs (running OpenWRT) that serve more than 70 devices. 
%Three servers are used as controller, middlebox, and traffic generator.
Through empirical evaluations, the authors showed the small delay overhead of vMB migration among MBs to support mobility, and they demonstrated the effectiveness of using WDTX rules to perform service differentiation. 
%Furthermore, a case study on video-on-demand optimizer is implemented by defining different per-flow WDTX rules with the aim of giving higher priority to video streaming flows. 

%REVISION 
%------------------------------------------------------------------ COLOR
\subsubsection{\textbf{Scalability}}
\label{arch_scalability}
Scalability depends on topology, as well as factors such as the overhead of south-bound protocol, the virtualization mechanisms employed, and the traffic and mobility pattern of clients.
The scalability of SDWLANs is particularly important due to the dynamic nature of these networks.
Specifically, an architecture with a high control plane delay cannot be used for quick reactions to network dynamics.

%%%%%%% AeroFlux 2014
\textbf{AeroFlux.}
This architecture \cite{AeroFlux,AeroFlux2} highlights that per-flow or per-packet control mechanisms require short and bounded control plane delay.
For example, implementing rate and power control mechanisms in a controller may result in an overloaded and high-latency control plane.
To address this concern, AeroFlux employs a 2-tier control plane: \textit{Global Control }(GC) plane, and \textit{Near-Sighted Control} (NSC) plane. 
GC handles non real-time tasks and operations that require a global network knowledge.
For example, authentication, large-scale mobility, and load balancing are handled by the GC.
NSCs are located close to the APs to perform time-critical operations, such as rate control and power adjustment, per packet.
For example, in the case of video streaming, AeroFlux can configure APs to use lower rates and higher transmission power values for key frames, compared to regular frames.

\textbf{CUWN.} This architecture \cite{Cisco} employs split-MAC as well as a multi-tier controller topology to address scalability.
Based on the RF locations of APs, CUWN establishes \textit{RF groups} and determines a leader controller for each RF group. %leader?
All APs connected to a controller belong to an RF group. 
In this way, the controllers are aware of RF location of APs and their interference relationship.

\textbf{Odin and CloudMAC.}
These architectures \cite{Odin,CloudMAC3} rely on the the split-MAC technique proposed by CAPWAP.
In other words, to mitigate the negative effect of controller-AP communication, time-critical operations are handled by APs, and non time-critical operations are offloaded to the controller.
In addition to utilizing split-MAC, during an association, CloudMAC prevents the overhead of moving VAPs between APs through configuring the switches of the distribution network to forward the client's data to another AP.

%------------------------------------------------------------------ COLOR
%REVISION : NOT SURE IF THIS SHOULD BE TRAFFIC SHAPING OR THE CURRENT TITLE 
\subsubsection{\textbf{Traffic Shaping}}
The architectures of this section offer more than the regular programmability used for configuring APs and switches.
In fact, these architectures extend data plane programmability by enabling traffic shaping, which is used for purposes such as scheduling and scalability.

%2009
\textbf{CENTAUR.} \cite{CENTAUR } can be added to SDWLAN architectures to improve the operation of the data plane in terms of channel access and contention resolution.
When data traffic passes through a controller or programmable middleboxes, data plane programmability can be supported without AP modification.
To this end, CENTAUR proposes central packet scheduling mechanisms to avoid hidden-terminal transmissions and exploit the exposed-terminal condition.
For example, when the transmission of two APs to their associated clients may cause a hidden-terminal collision, the controller carefully adjusts the interval between packet transmissions to avoid the problem.
A salient feature of CENTAUR is that it requires minor changes to APs and no client modification is required.
CENTAUR improves UDP and TCP throughput by around 46\% and 61.5\%, compared to DCF \cite{bianchi2005remarks}.

\textbf{CloudMAC}. This architecture \cite{CloudMAC} is capable of implementing downlink scheduling by configuring OpenFlow switches to use simple rate shaping or time division algorithms.
For instance, during a time slot, the switch may forward only the packets of a specific physical AP while queuing the packets of other APs. 
Switching rules may be changed per time slot to provide a time division protocol.

%%%%%% Ethanol 2015
\textbf{Ethanol.}
\label{EthanolArch}
%This work proposes APIs for handling client mobility, QoS, security and AP virtualization. 
Ethanol \cite{Ethanol} is similar to its predecessors (e.g., \cite{Odin,CloudMAC}) in terms of implementing slim APs and shifting most of the MAC functionalities to a controller.
However, Ethanol argues that the original OpenFlow protocol cannot be used for QoS provisioning in wireless networks.
Accordingly, to complement OpenFlow, the Ethanol architecture proposes a customized protocol, named \textit{Ethanol protocol}, to provide control interface for wireless components and QoS control.
Therefore, an Ethanol AP provides two interfaces: (i) Ethanol, through an \textit{Ethanol Agent,} and (ii) OpenFlow. 
The Ethanol Agent provides APIs for QoS control on the APs.
Ethanol uses the APIs provided by the Ethanol Agent to exploit Hierarchical Token Bucket (HTB) \cite{Pantou} scheduling in order to perform per-flow programmability.
Specifically, different queues are defined on Ethanol APs. 
When a flow arrives, it is assigned to the proper queue based on its class of service (e.g., voice, video).

Ethanol \cite{Ethanol} provides the details of API implementation following an object-oriented approach.
\textit{Entities} are defined as physical or virtual objects that could be configured or queried.
Each entity has its own properties.
For example, an AP is a physical entity that includes properties such as beaconing interval.
A flow is a virtual entity that includes properties such as a packet counter.
The controller communicates with entities through their get/set interfaces. 
Entities may also include events, which trigger the controller to take proper actions.
Ethanol API has been implemented on OpenWRT \cite{OpenWRT} by exploiting Pantou \cite{Pantou}, a software package enabling OpenFlow on OpenWRT.

%The authors conducted an experiment where three clients use an Ethanol AP.
%Three queues for three clients are defined with rates proportional to their traffic type.
%The different rate of queues enables bandwidth allocation to clients proportional to their required services. 

% %% ---------------------------------------------------------------------
% {\color{blue!50!black}
% %------------------------------------------------------------------ COLOR
% \subsubsection{\textbf{Home Networks}}
% \R{arch_home_class}
% %   REVISION : THERE ARE A FEW MORE PAPERS THAT WE COULD ADD HERE - REFER TO THE PAPER CITED IN THE FEW LINES BELOW
% The increase in the number of residential APs as well as the need to support emerging applications (e.g., video surveillance, medical monitoring, IoT) necessitate centralized control of these networks in order to achieve the QoS parameters desired.
% %For example, a recent study \cite{largeScaleMeas} shows that AP neighborhood size is around seventeen.
% %As a comprehensive review of residential networks has been recently published in \cite{wSDN1}, 
% In this section we review the state-of-the-art architectures proposed for home networks.
% }

%
%REVISION

%------------------------------------------------------------------ COLOR
% ** CONSIDER CHANGING THE TITLE TO "COMPARISON. LESSONS LEARNED, AND OPEN PROBLEMS"
\subsection{Architectures: Learned Lessons, Comparison, and Open Problems}
\label{archComp}
% REVISION:
% refer to the table here and add a summary of the objectives of the architectures and how did they achieve it

\begin{figure*}[!t]
	\centering
	\includegraphics[width=0.87\linewidth]{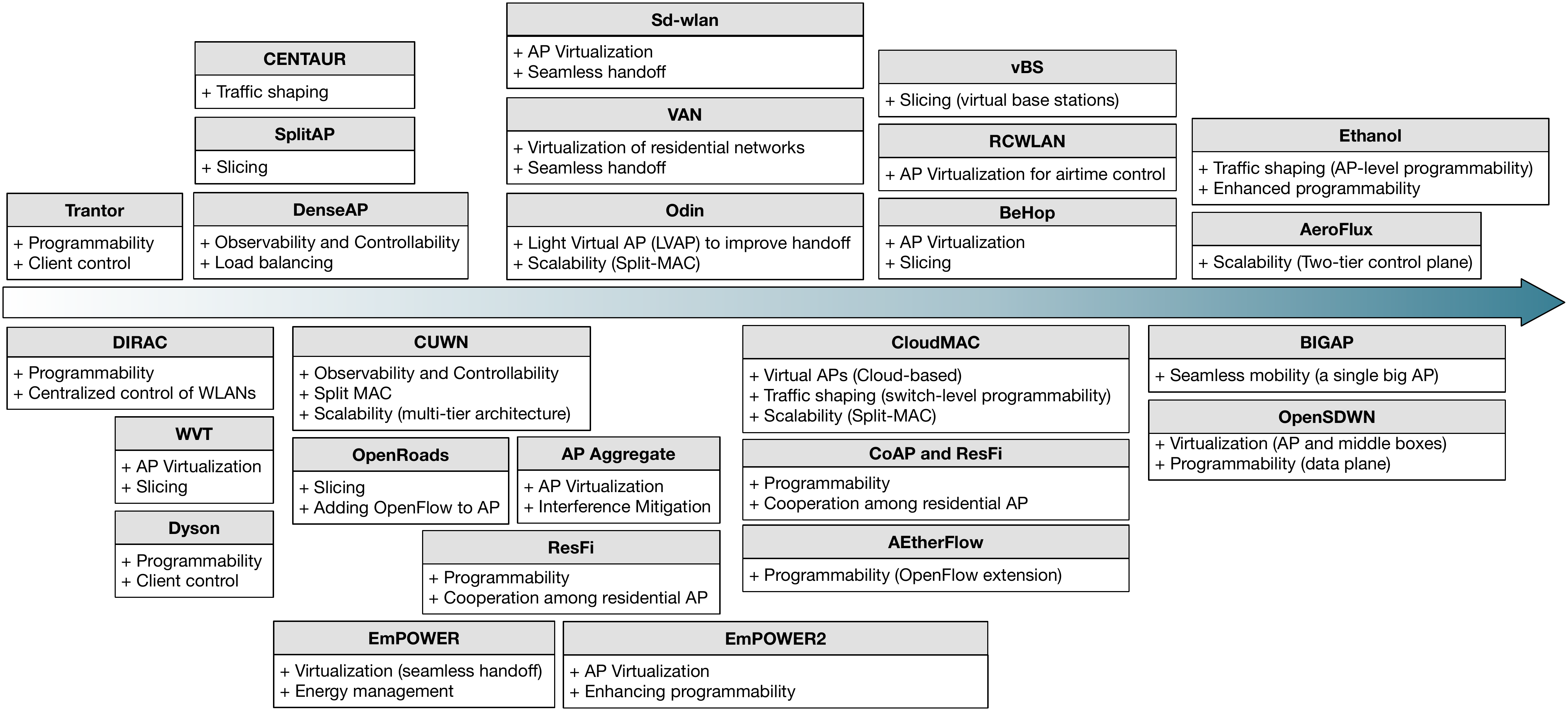}
	\caption{This figure summarizes the main features of SDWLAN architectures based on the categorization presented in Section \ref{Archs}.}
	\label{evolution}
\end{figure*}

Figure \ref{evolution} highlights the main features and Table \ref{ArchTable} summarizes the properties of reviewed architectures.
In this section, we summarize the learned lessons, study the proposed features and identify future research directions and potential solutions.
%
%

% PROGRAMMABILITY 
% REVISION minor changes

\subsubsection{\textbf{Programmability}}
\label{arch-program}
A \textit{reconfigurable architecture} only enables the adjustment of parameters pertaining to a set of predefined network control mechanisms. 
For example, CUWN's APs are like regular APs with added CAPWAP support; thereby, it is not possible to implement a new AsC mechanism as CUWN employs a set of proprietary mechanisms for network control.
In fact, we can argue that CUWN and DenseAP introduce a \textit{management plane} instead of a control plane.
On the other hand, a \textit{programmable architecture}, such as OpenSDWN and Odin, provides north-bound APIs through which control mechanisms running on the network operating system are developed.

%DenseAP is a centrally-controlled architecture which is not an open architecture, and Trantor is an open centralized architecture with a limited set of high-level interfaces, compared to Dyson. 

After the standardization of OpenFlow, most SDWLAN architectures have included this standard in their south-bound communication, as Table \ref{ArchTable} shows.
However, this protocol has been mainly designed to configure the flow tables of switches; therefore, it cannot be used for the configuration of wireless data plane equipment.
%For example, in addition to using OpenFlow, BeHop utilizes a proprietary protocol for channel, power and association control.
Furthermore, since OpenFlow is very low-level, too many implementation details are exposed to network programmers \cite{Primitives}.
Therefore, we can observe the introduction of proprietary protocols (e.g., Odin, Ethanol) and the extensions of OpenFlow (e.g., AeroFlux, $\AE$therFlow).
However, this results in interoperability issues.
Another shortcoming of OpenFlow is the lack of supporting transactions, which means the operation of a network device during receiving updates is unpredictable.
This has a particular implication on wireless network control mechanisms.
For example, when an AsC mechanism sends updates to multiple APs, network performance may significantly drop during the update process due to the inconsistency of APs' software.
Despite the significant number of studies on the performance and improvement of OpenFlow for wired networks \cite{Levin2012,jarschel2011modeling,vanbever2013hotswap,lara2014network}, there is a very limited study on the use of OpenFlow and other standard protocols (e.g., CWMP, SNMP, Netconf) for SDWLAN design\cite{Primitives,AEtherFlow,rao2015towards,slabicki2015performance}.
%In addition, the abstraction level provided by OpenFlow is unnecessarily complex.
In addition, it is not clear what programmability features are required to develop architectures that include recent high throughput standards.
For example, supporting the 802.11ad \cite{nitsche2014ieee} standard requires additional interfaces for central coordination and training of antennas.

Our review shows that data plane programmability is supported at four levels: controller (e.g., CENTAUR), switches (e.g., CloudMAC), APs (e.g., Ethanol), and clients (e.g., Dyson).
%For example, CENTAUR implements data plane scheduling at the controller.
While controller-based approaches provide higher flexibility, they pose scalability and processing challenges.
Specifically, the variations of data plane delay may inadvertently affect the accuracy of implemented mechanisms. 
For example, deterministic controller-AP delay is an essential requirement when packet scheduling is implemented at a controller in order to avoid collisions caused by concurrent downlink transmissions.
On the other hand, AP-based approaches require low and bounded control plane delay to ensure prompt enforcement of the decisions made centrally.
Meanwhile, the programmability of switching equipment can be employed to improve scalability and cope with overhead issues.
For example, when a client is associated with a new AP, CloudMAC configures the switches of the distribution network to forward the client's data to another AP, thereby preventing the overhead of moving VAPs between APs.
We believe that investigating the pros and cons of these design approaches is necessary, and mechanisms are required to integrate the benefits of these approaches.

AP programmability at layer-1 and layer-2 requires the use of software-defined radio (SDR) platforms such as OpenRadio \cite{OpenRadio}, Atomix \cite{Atomix} or Sora \cite{Sora}.
When integrated into a SDWLAN architecture, using SDRs enables us to offer a richer set of programmability options to network application developers.
Additionally, integrating SDRs with SDWLANs simplifies the upgrade of communication standards when the SDR hardware is capable of supporting the new technology.
For example, when the digital signal processor (DSP) used on a SDR is fast enough to support new modulation and coding schemes, the controller can update APs to match the capabilities of clients.
Furthermore, when signal processing operations are offloaded to a cloud platform, advanced signal processing techniques may be applied to the incoming signals to cope with interference.
While SDWLAN-SDR integration is an ongoing research trend, it is also important to study the implications of SDR platforms (e.g., FPGA, DSP, general-purpose processor) on architecture in terms of factors such as control plane delay and overhead.

Although most architectures only rely on the information collected from APs, some architectures, such as Dyson and SplitAP, also require client modification to collect information from the clients' point of view.
Extending data plane programmability to clients enables the use of more sophisticated control mechanisms.
For example, this would enable the control plane to adjust clients' MAC parameters based on network dynamics and user demands, which also simplifies resource allocation and network slicing.
However, in addition to client modification, which may not be desirable for public access networks, extending control plane to clients introduces traffic overhead.
Although newly proposed mechanisms such as FlashBack \cite{FlashBack} enable low-overhead exchange of control information with clients, the integration of these mechanisms with SDWLANs is not clear. 
Accordingly, we believe that client programmability has not been well studied in the literature.
Additionally, though the 802.11r and 802.11k amendments (which have been recently integrated into 802.11 standard \cite{802.11_2016}) realize richer interactions between clients and APs, the SDWLAN architectures did not benefit from these features.

\begin{table*}
	\centering
	\scriptsize
	\caption{Comparison of SDWLAN architectures}
	\label{ArchTable} 
	\def\arraystretch{1}
	\begin{tabular}{|c|c|c|c|c|c|c|c|c|c|}
		\Xhline{3\arrayrulewidth}
		\multirow{2}{*}{\textbf{Architecture}}&
		\multirow{2}{*}{\textbf{\makecell{Programmable \\Architecture}}}&
		\multirow{2}{*}{\textbf{\makecell{Details of \\Provided APIs}}}&
		\multirow{2}{*}{\textbf{\makecell{VAP}}}&
		\multirow{2}{*}{\textbf{\makecell{Network\\Slicing}}}&		
		\multirow{2}{*}{\textbf{\makecell{South-bound  \\Protocol}} }&
		\multirow{2}{*}{\textbf{\makecell{Split-MAC \\Support}}}&
		\multirow{2}{*}{\textbf{\makecell{Client \\Modification }}}&		
		\multicolumn{2}{c|}{Implemented Mechanisms}\\\cline{9-10}								
		&&&&&&&& \quad\quad \textbf{AsC} \quad\quad & \textbf{ChA}\\		\Xhline{3\arrayrulewidth}
		
		DenseAP \cite{DenseAP}&$\times$&$\times$&$\times$&$\times$&Proprietary&$\times$& $\times$ &$\checkmark$&$\times$\\\hline
        % SB X + Y    NB: Z
		CUWN \cite{Cisco}&$\times$&$\times$&$\times$&$\times$&CAPWAP&$\checkmark$&$\times$&$\checkmark$&$\checkmark$\\\hline

		DIRAC \cite{DIRAC} &$\checkmark$&$\checkmark$&$\times$&$\times$& Proprietary &$\times$& $\times$ &$\checkmark$&$\times$\\\hline
		
		Trantor \cite{Trantor}&\textbf{$\checkmark$}&$\checkmark$&$\times$&$\times$&Proprietary&$\times$&$\checkmark$&$\times$&$\times$\\\hline
		Dyson \cite{Dyson}&$\checkmark$&$\checkmark$&$\times$&$\times$&Proprietary&$\times$&$\checkmark$&$\checkmark$&$\times$\\\hline
		$\AE$therFlow  \cite{AEtherFlow}&$\checkmark$&$\checkmark$&$\times$&$\times$& Extended OpenFlow  &$\times$&$\times$&$\checkmark$&$\times$\\\hline		
        %it is not clear if they support VAP or not. they say mapping is possible, but the paper is very unclear
        %
		WVT \cite{smith2007wireless}&$\times$&$\times$&$\checkmark$& $\checkmark$ & Proprietary &$\times$&$\times$&$\times$&$\times$\\\hline
		AP Aggregate \cite{nagai2011framework}&$\times$&$\times$&$\checkmark$& $\checkmark$ & Proprietary &$\times$&$\times$&$\checkmark$&$\times$\\\hline
		OpenRoads \cite{Blueprints,yap2010openroads} & $\checkmark$&$\checkmark$&$\times$& $\checkmark$ & OpenFlow+SNMP &$\times$&$\times$&$\checkmark$&$\times$\\\hline
		Odin \cite{Odin,Odin2,OdinThor}&$\checkmark$&$\checkmark$&$\checkmark$&$\checkmark$& OpenFlow+Proprietary &$\checkmark$&$\times$&$\checkmark$&$\times$\\\hline
		SplitAP \cite{SplitAP}&$\times$&$\times$&$\checkmark$&$\checkmark$& Proprietary &$\times$&$\checkmark$&$\times$&$\times$\\\hline		
		CloudMAC \cite{CloudMAC,CloudMAC3}&$\checkmark$&$\times$&$\checkmark$&$\checkmark$& OpenFlow &$\checkmark$&$\times$&$\checkmark$&$\checkmark$\\\hline				
		EmPOWER \cite{EmPOWER}&$\checkmark$&$\times$&$\checkmark$&$\checkmark$& OpenFlow+REST &$\checkmark$&$\times$&$\checkmark$&$\times$\\\hline		
		EmPOWER2 \cite{Primitives}&$\checkmark$&$\checkmark$&$\checkmark$&$\checkmark$& OpenFlow &$\checkmark$&$\times$&$\checkmark$&$\checkmark$\\\hline				
		Sd-wlan \cite{SDWLAN}&$\checkmark$&$\times$&$\checkmark$&$\times$& Extended OpenFlow &$\checkmark$&$\times$&$\checkmark$&$\times$\\\hline
		BeHop \cite{BeHop}&$\checkmark$&$\times$&$\checkmark$&$\times$& OpenFlow+Proprietary &$\checkmark$&$\times$&$\times$&$\times$\\\hline
		%		% a configuration API for channel and power allocation --> No details about API !!!
		%
		RCWLAN \cite{nakauchi2012airtime}&$\times$&$\times$&$\checkmark$&$\checkmark$& Proprietary &$\times$&$\times$&$\times$&$\times$\\\hline	
		vBS \cite{vBS}&$\checkmark$&$\times$&$\checkmark$&$\checkmark$& OpenFlow &$\times$&$\times$&$\checkmark$&$\times$\\\hline				
		BIGAP \cite{BIGAP}&$\checkmark$&$\checkmark$&$\checkmark$&$\times$& Proprietary &$\times$&$\times$&$\checkmark$&$\checkmark$\\\hline		
		AeroFlux \cite{AeroFlux}&$\checkmark$&$\checkmark$&$\checkmark$&$\times$& Extended OpenFlow &$\checkmark$&$\times$&$\checkmark$&$\times$\\\hline		
		Ethanol \cite{Ethanol}&$\checkmark$&$\checkmark$&$\checkmark$&$\times$& OpenFlow+Proprietary &$\checkmark$&$\times$&$\checkmark$&$\times$\\\hline
		OpenSDWN \cite{OpenSDWN}&$\checkmark$&$\checkmark$&$\checkmark$&$\checkmark$& OpenFlow+REST  &$\checkmark$&$\times$&$\checkmark$&$\times$\\\hline
		VAN \cite{VAN} &$\checkmark$&$\checkmark$&$\times$&$\checkmark$& OpenFlow &$\times$&$\times$&$\times$&$\times$\\\hline	
		%		%		
		%		
		COAP \cite{COAP}&$\checkmark$&$\checkmark$&$\times$&$\times$& OpenFlow+Proprietary  &$\times$&$\times$&$\times$&$\checkmark$\\\hline
		%		
		%		%		%
		%
		%		%
		ResFi \cite{ResFi}&$\checkmark$&$\checkmark$&$\checkmark$&$\times$& Proprietary &$\times$&$\times$&$\checkmark$&$\checkmark$\\\Xhline{3\arrayrulewidth}		
		%\\
	\end{tabular}
\end{table*}

\subsubsection{\textbf{Mobility}}
\label{seam_mobility}
Mapping VAPs to physical APs is the main technique used by architectures to support seamless mobility.
As each VAP is represented by a simple data structure that holds client's status, exchanging VAPs between APs removes the burden of re-association, as offered by Odin.
To avoid the overhead of VAP migration, CloudMAC settles the APs in cloud servers, and programs the switches to route data between physical APs and VAPs.
However, performance evaluation of CloudMAC shows that, when traffic  is  transmitted  through  a  shared network, high-priority packets may be lost or delayed, thereby reducing the chance of connection establishment to 55\%.
%To remedy90this problem, CloudMAC extends OpenVSwitch [82] (running91on switches) to incorporate queue management strategies
Although queue management has been proposed as a remedy, the effectiveness of such approaches for different topologies, mobility, and traffic patterns, is an open research area.
Meanwhile, we believe that the match-action paradigm of OpenFlow must be exploited to improve scalability and handoff performance through designing traffic shaping and prioritization mechanisms that ensure the real-time delivery of time-critical packets.
However, the literature does not present any relevant contribution.
Another solution is to use network slicing, as we will explain in Section \ref{net_vir_dis}.
On the other hand, most of the reviewed architectures (except CUWN and AeroFlux) use a single-tier controller topology, and unfortunately, they do not present an evaluation of handoff performance in a real-world scenario with tens of APs, background traffic and variable mobility patterns.
For example, Odin, 	$\AE$therFlow and BIGAP use testbeds with 10, 2 and 2 APs, respectively.
Therefore, a study of architectural implications on handoff performance is missing.
We present further discussion about the challenges of mobility management after the review of central control mechanisms in Sections \ref{AscProblems} and \ref{ChAProblems}.

% VIRTUALIZATION 
% REVISION

%%%%%
\subsubsection{\textbf{Network Virtualization}}
\label{net_vir_dis}
%REVISION
Our review shows that the most common approach of network slicing is through the use of VAPs.
For example, in Odin, a slice is defined as a set of APs (Odin Agents), LVAPs corresponding to clients, SSIDs, and network applications.
Each SSID may belong to one or multiple slices, and each client joins the slice to which the SSID belongs to.
Since each application can only see the LVAPs belonging to its slice, slice isolation is logically provided through the concept of LVAPs.

Although VAP migration can be used for slicing airtime, bandwidth and processing power, it does not provide a fine granularity for resource allocation.
For example, if the controller moves a client's VAP to another AP in order to increase its available bandwidth, the actual bandwidth offered depends on the activity of other associated clients.
In order to achieve high granularity, layer-1 and layer-2 resource slicing techniques are required to support airtime, frequency and space multiplexing.
For example, a middleware should manage the access of VAPs to the AP's physical resources through TDMA scheduling.
However, compared to wired networks, the highly variable nature of wireless communications makes predictable slicing a challenging task.
Specifically, it is hard to predict and manage the effect of one slice on another. 
For example, user mobility and variations of traffic in one slice affect the available bandwidth experienced by the users in another slice \cite{nakauchi2012airtime}.
Compared to LTE networks, resource slicing is more challenging in 802.11 networks because 802.11 relies on random access mechanisms (i.e., CSMA) and does not utilize a dedicated control plane to communicate with clients.
The problem is exacerbated when multiple networks managed by different entities coexist.
To cope with these challenges, the existing network slicing techniques heavily rely on a localized network view \cite{SplitAP,ViFi}.
For example, while SplitAP supports central network control, each AP decides about its uplink airtime allocation individually and through a distributed algorithm.
Based on this discussion, utilizing algorithms that rely on global network view, and the integration of centralized AsC and virtualization are necessary towards improving the capacity and QoS of virtualized networks.

Network slicing can also be used to tackle the challenges of control and data plane communication.
To this end, an abstraction layer slices the resources of switching elements based on the communication demands of higher layer control mechanisms.
For example, based on the mobility pattern and number of clients, an AsC mechanism may request the abstraction layer for the allocation of switching resources.
This is an open research area.

%Although mechanisms such as ViFi \cite{ViFi} address these challenges in a distributed manner, centralized performance measurement and dynamic resources slicing and allocation mechanisms are open research challenges.
%Layer-1 isolation mechanisms for SDWLANs is an open research challenge.

Network slicing would also enable the coexistence of M2M communication with regular user generated traffic, similar to the 5G vision given in \cite{alliance20155g}.
For example, an IoT device may utilize multiple slices of a network to transmit flows with different QoS requirements.
Developing such architectures and their associated control mechanisms is an open research area.
We present further discussion about this after the review of control mechanisms, in Section \ref{ch_disc_int_virt}.

%In addition, to achieve client-level virtualization, the  

\subsubsection{\textbf{Network Function Virtualization}}
\label{nfv_dis}
%---
The most common usage of NFV in SDWLANs is split-MAC, which is mainly implemented through VAP.
Split-MAC was first introduced by the split-MAC protocol implemented in CAPWAP. 
As an another example, CloudMAC separates MAC functionalities between physical AP and cloud platform.
Using the split-MAC strategy, time critical operations run on APs, while other MAC operations are handled by the controller. 
This strategy enables exploiting the resources of remote computing platforms and simplifies MAC and security updates without requiring to replace APs as either no or minimal changes to APs is required.
Despite the benefits of VAPs, it is important to establish a balance between flexibility and communication overhead because, when more functionalities are shifted to a controller, the overhead and delay of distribution system increase.
For example, even for small networks, the importance of utilizing queuing strategies with OpenFlow switches has been demonstrated \cite{CloudMAC3,Vestin2015b,Ethanol}.
Consequently, studies are required to show the tradeoffs between centralization and scalability.

SDR is an another example of NFV.
SDR platforms transfer radio signal processing operations into a general-purpose processor on the same board or a remote server.
In addition to improving programmability, this enables the use of powerful processors for centralized signal processing, which can be used to improve signal decoding probability and coping with interference challenges.
As mentioned in Section \ref{arch-program}, the integration of SDRs with SDWLANs is in its early stages.

\subsubsection{\textbf{Home WLANs}}
\label{arch_discus_home_net}
The importance of utilizing SDN-based mechanisms in home networks is justified given the facts that the number of 802.11 connected devices is being increased (e.g., home appliances, lighting and HVAC, medical monitoring) and some of these devices require timely and reliable data exchange with APs \cite{Qin2014,Tozlu2012,REWIMO}.
In addition, the emergence of 802.11 mesh technologies that rely on the installation of multiple APs per home \cite{jing2016multi} makes central control of home networks even more important.
Compared to enterprise networks, the topology of AP placement in home networks is random and variable.
Topology is even more randomized when users or distributed algorithms modify the transmission power and channel of their APs.
This has serious implications:
While ISP-based control (e.g., VAN) simplifies the installation and operation of home networks, the ISP may not be able to collect enough information from the home network because not all the neighbors choose the same ISP.
On the other hand, collaboration between homes (e.g., ResFi) introduces serious security challenges.
Finally, cloud-based control may not be very effective because neighboring users may not subscribe to the same service.
Therefore, the opportunities and challenges of home SDWLANs requires further research.

\subsubsection{\textbf{Security}} 
\label{arch-security}
In addition to running new network control mechanisms such as AsC and ChA, programmability (see Section \ref{arch-program}) simplifies security provisioning.
For example, when a security mechanism does not require hardware replacement (e.g., 802.11i's WPA), it can be implemented by reprogramming APs and clients.
However, the decomposition of control and data plane exposes security threats due to enabling of remote programming of network equipment.
For example, a DoS attack can be implemented by instructing a large number of clients to associate with an AP.
Therefore, it is important to identify security threats and take them into account during the architecture design process.

As mentioned earlier, virtualization through network slicing is an effective approach to control and isolate clients' access to resources.
However, as users will be sharing an infrastructure, achieving a strict isolation becomes more challenging.
For example, a client may simply switch to the frequency band of another slice to disrupt the capacity offered in that slice.
Unfortunately, secure virtualization has not been investigated by the research community.

SDWLANs enable network applications and administrators to detect abnormal activities and network breaches.
In the architectures that extend their control plane to clients in order to exchange monitoring and control packets (e.g., Trantor, Dyson, SplitAP), the reports received from malicious clients result in control decisions that negatively affect network performance.
To cope with this challenge, a malicious client may be detected using the flow statistics collected from APs and switches \cite{OpenSketch}.
In the next step, client location may be found through analyzing RSSI measurements \cite{SpotFi}.
Rogue (unauthorized) APs can be detected through collaboration among APs. 
For example, an abnormal interference level or sniffing the packets being exchanged with a rogue AP can be used by the detection algorithm.
Depending on the architecture in use, a network application can block the access of such APs through various mechanisms.
For example, using Odin, client access can be blocked through updating OpenFlow switches.
Proposing security mechanisms that benefit from the features of SDWLAN architectures is an important future direction, especially for large-scale and public networks.

%[Trantor]: However, it is a much harder problem to determine if a malicious (or faulty) client is sending spurious reports. 
%One potential way to address this problem is to verify such reports with reports from other APs and clients in the neighborhood, but this remains an open issue.

% ----->>>>
% REVISION
% consider writing a section on using millimeter wave communication

%------------------------------------------------------------------ COLOR

% REVISION

\section{Centralized Association Control (AsC)}
\label{AMmech}
In this section we review centralized AsC mechanisms. 
We focus on the metrics employed as well as the problem formulation and solving approaches proposed. 
This section also highlights the impact of architecture on design when details are available.
We employ a consistent notation to ease the understanding and comparison of metrics and formulations.
Our review summarizes the performance improvements achieved to reveal the benefits of these centralized mechanisms compared to distributed approaches.\footnote{Note that in this section and the next section (Section \ref{CMmech}) we do not study all the AsC and ChA mechanisms implemented by the architectures reviewed in Section \ref{Architectures}. 
This is because some of these architectures only show the feasibility of implementing control mechanisms, and they do not propose any \textit{new} mechanism to benefit from the features of SDWLANs.}
Although we mostly focus on state-of-the-art centralized mechanisms, we review the seminal distributed mechanisms as well because of their adoption as the baseline to evaluate the performance of centralized mechanisms.

%------------------------------------------------------------------ COLOR
At a high level, we categorize AsC mechanisms from two perspectives: 
\begin{itemize}
	\item \textit{\textbf{Seamless handoff}}: refers to the mechanisms that their objective is to reduce the overhead and delay of client handoff,
	\item \textit{\textbf{Client steering}}: refers to the mechanisms that adjust client-AP associations to optimize parameters such as the load of APs and the airtime allocated to clients.
\end{itemize}
Supporting seamless handoff is usually addressed by proposing architectures that reduce the overhead of re-association.
In contrast, client steering is performed through proposing AsC mechanisms that run on the control plane.
For example, an AsC mechanism may propose an optimization problem to balance the load of APs, while handoff delays depend on the architectural properties.
From the client steering point of view, AsC is particularly important in dense topologies because there are usually multiple candidate APs for a client in a given location. 
Therefore, using a simple RSSI metric may result in hot spots, unbalanced load of APs and unfair resource allocation to clients.
Hence, one of the main goals of AsC is to achieve \textit{fairness} among clients and APs.

%Client mobility is handled through dynamic AsC. 
%To this end, low time-complexity AsC mechanisms must be run periodically to decide about client re-association.
%Metrics such as RSSI, client demand and AP load may be considered by AsC mechanisms to choose a destination AP.

%There are numerous AsC mechanisms proposed in the literature.
%In this paper, however, we only focus on mechanisms that rely on global network knowledge.
%Before the overview of these mechanisms, we first explain two widely-adopted distributed AsC mechanisms that have been used as baselines to evaluate the performance of centralized mechanisms.

The distributed AsC mechanism employed by 802.11 standard is \textbf{strongest signal first (SSF)} \cite{SSF,802.11}. 
Using SSF, each client decides about its association based on the RSSI of probe response and beacon messages.
Each client associates with the AP from which highest RSSI has been received.
%Most of association management techniques use SSF as a baseline for their comparison.
%maybe add about 802.11k and r
\textbf{Least load first (LLF)} \cite{LLF} is an another widely-adopted distributed mechanism where APs broadcast their current load through beacon messages to help the clients include AP load when making association decisions.
The load of an AP is represented through various metrics, such as the number of associated clients\footnote{The traffic indication map (TIM) of a beacon packet represents a bitmap that indicates the clients for which the AP has buffered packets.}.
In the following, we review AsC mechanisms.

%------------------------------------------------------------------ COLOR
\subsection{Seamless Handoff}
\label{seam-handoff}
Some of the architectures reviewed in Section \ref{Archs} propose mechanisms to reduce the overhead of client handoff.
Handoff overhead refers to: (i) the packets exchanged between client and AP to establish a connection, and (ii) the delay incurred by the client during this process \cite{pack2007fast}.
In this section we study the contributions of SDWLAN architectures in terms of supporting seamless handoff.

% 2010
% Individual - Seamless handoff
\textbf{Cisco unified wireless network (CUWN).}% demand-agnostic
\label{CUWN_AM}
\cite{Cisco} enables the central configuration of RSSI threshold with hysteresis to perform seamless handoff of CCX \cite{CiscoCCX} compatible clients.
When the RSSI received from associated AP drops below the \textit{scan threshold}, the client increases its AP scanning rate to ensure fast handoff to another AP when the difference between the RSSI of the associated and new AP is equal or greater than the \textit{hysteresis} value specified.
In addition to hysteresis, specifying the \textit{minimum RSSI} value forces the clients to re-associate when their RSSI drops below a minimum RSSI. 

Three types of client roaming scenarios are handled by CUWN: (i) intra-controller roaming, (ii) inter-controller layer-2 roaming, and (iii) inter-controller layer-3 roaming.
The controller simply handles an intra-controller roaming by updating its client database with the new AP connected to the roaming client. 
Inter-controller layer-2 roaming occurs when a client associates with an AP that is controlled by a different controller belonging to the same subnet. 
In this type of roaming, mobility messages are exchanged between the old controller and the new one. 
%({\color{red}e.g.,} {\color{blue} There is no details in \cite{Cisco} about these messages!}) 
Then, the database entry related to the roaming client is moved to the new controller. 
Inter-controller layer-3 roaming occurs when a client is associated with an AP that is controlled by a controller belonging to a different subnet. 
In layer-3 roaming, the database entry of the roaming client is not moved to the new controller. 
Rather, the old controller marks the client's entry as \textit{anchor entry}. 
The entry is copied to the new controller and marked as \textit{foreign entry}. 
Therefore, the original IP address of the roaming client is maintained by the old controller. 
CUWN enables network administrators to establish mobility groups, where each group may consist of up to 24 controllers. 
A client can roam among all the controllers in a mobility group without IP address change, which makes seamless and fast roaming possible.

%\todo[inline,color=cyan]{any details about how the association algorithm works? \\ \textit{The previous paragraphs describe the association algorithms.} }  

%2012 August, 2014 June
% Individual - Seamless handoff
\textbf{Odin's Mobility Manager (OMM).}
\label{AMOdin}
%In this section we review the mobility and load balancing algorithm proposed by Odin \cite{Odin,Odin2} (see Section \ref{Odin_arch} ).
%As mentioned earlier, Odin introduces the concept of Light Virtual AP (LVAP), which is a small data structure residing on APs to indicate client association.
Odin  \cite{Odin,Odin2} enables seamless handoffs through LVAP migration between APs. 
The Odin controller maintains a persistent TCP connection per Odin Agent running on AP, thereby, switching among agents does not require connection reestablishment. 
In addition, the delay of a re-association equals the delay of sending two messages from the Odin controller to the old and new APs. 
The first message removes an LVAP from the old AP, and the second message adds an LVAP to the new AP. 
Assuming that the messages are sent successfully, the delay equals the longest round trip time (RTT) between the controller and the two APs, which depends on network size. 
%note: the controller may communicate with the to APs concurrently

To show the effectiveness of LVAPs, Odin employs a simple RSSI-based AsC mechanism.
The mobility application (running on the controller) selects the Odin Agent with highest RSSI if client movement is detected by the controller. 
The effect of handoff on TCP throughput has been evaluated using two APs and a client. 
While a period of throughput drop in regular 802.11 layer-2 handoff is observable, Odin does not show any throughput degradation.
%The authors also reported the maximum number of LVAP handoffs that does not cause throughput drop.
Performance evaluations also show that executing 10 handoffs per second results in a negligible reduction in TCP throughput.

%It is worth mentioning that, before Odin, seamless mobility support through VAP migration was proposed in \cite{Grunenberger2010a}.
%We do not discuss about the details of this mechanism due to its similarity to Odin.

%%%%%%%%%%%%%%%%%%%
%2015 Nov
% Individual - Seamless handoff

\label{AEtherflow_AM}
\textbf{$\AE$therFlow.} \cite{AEtherFlow} argues that handoff support through LVAP migration (e.g., Odin \cite{Odin2} and OpenSDWN \cite{OpenSDWN}) imposes high computational and communication overhead, especially in large networks with many mobile clients.
They propose a predictive handoff strategy by relying on the extended OpenFlow protocol proposed in this work (see Section \ref{AEtherFlow}).
The handoff mechanism works in three phases: 
\begin{itemize}
	\item \textit{Prediction}: %The controller collects the RSSI of the links between all clients and APs. 
	The controller predicts an association when the RSSI of a client to its associated AP declines while its RSSI to another AP increases. 
	\item \textit{Multicasting}: By updating OpenFlow tables, the controller multicasts the packets of the client to both the current AP and the predicted AP.
	\item \textit{Redirection}: The controller will redirect the client's traffic to the new AP after the handoff completion. If no handoff occurs, then the multicasting is stopped.
\end{itemize}

Experiments using two APs and a client shows that the handoff delay of $\AE$therFlow is around 5.3 seconds, compared to the 7.1 seconds delay of 802.11 standard.
%A 9Mbps UDP traffic is sent to the client.
%Handoff duration is measured as the time interval during which throughput drops below 8Mbps.
% No comparison with Odin and OpenSDWN?

% 2016 April
% Individual - Seamless handoff
\textbf{BIGAP.}
\label{BIGAPhandoff}
\cite{BIGAP} uses the 25 non-overlapping channels of 5GHz band to form disjoint collision domains for handoff.
As explained in Section \ref{Archs}, a separate NIC is used to periodically overhear packets on all channels, which enables the controller to compute the potential SNR values of client-AP links.
A handoff happens when a higher SNR would be achievable for a client.
However, to avoid the ping-pong effect, an 8dB hysteresis value is used. 

BIGAP performs handoff through client channel switching.
Since all APs share the same BSSID, in order to handoff a client from $AP_{1}$ to $AP_{2}$, the controller instructs $AP_{1}$ to send a channel switching command to the client.
The client then switches to the channel being used by $AP_{2}$.
Since both APs use the same BSSID, the client does not notice handoff.

Performance evaluations (using two APs) show that the BIGAP handoff is about 32 times shorter than the regular 802.11 handoff.
In addition, BIGAP results in lower energy consumption because, it moves the overhead of handoff to APs and there is no need for the clients to scan the channels.
In addition, while 802.11 results in zero throughput for about 4 seconds, BIGAP shows only 5\% throughput reduction during the handoff.

%------------------------------------------------------------------ COLOR
\subsection{Client Steering}
\label{client-steering}
Based on the scope of the optimization problem employed, we categorize client steering mechanisms into two groups: \textit{\textbf{centrally-generated hints}} and \textit{\textbf{centrally-made decisions}}.
In AsC mechanisms using centrally-generated hints, the controller relies on the global network view to generate hints for the association of clients.
In other words, the controller does not make the final decision about associations and instead, enables the clients to make more informed decisions through the hints conveyed.
On the other hand, in AsC mechanisms using centrally-made decisions, the controller makes the association decisions and enforces the clients to apply them.
We will discuss in Sections \ref{AM-InvidualOpt} and \ref{AM-GlobalOpt} the two sub-categories of mechanisms based on centrally-made decisions.

\subsubsection{\textbf{Centrally-Generated Hints}} 
\label{Per-clientAM}
In this section we review AsC mechanisms that employ client steering through hints generated centrally.

\textbf{BestAP.}
\label{BEST-AP}
\cite{BEST-AP} proposes an AsC mechanism based on the estimation of \textit{available bandwidth} (ABW) for each client at every AP in its vicinity. 
The available bandwidth depends on channel load, which varies with packet loss and PHY rate. 
The estimated available bandwidth at PHY rate $r$ is computed as,
\begin{equation}
E[ABW(r)]=\frac{8S_{data}(1-B)}{\sum_{k=0}^{n}(1-p_s(r))^kp_s(r)T_{tx}(r,k)},
\end{equation}
where $S_{data}$ is data size (bytes), $p_s$ is the probability of successful packet transmission, $T_{tx}(r,k)$ is the transmission time of a packet during the \textit{k}th transmission attempt with PHY rate $r$, and $B$ is the fraction of channel busy time. 
$B$ is measured through using the CCA register of NIC.
%If there is no such hardware support, $b$ can be calculated using the method proposed in \cite{ChannelLoad}, which is based on measuring the airtime consumed by each packet. 
$p_s(r)$ is measured using the statistics provided by the rate adaptation algorithm. 
All other parameters are configured statically. 

A scheduler is run on each client to allocate a measurement period (e.g., 50ms every 2s) during which the client sends data to all nearby APs in order to update packet loss and channel busy fraction. 
%This updates the ABW of each client at their reachable APs.
A monitoring service is run on APs to collect the statistics from clients, calculate ABW, and send a report to the controller.
The controller sends the estimated ABWs of the best five APs to each client periodically. 
%The report is also sent when the ABW of a client has been changed more than $x\%$ ($x=10$ is used during the experiments). 
BEST-AP only considers the ABW of downlinks.

%Experimental results show that BEST-AP's ABW estimation is more accurate than WBest \cite{WBest}. 
Testbed evaluations show that the delay overhead of ABW estimation is less than 50ms, which is appropriate for a dynamic AsC mechanism.
%\todo[inline, color=cyan]{how do you claim it is appropriate for a DAC mechanism? \\ \textit{It is not my claim. It is the claim of authors !!} }  
Compared to SSF \cite{SSF}, the proposed AsC mechanism shows 81\% and $176\%$ improvement in throughput for static and mobile clients, respectively.
%\todo[inline, color=cyan]{the above sentence needs work- please correct it}  

%2015
% Global - Client steering - equal number of clients per AP
\textbf{Ethanol.}
This architecture \cite{Ethanol} (see Section \ref{EthanolArch}) has been evaluated through running a load-aware AsC mechanism that aims to balance the number of associated clients among APs.
When a client requests to join an AP with higher load, the controller drops the request to force the client look for another AP.
A simple testbed with two APs and up to 120 clients shows that the maximum difference between the number of clients associated to APs is two, which is due to the concurrent arrival of association requests.

%\textbf{vBS.}
%The vBS \cite{vBS} architecture (please see Section \ref{Archs}) also proposes a fast handoff mechanism to reduce reconnection delay.
%Experimental results show that vBS can perform handoff in less than 65ms without any packet drop.

%------------------------------------------------------------------ COLOR
\subsubsection{\textbf{Individual Optimization through Centrally-Made Decisions}}
\label{AM-InvidualOpt}
In this section we review AsC mechanisms that employ client steering through decisions generated centrally. 
These mechanisms, however, do not define a global optimization problem; thereby they do not take into account the effect of an association on other clients/APs.
Due to the individual nature of association control, these mechanisms only improve the overall network performance, and cannot be used to enforce fairness.

%2008
% Individual - Client steering - ap load balancing
\textbf{DenseAP.}
\label{DenseAP-AM}
The AsC mechanism proposed by DenseAP \cite{DenseAP} (see Section \ref{DenseAP_arch}) works as follows.
The \textit{available capacity} metric is defined to rank all the APs a client could be associated with. 
A client associates to the AP with highest available capacity. 
The available capacity of $AP_{i}$ operating on channel $ch_{i}$ is defined as follows,
\begin{equation}
AC_{AP_{i},c_{j}}^{ch_{i}} = F_{AP_{i}}^{ch_{i}} \times r_{AP_{i},c_{j}}^{ch_{i}},
\end{equation} 
where $F_{AP_{i}}^{ch_{i}} $ is the \textit{free airtime} of $AP_{i}$ on channel $ch_{i}$, and $r_{AP_{i},c_{j}}^{ch_{i}}$ is the \textit{expected transmission rate} of $c_{j}$ when communicating with $AP_{i}$.
Free airtime is estimated by measuring the amount of time that the MAC layer contends for channel access to send a high-priority packet.
The expected transmission rate is estimated using the RSSI of probe request frames received at the AP. 
A mapping table is used for this purpose.
APs hear the probe requests of clients and send reports to a controller. 
APs also measure their free airtime and report it to the controller. 
The controller selects the AP with highest available capacity for each client to associate with. 
To instruct a client associate with the selected AP, only the selected AP responds to the client's probe message. 

DenseAP also proposes a dynamic load balancing algorithm that periodically decides about associations.
In particular, the controller checks the free airtime of APs every minute. 
An AP is \textit{overloaded} if its free airtime is less than 20\% and it has at least one associated client. 
If an overloaded AP exists, the controller considers its clients as the potential candidates for association with APs experiencing lower load.
% REMOVAL CANDIDATE
A candidate AP must satisfy these conditions: 
(i) the expected transmission rate of clients when associated with the new AP must not be lower than the current transmission rate, and (ii) the free airtime of new AP must be at least 25\% more than the current AP. 
During each decision period, at most one client is allowed to be associated with a new AP, and two consecutive associations for a client is prohibited to avoid the ping-pong effect.

%The performance of DenseAP's AsC mechanism is evaluated through empirical experiments using the 5GHz band (802.11a) with 8 channels. 
Empirical evaluations show a 40\% to 70\% increase in per-client throughput, compared to SSF. 
Moreover, the authors conduct a small experiment using three clients and two APs to show that the load balancing algorithm improves the throughput of clients by more than 200\%.

% 2012
% Individual - Client steering - fair bandwidth allocation 
\textbf{Odin's Load Balancing (OLB).}
This load-balancing mechanism \cite{Odin2} periodically (every minute) inquires APs to collect the RSSI relationship between clients and APs, and then LVAPs are evenly distributed between APs to balance their loads.
%Using 10 APs and 32 clients, the efficiency of this mechanism on TCP throughput has been evaluated. %when a fixed rate 6Mbps is used.
The evaluations on a testbed with 10 APs and 32 clients show that around 50\% and 15\% of clients were able to receive a fair amount of throughput when the proposed mechanism was enabled and disabled, respectively.

\begin{figure}[!t]
	\centering
	\includegraphics[width=0.6\linewidth]{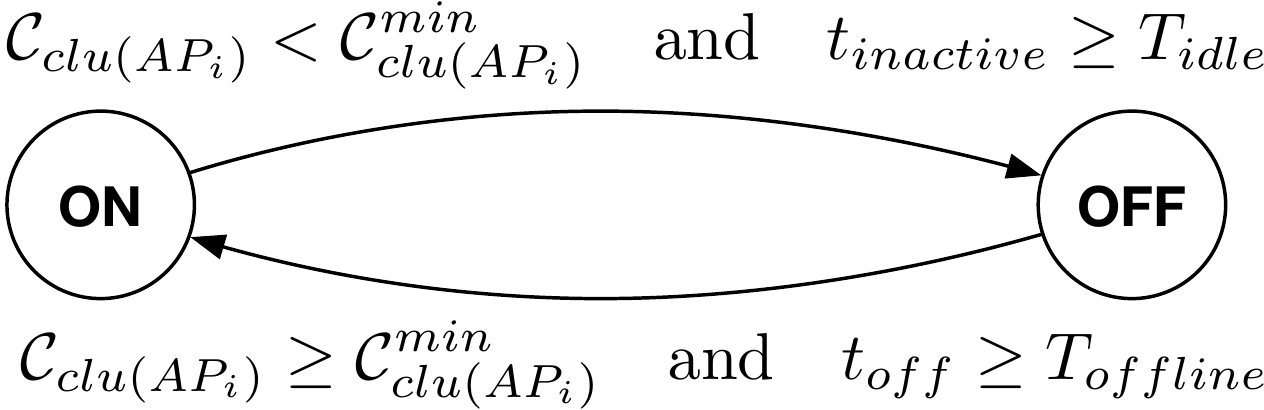}
	\caption{Finite state machine (FSM) of energy manager for a slave AP in EmPOWER \cite{EmPOWER}.}
	\label{fig_FSM_EmPOWER}
\end{figure}

%2013 Nov, 2015 June
% Individual - Client steering - energy efficiency
\textbf{EmPOWER.} 
\label{EmPowerAM}
This mechanism \cite{EmPOWER,Primitives} relies on client steering to improve the energy efficiency of APs.
APs are partitioned into clusters, where each cluster has one master and multiple slaves. 
Master APs are always active, and they are manually selected during the network deployment phase to provide a full coverage.
Slave APs are deployed to increase network capacity.
These APs are turned on/off using the finite state machine (FSM) depicted in Figure \ref{fig_FSM_EmPOWER}.  
%
%\todo[inline, color=cyan]{what is the clustering strategy? \\ \textit{There is no specific clustering strategy in the paper. I added more explanation to clarify it a little more.}}  
In the ON mode, all wireless interfaces of the AP are on. 
In OFF mode, only the Energino \cite{Energino} module of the AP is on.  

Two metrics are defined for a slave $AP_i$ belonging to cluster $clu(AP_i)$: $\mathcal{C}_{clu(AP_i)}$ is the number of clients in the cluster, and $\mathcal{C}^{min}_{clu(AP_i)}$  is the minimum required number of clients in cluster $clu(AP_i)$ to keep $AP_i$ active. 
A slave $AP_i$ transitions from ON mode to OFF mode if: (i) the number of its cluster's clients is less than $\mathcal{C}^{min}_{clu(AP_i)}$, and (ii) $AP_i$ has been inactive (i.e., no client associated) for at least $T_{idle}$ seconds. 
Also, $AP_i$ transitions from OFF mode to ON mode if: (i) the number of its cluster's clients is at least $\mathcal{C}^{min}_{clu(AP_i)}$, and (ii) $AP_i$ has been OFF for at least $T_{offline}$ seconds. 

The mobility manager associates a client to a new AP with higher SNR. 
However, to establish a balance between performance and energy consumption, the mobility manager may associate a client to an AP with lower SNR but smaller $\mathcal{C}^{min}_{clu(AP_i)}$.  
In this way, the energy manager is able to turn off the APs with higher $\mathcal{C}^{min}_{clu(AP_i)}$  in order to decrease energy consumption. 
Re-association is performed if there is a better AP in terms of SNR and $\mathcal{C}^{min}_{clu(AP_i)}$. 
By relying on the Odin \cite{Odin2} APIs, the authors showed that this AsC mechanism was implemented as a Java network application with only 120 lines of codes.

%2016 July
% Individual - Client steering - ap load balancing
\textbf{Adaptive mobility control (AMC).} 
\cite{mob-Essex-2016} shows that using a fixed RSSI threshold by clients results in an unbalanced load of APs.
% idea
They propose a heuristic algorithm which uses RSSI and traffic load of APs to provide dynamic hysteresis margins on AP traffic load level. 
The load of $AP_{i}$ is defined as,
\begin{equation}
\label{eq:eq18}
L_{AP_{i}}=\left\{\begin{array}{ll}
B\textrm{\ \ \ \ \ \ \ \ \ \quad\quad\quad\quad\quad\quad\quad if  $|\mathcal{C}_{AP_{i}}|=0$}\\
0.8\times B +0.2\times |\mathcal{C}_{AP_{i}}| \textrm{\quad\ \  if $|\mathcal{C}_{AP_{i}}|>0$}
\end{array}\right.
\end{equation}
where $|\mathcal{C}_{AP_{i}}|$ is the number of clients associated with $AP_{i}$, and $B$ is channel busy time.
The algorithm defines three thresholds on AP load and RSSI: low, medium, and high.
%These thresholds are determined through applying experimental tests to measure the effect of AP load and RSSI value on network throughput. 
Using these thresholds, a client is associated with a new AP that satisfies one of these conditions: (i) higher signal strength and lower load, (ii) significantly higher signal strength and slightly higher load, or (iii) significantly lower load and slightly lower signal strength.
% QUESTIONS (Second phase) --> COMPLETED
% ^^- AP load metric  
% ^^ - hysteresis margins
% ^^ - How is the algorithm working?
% ^^ - adaptive margins ==> based on which parameter, it is adaptive?
% QQQQQQQQQQQQQQQQQQQQQQQQQQQQQQQQQQQQQQQQQQ

The proposed adaptive mobility manager has been implemented using Odin.
Empirical results show more than 200\% improvement in TCP throughput, compared to SSF.

%2016 June
% Individual - Client steering - reduce delay
\textbf{Associating to good enterprise APs (AGE).}% demand-agnostic
\label{AGE}
The main objective of AGE \cite{WiFiSeer} is the reduction of clients' packet exchange delay over wireless links through client steering.
AGE has two main phases: learning and AP selection. 
%The learning phase uses a training set obtained during network operation for a week every three months.
During the learning phase, the performance metrics and environmental factors are pulled every minute from all APs using SNMP \cite{SNMP}. 
These metrics include AP-client RTT, RSSI, SNR, number of associated clients, channel number, frequency band, AP location, day of week, and time.
%The wireless latency of each client is calculated every minute through a technique called \textit{ping2}. 
%The authors argue that measuring wireless latency through sending only one ping packet might be inaccurate due to the wake up delay of NIC. 
%Therefore, ping2 uses two consecutive ping packets.
%The first ping is used to wake up the client's NIC, if it is in energy-saving mode. 
%The second ping is sent to calculate latency. 

\begin{figure}[!t]
	\centering
	\includegraphics[width=0.9\linewidth]{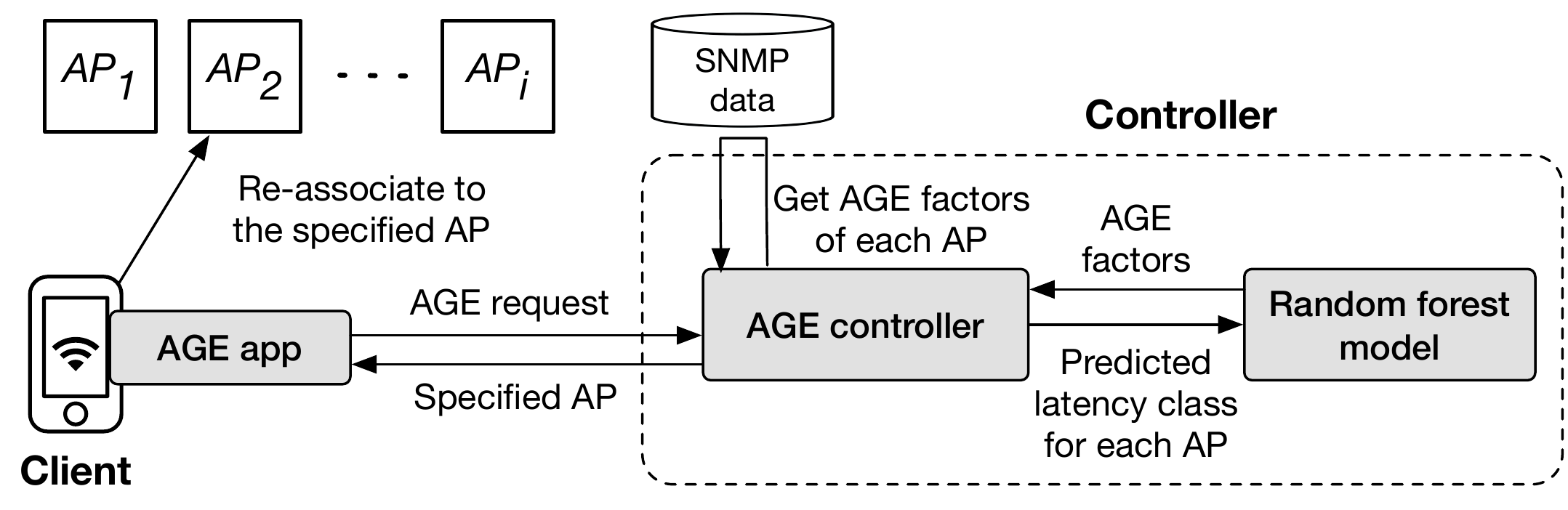}
	\caption{Associating to Good Enterprise APs (AGE)\cite{WiFiSeer}. Each client is instructed to connect to the AP that provides minimum delay.}
	\label{fig_AGE}
\end{figure}

Using the collected training set, the authors use the \textit{random forest} \cite{random-forest} technique to generate a two-class learning model for classifying APs into \textit{high latency} and \textit{low latency}.
Figure \ref{fig_AGE} shows an overview of AGE and its two main components: (i) the \textit{AGE app} that is installed on each client's mobile phone, and (ii) the \textit{AGE controller} that uses the random forest model to classify the APs in the vicinity of each client.
AGE operates as follows:
\begin{itemize}
	\item The AGE application on a client device sends an AGE request (including the list of achievable APs) to the AGE controller, periodically. 
	%AGE requests are sent every 5 and 20 minutes when the device's screen is on and off, respectively.
	\item The AGE controller pulls the SNMP data of the APs requested by AGE app.
	\item The latency class of each AP is predicted using the learning technique mentioned earlier.
	\item The AGE controller informs the AGE application about the best AP nearby.
	\item The client re-associates with a new AP using AGE app. 
\end{itemize}

The authors deployed AGE at Tsinghua University campus where over 1000 devices used the network for 2.5 months. 
Measurements confirmed that the wireless data exchange delay of more than 72$\%$ of clients has been reduced by over 50$\%$.

%------------------------------------------------------------------ COLOR
\subsubsection{\textbf{Global Optimization through Centrally-Made Decisions}}
\label{AM-GlobalOpt}
In this section we review AsC mechanisms that employ client steering by formulating a problem that aims to optimize performance parameters globally.
These mechanisms propose heuristics to solve the NP-hard problems that usually aim to achieve network-wide fairness.

%2004-2007
% Global - Client steering - fair bandwidth allocation
\textbf{Association Control for fairness and load balancing (ACFL).} 
In \cite{F-LB-AsscCtrl-2004} and \cite{F-LB-AsscCtrl-2007}, the authors address the unbalanced load of APs as a result of using SSF or LLF.
%However, their fairness measure is max-min fairness.Max-min throughput fairness can significantly reduce aggregate throughput in multi-rate WLANs. The max-min time fairness problem they consider is intended for single-rate WLANs.
%from the paper:  The collected infor-mation is reported to a network operation center (NOC) which runs our algorithm to come up with the user-AP association de-cisions.
%They consider the bandwidth constraints of APs in both wireless and wired links.
To address this challenge, an AsC problem is formulated to establish the max-min bandwidth fairness among APs. 
Intuitively, the load of a client on its associated AP is inversely proportional to the effective bit rate of AP-client link. 
The load of $AP_{i}$ (denoted as $L_{AP_{i}}$) is modeled as the maximum of the aggregated loads of its wireless and wired links generated by all clients $c\in \mathcal{C}$, as follows,
\begin{equation}
L_{AP_{i}}=\max\left\{ \sum_{\forall c_{j}\in \mathcal{C}} \frac{\omega_{c_{j}} \times X_{AP_i, c_j}}{r_{AP_i, c_j}},  \sum_{\forall c_{j}\in \mathcal{C}} \frac{\omega_{c_{j}} \times X_{AP_{i},c_j}}{R_{AP_{i}}} \right\}
\end{equation}
where $r_{AP_i, c_j}$ is the transmission rate between $AP_{i}$ and $c_{j}$, $R_{AP_{i}}$ is the transmission rate of wired interfaces of $AP_{i}$, $X_{AP_{i},c_{j}}\in\{0,1\}$ is the association state of $c_{j}$ to $AP_{i}$, and $\omega_{c_{j}}$ is the traffic volume size of client $c_{j}$. 
In other words, the load of $AP_{i}$ is defined as the period of time this AP requires to handle the traffic of its associated clients. 
$AP_{i}$ provides bandwidth $X_{AP_{i},c_{j}} \times \omega_{c_{j}}/L_{AP_{i}}$ to client $c_{j}$. 

Two approximation algorithms are introduced to solve the formulated NP-hard max-min fair bandwidth allocation problem. 
The first algorithm solves the problem for unweighted greedy clients. 
The second algorithm proposes a solution for weighted and limited throughput demand of clients.
The algorithms are run periodically by a controller to update client associations. 
Simulation results show over 20$\%$ improvement in terms of average per-client bandwidth, compared to SSF and LLF.

%\todo[inline,color=cyan]{sentence not clear- please rewrite}  
%\todo[inline,color=cyan]{why did you remove the formulas? can you make the explanation more complete?\\ \textit{this paper is so confusing. I tried to add more explanation.}}  

%2014
% Global - Client steering - fair bandwidth allocation 

\textbf{Association control for proportional fairness (ACPF).}
\cite{Proportional-Fairness-AP-2014} argues that using throughput-based fairness (e.g., \cite{Time-based-basic-1, Time-based-basic-2}) in multirate networks leads to low overall network throughput because clients with a low bit rate can occupy the channel longer than those with higher rate. 
The authors investigate the problem of achieving proportional fairness by introducing the following objective function formulated based on the effective bandwidth of clients:
\begin{equation}
\sum_{\forall c_{j}\in \mathcal{C}} \rho_{c_{j}} \times  \log (\beta_{c_{j}}),
\end{equation}
where $\rho_{c_{j}}$ and $\beta_{c_{j}}$ are the priority and effective bandwidth of client $c_{j}$, respectively. 
$\beta_{c_{j}}$ is defined as 
\begin{equation}
\sum_{\forall AP_{i}\in \mathcal{AP}}X_{AP_{i},c_{j}}\times t_{AP_{i},c_{j}}\times r_{AP_{i},c_{j}}, 
\end{equation}
where $X_{AP_{i},c_{j}}\in\{0,1\}$ is the association index, $t_{AP_{i},c_{j}}\in[0,1]$ is the effective normalized time, and $r_{AP_{i},c_{j}}$ is the PHY rate between $c_{j}$ and $AP_{i}$. 
Achieving fairness through association is formulated as a linear-programming problem to maximize the objective function.
%\todo[inline, color=cyan]{"The client traffic-based proportional fairness"- this is not a correct sentence. not clear.}  
A centralized algorithm, called \textit{non-linear approximation optimization for proportional fairness} (NLAO-PF), is proposed to solve the optimization problem.
%This algorithm computes client-AP association relations and transmission time of clients. 
It has been proven that the approximation ratio of NLAO-PF is 50\%.

The authors used the OMNet++ \cite{OMNET} simulator to evaluate the effectiveness of NLAO-PF compared to cvapPF \cite{cvapPF} in terms of average throughput. 
The average improvement is reported as 18.8$\%$ and 35$\%$ for uniformly distributed clients as well as clients in a hotspot area, respectively.

%2014
% Global - Client steering - fair bandwidth allocation 

\textbf{Association control with heterogeneous clients (ACHC).}  % demand-agnostic?
Experimental analysis of the negative impacts of legacy clients (802.11a/b/g) on 802.11n clients has been reported by \cite{802.11n-AP-Association-2014}.
The authors propose a two-dimensional Markov model to calculate the uplink and downlink throughput of heterogeneous clients. 
MAC efficiency of a client $c_{j}$ is defined as follows,
\begin{equation}
\label{MACeffMetr}
\Upsilon({c_{j}}) = \frac{Th_{c_{j}}^{up}+Th_{c_{j}}^{down}}{\min\{1,\; p_{c_{j}}^{up}+p_{c_{j}}^{down}\}\times r_{AP_{i},c_{j}}},
\end{equation}
where $Th_{c_{j}}^{up}$ and $Th_{c_{j}}^{down}$  are the estimated uplink and downlink throughput of client $c_{j}$, respectively;
$p_{c_{j}}^{up}$ and $p_{c_{j}}^{down}$ are, respectively, the uplink and downlink traffic probabilities of $c_{j}$; 
$r_{AP_{i},c_{j}}$ is the optimal PHY rate between $c_{j}$ and $AP_{i}$. 
AsC is formulated through an optimization problem to provide a bandwidth proportional to each client's achievable data rate, which is obtained by maximizing $\sum_{\forall c_{j}\in \mathcal{C}} \log \Upsilon({c_{j}})$.

Two heuristic dynamic AsC algorithms are proposed to solve the optimization problem: \textit{FAir MAC Efficiency} (FAME) and \textit{Categorized}. 
Instead of maximizing all MAC efficiencies, these algorithms maximize the minimum MAC efficiency.
To run FAME, the data rate and traffic load of all clients in nearby BSSs must be collected from each client's point of view.
With \textit{Categorized}, the type of all clients in nearby BSSs must be collected as well.
However, \textit{Categorized} is less sensitive to network dynamics because it tries to maximize the number of similar-standard 802.11 clients connected to each AP.
Therefore, \textit{Categorized} takes advantage of dense deployment and alleviates the performance degradation of 802.11n clients that is caused by the presence of legacy clients. 

Simulation and testbed experiments confirmed the higher performance of \textit{FAME} and \textit{Categorized} compared to SSF and ACFL in terms of TCP and UDP throughput, MAC efficiency and aggregated throughput of clients supporting different standards.
Although the conducted experiments use distributed execution of \textit{FAME} and \textit{Categorized}, collecting the information required by these algorithms imposes a high burden on clients. 
Therefore, these algorithms are mostly suitable for SDWLANs. 
%Furthermore, it is required to change beacon packet format as data collection is performed through exchanging beacon packets.

% demand-agnostic
%2015
% Global - Client steering - fair bandwidth allocation 

\textbf{Association control and CCA adjustment (ACCA).} 
Two central algorithms for AsC and CCA threshold adjustment are proposed in \cite{Association-CCA-2015} for dense AP deployments.
%\todo[inline, color=cyan]{why did you call it a joint approach while the two algorithms are separate? do they work together? \\ \textit{thanks for your precise comment. It is not joint mechanism.}}  
The basic idea of the AsC mechanism is to utilize the SINR of a client perceived by an AP in order to measure the interference (congestion) level of APs. 
The network is modeled as a weighted bipartite graph (WBG), and the AsC problem is formulated as a maximum WBG matching combinatorial optimization problem. 
The weights of edges in WBG are the uplink SINR values. 
The optimization problem's objective is to maximize the sum of the clients' throughput, where the throughput upper-bound is computed using the Shannon-Hartley formula \cite{rappaport1996wireless}.
Per-BSS CCA level is adjusted by using a constraint of the optimization problem based on the cell-edge SINR of each BSS. 
The optimization problem has been solved using Kuhn-Munkres \cite{Kuhn} assignment algorithm.
%The proposed centralized technique requires global knowledge about all clients and APs, which is collected by the functionalities of IEEE 802.11k \cite{802.11k}. 
% Details of graph modeling
% Details of optimizaiton problem

MATLAB simulations show a 15 to 58\% improvement at the 10th percentile of cumulative distribution of clients throughput, compared to SSF.
%In addition, the improvement level is higher for networks with randomly deployed APs, compared to uniformly-spaced AP placement.
Furthermore, the CCA adjustment algorithm significantly improves the throughput of cell-edge clients, compared to the network with fixed CCA threshold.
%\todo[inline, color=cyan]{please correct the last sentence, not clear at all. any numerical values? \\ \textit{More explanations are added.}}  

%2015
% Global - Client steering - fair time fairness allocation 

\textbf{Demand-aware load balancing (DALB).} % more focus on scheduling rather than association (transmission time adjustment) --> it's not based on CSMA/CA
\cite{Demand-aware-LB-Association-15} formulates joint AP association and bandwidth allocation as a mixed-integer nonlinear programming problem that includes the bandwidth demand of clients. 
The objective is to maximize the aggregated bandwidth of clients, which establishes a trade-off between throughput gain and time-based user fairness. 
The key idea of this work is the inclusion of clients' bandwidth demands in the computation of APs' load. 
For a client $c_{j}$, bandwidth demand is defined as $\beta_{c_{j}}=S_{c_{j}}/{T}$, where $S_{c_{j}}$ is the size of data received for client $c_{j}$ by its associated AP during the allocable transmission time of the AP, i.e., $T$. 
The transmission time demand of $c_{j}$ associated to $AP_{i}$ is defined as $T_{AP_{i},c_{j}} = \beta_{c_{j}}\times T/r_{AP_{i},c_{j}}$.

The optimization problem's objective is defined as follows,
\begin{equation}
\label{demandEqu}
\sum_{\forall c_{j} \in \mathcal{C}} \log\sum_{\forall AP_{i} \in \mathcal{AP}} \frac{X_{AP_{i},c_{j}}\times r_{AP_{i},c_{j}} \times t_{AP_{i},c_{j}}}{T},
\end{equation}
where the transmission time allocated to $c_{j}$ associated with $AP_{i}$ is expressed by $t_{AP_{i},c_{j}}$. 
The authors proved that maximizing Equation \ref{demandEqu} subject to a constraint on the transmission time demand of clients, i.e., $0 \le t_{AP_{i},c_{j}} \le T_{AP_{i},c_{j}}$, leads to proportional time fairness among clients.

A two-step heuristic algorithm is introduced to solve the optimization problem. 
The first step associates a client to an AP with the lowest allocated transmission time at each iteration until all clients are associated. 
At each iteration, the algorithm selects a client with the largest bandwidth demand. 
The second step schedules the transmission time of APs to establish proportional transmission time fairness among clients. 

Simulations show a 23$\%$ improvement in clients' throughput compared to SSF, ACPF and NLB \cite{NLB-13}.
In addition, this improvement has been achieved without sacrificing fairness. 
In particular, while bandwidth fairness is slightly better than that of SSF, ACPF and NLB, time fairness is significantly improved compared to these algorithms.

%2016
% Global - Client steering - maximizing throughput of clients

\textbf{Migration-cost-aware association control  (MCAC).} 
\label{Migration-DAM}
\cite{Migration-DAM} investigates the problem of max-min fairness subject to a migration cost constraint.
An integer linear programming problem is formulated with the aim of maximizing the minimum throughput of clients in order to establish max-min fairness.
The migration cost of clients is the main constraint of the formulation.
The load of $AP_{i}$ is defined as $L_{AP_{i}}=\sum_{\forall c_{j} \in \mathcal{C}} X_{AP_{i},c_{j}}/r_{AP_{i},c_{j}}$.
Assuming the achievable throughput of clients is proportional to the inverse of the load of the associated AP (i.e., $1/L_{AP_{i}}$), maximizing the minimum throughput of clients is equivalent to minimizing $\max_{\forall AP_{i} \in \mathcal{AP}}L_{AP_{i}}$. 
%Consequently, the objective of the optimization problem is defined as $\min \max_{i} H_i$, where $H_i$ is the load of AP $i$ after client re-association. 
The migration cost constraint is defined as, 
\begin{equation}
\sum_{\forall AP_{i} \in \mathcal{AP}}\;\sum_{\forall AP_{l} \in \mathcal{AP}} (Y_{AP_{i},AP_{l}}\times M_{c_j}) \le M,
\end{equation}
where $Y_{AP_{i},AP_{l}}\in\{0,1\}$ indicates whether a client $c_{j}$ currently associated with $AP_{i}$ will be associated with $AP_{l}$, $M_{c_{j}}$ is the migration cost of client $c_{j}$ from $AP_{i}$ to $AP_{l}$, and $M$ is the total permissible migration cost.

The authors showed that the proposed problem is NP-hard, and they proposed an approximation algorithm, named the \textit{cost-constrained association control algorithm} (CACA), to solve the optimization problem. 
In this algorithm, the problem is divided into two sub-problems: client removal and client re-association. 
In the first sub-problem, a subset of clients are removed from their current AP with the aim of minimizing the maximum AP load, considering the migration cost constraint. 
%When migration costs are equal, the migration cost constraint is simplified as a limitation on the number of migrating clients, i.e., no more than $K^{\prime}=K/c$ clients can be removed. 
%These clients impose the highest load on the maximum-loaded AP.
In the second sub-problem, the algorithm performs client re-association to minimize the load of maximum-loaded AP.

NS3 \cite{NS3}  simulations and experimental results show improvements in terms of the number of re-associations, throughput, loss ratio and end-to-end delay, compared to SSF, ACPF and GameBased \cite{GameBased}. 

\textbf{QoS-driven association control (QoSAC).}
\label{QoS-DAM}
\cite{flow-level-DAM} investigates the problem of flow-level association to address the QoS demands of clients.
Specifically, the backhaul capacity of APs is included in the association decision making.
This work proposes a mechanism for concurrent association of a client to multiple APs and supports flow-level routing of traffic. 
This mechanism facilitates dedicated management of each flow. 
For example, a client may use $AP_{1}$ for video streaming while using $AP_{2}$ to upload a file.

The main objective is to minimize the \textit{average inter-packet delay} of individual download flows. 
The inter-packet delay for a client associated to an AP is formulated based on the bi-dimensional unsaturated Markov model proposed in \cite{802.11n-AP-Association-2014}. 
An optimization problem is formulated to minimize the sum of the average inter-packet delay of all APs. 
The backhaul capacity of an AP should not be smaller than the sum of the arrival traffic rates of the AP, i.e., the sum of all download flow rates. 
This is included as a constraint of the optimization problem.
%\todo[inline, color=cyan]{not clear what is has to do with flow level? }
%\todo[inline, color=cyan]{not clear how it benefits from multiple associations \\ I added one sentence to clarify it more.}    
The proposed optimization problem is interpreted into a supermodular set function optimization \cite{supermodular}, which is an NP-hard problem. 
This problem has been solved using two heuristics: (i) greedy association, and (ii) bounded local search association. 
The greedy algorithm associates a client with the AP that minimizes the total inter-packet delay in each iteration.
In this algorithm, the association loop is continued until all clients are associated. 
The bounded local search is a polynomial-time algorithm similar to the algorithm proposed in \cite{boundedAlg}.
%\todo[inline, color=cyan]{add a high level explanation of these algorithms}  

Simulation results confirm the reduced average inter-packet delay achieved with this mechanism, compared to SSF and FAME \cite{802.11n-AP-Association-2014}. 
The delay is considerably reduced when the backhaul capacity is limited and clients form hotspots.

%From their paper:
% With the above features, we further consider client association in the high-density scenario. In this paper, “high-density”
%mainly implies the overlapped basic service set (OBSS) case, in which all the APs and clients are located within a certain coverage area and operate on the same channel.

%TABLE DAM
\begin{table*}
	\centering
	\scriptsize
	\caption{  Comparison of Association Control (AsC) Mechanisms  }
	\label{DAMtable} 
	\def\arraystretch{1}
	\begin{tabular}{|c|c|c|c|c|c|c|c|c|c|c|c|}
		\Xhline{3\arrayrulewidth}
		%	\textbf{Ref.} & \multicolumn{2}{|c|}{OneTwoThree} & \multicolumn{2}{|c|}{OneTwoThree}&\textbf{Channels} & \textbf{Dynamic/Static} & \textbf{\multicolumn{2}{|c|}{OneTwoThree}}\\ \Xhline{3\arrayrulewidth}
		%	
		\multirow{3}{*}{\textbf{Mechanism}}& \multicolumn{2}{c|}{Objective} & \multicolumn{3}{c|}{Optimization Scope}&\multirow{3}{*}{\textbf{Decision Metric}}& \multicolumn{2}{c|}{Traffic Awareness}&   \multicolumn{2}{c|}{ Performance Evaluation }\\ \cline{2-6}\cline{8-9}\cline{10-11}
		
		&\textbf{\makecell{Seamless \\ Handoff}}&\textbf{\makecell{Client \\Steering}}&\textbf{\makecell{Cent. Gen.\\ Hints}}&\textbf{\makecell{Indv. Opt. \\Cent. Made\\ Decisions }}&\textbf{\makecell{Glob. Opt. \\ Cent. Made\\ Decisions }}&& \textbf{Downlink}& \textbf{Uplink}& \textbf{Simulation} & \textbf{Testbed}\\ \Xhline{3\arrayrulewidth}
		%1
		%SSF\cite{SSF}&$\checkmark$&$\times$&RSSI&$\checkmark$&$\times$&$\times$&$\times$&$\checkmark$\\\hline
		%	
		\makecell{CUWN \cite{Cisco}} & $\checkmark $ & $\times $ & $\times $ & & $\times$&RSSI&$\times$&$\times$ &$\times$&$\times$\\\hline
		\makecell{OMM \cite{Odin2}} & $\checkmark $ & $\times $ &$\times $& & $\times$&RSSI&$\times$&$\times$ &$\times$&$\checkmark$\\\hline
		%2
		$\AE$therFlow \cite{AEtherFlow} & $\checkmark $ & $\times $ &$\times $& & $\times$&RSSI&$\times$&$\times$&$\times$&$\checkmark$\\\hline		
		BIGAP  \cite{BIGAP} & $\checkmark$ & $\times $ &$\times $ &  & $\times$&RSSI&$\times$&$\times$&$\times$&$\checkmark$ 	  \\\hline  		
        %        %        
        %        
        %			
		BestAP \cite{BEST-AP}& $\times$ &  \makecell{Bandwidth\\ Improvement }  &$\checkmark$& & $\times$&\makecell{Available\\ bandwidth}&$\checkmark$&$\times$&$\times$&$\checkmark$\\\hline
		Ethanol \cite{Ethanol}& $\times$ & \makecell{AP Load \\ Balancing}  &$\checkmark $ &  & $\times$ &Number of clients &$\times$&$\times$&$\times$&$\checkmark$\\\hline		
		DenseAP \cite{DenseAP}& $\times$ & \makecell{AP Load \\ Balancing} &$\times$ & \makecell{Overall\\ performance} & $\times$  & \makecell{Available\\ bandwidth}&$\checkmark$&$\times$&$\times$&$\checkmark$\\\hline
        \makecell{OLB \cite{Odin2}} & $\times$ & \makecell{AP Load \\ Balancing} &$\times$& \makecell{Overall\\ performance} & $\times$ &\makecell{RSSI,\\Number of clients}&$\times$&$\times$ &$\times$&$\checkmark$\\\hline   
		EmPOWER \cite{EmPOWER}& $\times $ & \makecell{Energy\\Efficiency} &$\times $ & \makecell{Overall\\ performance}& $\times$ &\makecell{RSSI,\\Number of clients}&$\times$&$\times$&$\times$&$\checkmark$\\\hline
		AMC \cite{mob-Essex-2016}& $\times$ &  \makecell{AP Load \\ Balancing}   &$\times$ & \makecell{Overall\\ performance} & $\times $ &\makecell{RSSI, Free airtime}&$\checkmark$&$\times$&$\times$&$\checkmark$\\\hline
		AGE \cite{WiFiSeer}& $\times$ & \makecell{Reducing Packet \\ Exchange Delay} &$\times$ & \makecell{Overall\\ performance}& $\times$&\makecell{Latency}&$\times$&$\times$&$\times$&$\checkmark$\\\hline
		ACFL \cite{F-LB-AsscCtrl-2007}& $\times$ & \makecell{AP Load \\ Balancing}  &$\times$&  & \makecell{Max-min\\ fairness} &\makecell{Free airtime, \\Transmission rate}&$\checkmark$&$\times$&$\checkmark$&$\times$\\\hline
		%3
        %		
        %		
        %        
        %4        %
		ACPF \cite{Proportional-Fairness-AP-2014}& $\times$ & \makecell{Fair Bandwidth\\ Allocation}  &$\times$&  & \makecell{Proportional\\ Fairness} &\makecell{Available\\ bandwidth}&$\checkmark$&$\times$&$\checkmark$&$\times$\\\hline
		ACHC \cite{802.11n-AP-Association-2014}& $\times$ & \makecell{Fair Bandwidth\\ Allocation}  &$\times$&  & \makecell{Proportional\\ Fairness} &MAC efficiency&$\checkmark$&$\checkmark$ & $\checkmark$& $\checkmark$\\\hline   
		ACCA \cite{Association-CCA-2015}& $\times$ & \makecell{Maximize\\ Throughput} &$\times$&  & \makecell{Overall\\ Performance} &Uplink SINR&$\times$&$\checkmark$&$\checkmark$&$\times$\\\hline
		%6
		DALB \cite{Demand-aware-LB-Association-15}& $\times$ & \makecell{Maximize\\ Throughput}  &$\times$&  & \makecell{Proportional\\ Fairness} &Transmission time&$\checkmark$&$\times$&$\checkmark$&$\times$\\\hline
		%
		%8
		MCAC \cite{Migration-DAM}& $\times$ & \makecell{Fair Bandwidth\\ Allocation}  &$\times$&  & \makecell{Max-min\\ Fairness}  &\makecell{Throughput,\\Re-association cost}&$\checkmark$&\tiny$\times$&$\checkmark$&$\checkmark$		\\\hline
		%13
		QoSAC \cite{flow-level-DAM} & $\times$ & \makecell{Reducing Packet \\ Exchange Delay} &$\times$ &  & \makecell{Overall\\ Performance}  &Inter-packet delay&$\checkmark$&$\times$&$\checkmark$&$\times$
		\\\Xhline{3\arrayrulewidth}
	\end{tabular}
\end{table*}

\subsection{Association Control: Learned Lessons, Comparison, and Open Problems}
\label{AscProblems}
Table \ref{DAMtable} presents and compares the features of the AsC mechanisms.
Although most of the AsC mechanisms focus on client steering, we should note that client steering and seamless handoff are interdependent.
For example, when association decisions are made centrally to balance the load of APs, a quick reaction to network dynamics requires seamless handoff; otherwise, the effect of load balancing will be compromised by the overhead and delay of re-associations.
In the following, we study the features of AsC mechanisms and identify research directions.

%<><><><><><><><><><><><><><><><><><><><><><><><><><><><><><><><><><><><><><><><><><><><><><><><>

\subsubsection{\textbf{Optimization Scope of Client Steering}} 
\label{asc_opt_scope_cl_ste}
Based on our review, we have classified the client steering approaches into three categories: (i) centrally-generated hints, (ii) individual optimization through centrally-made decisions, and (iii) global optimization through centrally-made decisions. 
Our review shows that most of the AsC mechanisms fall into the last two categories, where their optimization problems is classified as follows:
\begin{itemize}
	\item \textit{Overall network performance}: To improve the overall performance of clients and APs. 
	For example, QoSAC minimizes the sum of inter-packet delay for all APs, and ACCA  maximizes the sum of clients' achievable throughput.
	Note that improving overall performance does not necessarily require defining an optimization problem, as Section \ref{AM-InvidualOpt} shows.
	\item \textit{Max-min fairness}: To maximize the minimum performance. In other words, \textit{max-min} fairness indicates providing a client with more resources would not be possible without sacrificing the resources of other clients \cite{Bonald2006}. 
	For example, the objective function of ACFL and MCAC maximizes the throughput of minimum-throughput client.
	\item \textit{Proportional fairness}: Clients occupy the channel proportional to their transmission rate.
	Our review shows that proposing these approaches is motivated by the challenges of achieving fairness in multi-rate networks. 
	Multi-rate is caused by both client heterogeneity (e.g.,  802.11a/b/n) and the rate adaptation mechanism of 802.11 standards.
	Since the 802.11 DCF protocol aims to provide an equal channel access probability for all contending clients, low-rate clients occupy the medium longer than high-rate clients.
	In this case, the max-min throughput fairness results in sacrificing the performance of high-rate clients, which leads to a problem called \textit{performance anomaly}. 
	Performance anomaly in multi-rate networks may result in total throughput degradation, even in comparison to SSF \cite{DAM-G-5,bellalta2016interactions,802.11n-AP-Association-2014,SSF}. 
	A well-known solution is proportional fairness, which is usually achieved by defining a logarithmic objective function and scheduling the transmission time of clients to be proportional to their transmission rate. 
	ACPF, ACHC and DALB are examples of AsC mechanisms that establish proportional fairness. 
\end{itemize}

Using proportional fairness in multi-rate networks decreases the performance of low-rate legacy clients and may dissatisfy their QoS requirements. 
We suggest two approaches to address this problem: (i) hybrid fairness, and (ii) demand-aware proportional fairness. 
In hybrid fairness, a weighted fairness metric is defined by assigning weights for the max-min and proportional fairness metrics. 
Therefore, it is possible to adjust the weights based on the demand of low-rate legacy clients. 
In the second approach, the demand of clients is merged with their rates when formulating the problem.

\subsubsection{\textbf{Decision Metrics}}
\label{asc_dec_met}
The most important component of an AsC mechanism is its decision metrics, as they affect on both accuracy and overhead.
When using the AsC mechanism of 802.11 standard (i.e., SSF), a client associates with the AP that is providing the highest RSSI. 
The main shortcoming of SSF is that it does not recognize the load of APs and the demand of clients.
Therefore, AsC mechanisms introduce additional metrics in their decision process. 

The "Decision Metric" column of Table \ref{DAMtable} summarizes the decision metrics employed.
Generally, our review shows that the most popular decision metrics are RSSI, AP load, throughput of clients, and packet exchange delay experienced by clients. 
Balancing the load of APs is a widely-adopted optimization metric as it implicitly results in an improved performance of clients.
In the mechanisms relying on this metric, load is modeled through parameters such as the number of associated clients, AP throughput, and channel busy time. 
Although these mechanisms implicitly improve the performance of clients, explicit consideration of clients' traffic results in higher performance.

Based on this, we can classify AsC mechanisms into two groups:
\begin{itemize}
	\item \textit{Demand-agnostic.} 
	Refers to the AsC mechanisms that associate each client with the best AP nearby, without taking into account clients' demands.
	For instance, DenseAP, ACPF and BestAP associate each client with the nearby AP that is providing the highest available bandwidth, independent of the current demand of clients. 
	Similarly, AGE and MCAC associate each client with the AP that is providing the minimum delay, without considering the tolerable delay of clients in the decision making process. 	
	As a result, demand-agnostic approaches may perform unnecessary associations, which incur handoff overheads that result in performance degradation.	
	
	\item \textit{Demand-aware.}
	Refers to the AsC mechanisms that explicitly include the demand of clients in their decision making process.
	Unfortunately, the number of demand-aware AsC mechanisms (e.g., ACHC, and DALB) is very limited.
\end{itemize}
%

 %these works take into account the throughput demand

Throughput demand, in particular, is composed of uplink and downlink traffic.
Unfortunately, most of the AsC mechanisms assume downlink traffic is dominant when calculating their association decision metrics, as Table \ref{DAMtable} shows. 
However, the emerging applications of SDWLANs justify the importance of uplink traffic.
For example, medical monitoring, industrial process control, surveillance cameras, and interaction with cloud storage, require timely and reliable uplink communication \cite{MARS,RTwifi,REWIMO,IOT-future1, IOT-future2, singh_survey_2014}.

These discussions reveal that designing AsC mechanisms to support both uplink and downlink requirements is an open problem. 
Additionally, due to the resource-constraint nature of IoT devices, it is important to define, compute and include metrics, such as channel contention intensity and packet loss rate, in the decision making process.

In addition to the mentioned challenges, it is also important to identify and address the effect of hardware on the metrics used.
For example, the RSSI perceived by a tiny IoT device would be different from the value perceived by a multi-antenna smartphone, under the same condition.
Therefore, the controller cannot generate a realistic network map based on the information collected from clients.
Similarly, for environments with highly asymmetric links (for example caused by shadowing), the information collected in the controller based on the RSSI perceived at the APs does not reflect the real connectivity and interference relationship.
Understating the effect of metric evaluation irregularity on performance and the calibration of metrics and inclusion in the AsC mechanisms are open research areas.

\subsubsection{\textbf{Dynamicity and Overhead}}
\label{asc_dyn_overhead}
Dynamic and scalable AsC requires minimizing overhead from the following point of views:
\begin{itemize}
\item  \textit{Measurement delay}: the delay of measuring decision metrics must be short to reflect network dynamics;
\item  \textit{Bandwidth requirements}: the amount of bandwidth consumed for exchanging control data should be minimized to enhance scalability. 
Each association command requires communication between the controller and APs.
In addition, in a VAP-based architecture such as Odin, each association requires a VAP transfer, where the number of transfers depends on the number of associations.
Therefore, both the frequency and the scale of reassociation should be taken into account for various types of architectures;
\item \textit{Deployment cost}: it is ideal to define decision metrics that do not require AP or client modification.
For example, although BestAP  claims low measurement overhead in terms of delay and bandwidth, it requires the modification of APs and clients. 
\end{itemize}

Unfortunately, the existing AsC mechanisms do not take into account the effects of these three factors, the scalability of the proposed mechanisms has not been studied thoroughly, and their effect on the energy consumption of smartphones and resource constraint devices is unknown.
%For example, although DenseAP and Odin report low measurement overhead, a small testbed is used for their evaluation.

%Therefore, investigations are required to establish a balance between the accuracy and overhead of measuring network status metrics.
Although some AsC mechanisms propose their own measurement techniques, continuous and low-overhead network monitoring should be provided as a SDWLAN architectural service to facilitate the development and interoperability of control mechanisms.
For example, as we will see in Section \ref{CMmech}, dynamic ChA mechanisms share some of the metrics used by AsC mechanisms.
%To this end, a shared repository (managed by the OS) could be inquired by network applications.
We propose the followings to provide network applications with low-overhead and continuous measurement of metrics:
(i) designing metrics that can be used by multiple control mechanisms; 
(ii) efficient encoding of control data conveyed by a south-bound protocol, 
(iii) novel hardware/software techniques deployed on network infrastructure devices for passive (i.e., without introducing extra traffic) measurement of network status, 
and (iv) inference and prediction algorithms that extract new metrics and provide insight into the future status of the network.
Compared to distributed implementations, SDWLANs enable in-depth analysis of clients' traffic type and pattern by looking into packet headers \cite{gibb2013design,akyildiz2014roadmap,ng2015developing,Yoon2017,oliveira_characterizing_2016}.
This information could be employed to improve the accuracy of traffic prediction and interference modeling.
Specifically, while the existing approaches rely on the saturated demands of clients when estimating load and interference levels, the throughput and airtime of clients and APs strongly depend on the protocols used at the transport and application layers \cite{sinky_analysis_2015,IOT-future1,hobfeld_challenges_2012}.
For example, the periodic nature of an IoT device's uplink traffic could be exploited to improve AsC performance.
We believe that learning and prediction mechanisms must be employed by SDWLANs to provide network control applications (such as AsC and ChA) with meaningful, realistic, and predictive knowledge regarding network operation.

%In particular, interference modeling is usually done based on physical and MAC layer features in a saturated condition (nodes always have a packet to send). 
%On the other hand, traffic load modeling of APs is performed based on the measurement of throughput or airtime during a specific period and prediction of future demands. 

% REVISION

\textit{Hybrid design} is another approach towards achieving timely reactions against dynamics.
The overhead of network monitoring and control can be improved through hybrid control mechanisms, as proposed by SoftRAN \cite{gudipati2013softran}.
Due to the high dynamics of wireless networks, it is desirable to design control mechanisms that can make control decisions locally, while improving decisions based on the global network view as well.
This solution could also benefit from hierarchical controller topologies to make decisions at multiple levels, depending on the dynamicity of variations.
For example, AeroFlux (Section \ref{Archs}) proposes an architecture to enable the design of multi-level control mechanisms.
Such mechanisms make control decisions in a distributive manner, at local controllers or by a central controller, based on network dynamics, time constraints, and overhead of exchanging control messages.

% {\color{blue!50!black}
% \subsubsection{\textbf{Effects of Virtualization}}
% %REVISION
% Mobility management becomes more challenging when network virtualization is employed.
% When a client requires a new point of association due to its mobility, in addition to parameters such as fair bandwidth allocation, the available resources of APs should also be taken into account. 
% More specifically, for a client belonging to slice $n$, the AsC mechanism should ensure that after the association of this client with a new AP, the the QoS provided by slice $n$ and other slices is not violated.
% However, as this may require client steering, the cost of re-associations should be minimized.

% % In fact, when a user switches between different slices of a network, the controllers should synchronize the operation of these slices to ensure seamless connectivity.
% % However, achieving a low-overhead and fast collaboration among these controllers is a challenging task.
% % In addition, as mobility results in variations of link quality and capacity, dynamic resource allocation mechanisms become necessary to ensure fair allocation of resources.
% }

\subsubsection{\textbf{Security Considerations}}
%Although such metrics enable network applications to implement centralized control mechanisms, they should not compromise network performance because of security breaches.
As centralized mechanisms rely on global network information and usually aim to achieve a network-wide optimization, receiving malicious reports has a more severe effect on performance, compared to distributed control.
Verifying the legitimacy of network status reports and clients' demands have not yet been addressed by SDWLAN architectures and AsC mechanisms.
%

%

%%%%%%%%%%%%%%%%%%%%%%%%%%%%%%%%%%%%%%%%%%%%%%%%%%%%%%%%
%%%%%%%%%%%%%%%%%%%%%%%%%%%%%%%%%%%%%%%%%%%%%%%%%%%%%%%%
%%%%%%%%%%%%%%%%%%%%%%%%%%%%%%%%%%%%%%%%%%%%%%%%%%%%%%%%
%%%%%%%%%%%%%%%%%%%%%%%%%%%%%%%%%%%%%%%%%%%%%%%%%%%%%%%%
%%%%%%%%%%%%%%%%%%%%%%%%%%%%%%%%%%%%%%%%%%%%%%%%%%%%%%%%
%%%%%%%%%%%%%%%%%%%%%%%%%%%%%%%%%%%%%%%%%%%%%%%%%%%%%%%%
%%%%%%%%%%%%%%%%%%%%%%%%%%%%%%%%%%%%%%%%%%%%%%%%%%%%%%%%
%%%%%%%%%%%%%%%%%%%%%%%%%%%%%%%%%%%%%%%%%%%%%%%%%%%%%%%%

% REVISION

\section{Centralized Channel Assignment (ChA)}
\label{CMmech}
In this section, we review centralized ChA mechanisms.
To be consistent with Section \ref{AMmech}, we focus on the metrics employed as well as the problem formulation and solution proposed by these mechanisms.
In addition, we highlight the performance improvements achieved, compared to distributed mechanisms.

Considering the broadcast nature of wireless communication as well as the dense deployment of APs, ChA is an important network control mechanism to reduce co-channel interference and improve capacity.
Basically, a WLAN can be modeled as a conflict graph where the vertices represent APs, and edges reflect the interference level between APs. 
The color of each vertex is the channel assigned to that AP. 
The objective of a ChA mechanism is to color the graph using a minimum number of colors in order to improve channel reuse.

A typical channel assignment algorithm relies on all or a subset of the following inputs: (i) the set of APs and clients, (ii) AP-client associations, (iii) interference relationship among APs and clients, and (iv) total number of available channels. 
Using the aforementioned information, optimization metrics, such as interference, throughput, spectral efficiency, and fairness, are used to define an optimization problem. 
In this section we study centralized ChA mechanisms.

\begin{figure}[!t]
	\centering
	\includegraphics[width=0.85\linewidth]{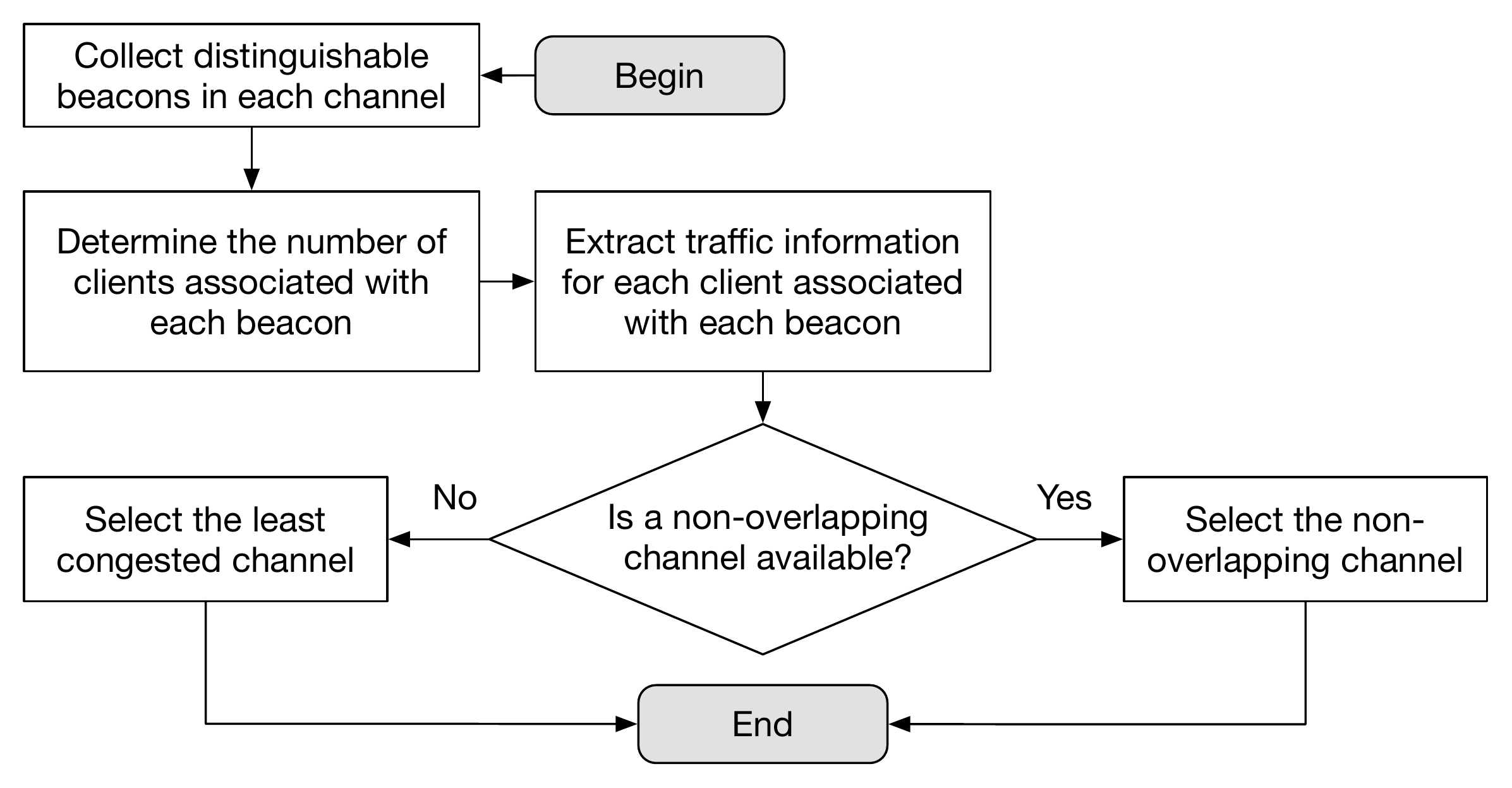}
	\caption{Least Congested Channel Search (LCCS) algorithm \cite{LCCS}.}
	\label{fig_LCCS}
\end{figure}

Figure \ref{fig_LCCS} represents the flowchart of the \textbf{least congested channel search (LCCS}) \cite{LCCS} algorithm. 
Although LCCS is not a central ChA mechanism, we briefly discuss its operation as it has been used by commercial off-the-shelf APs and adopted as the baseline for performance comparison of various ChA mechanisms.
In this algorithm, APs scan all channels passively after startup. 
Each AP collects the beacon frames from its neighboring APs and extracts the number of connected clients and their traffic information. 
Then, the AP selects a non-overlapping channel if it is available. 
Otherwise, the AP selects the channel with lowest traffic sum.
To obtain client and traffic information, LCCS relies on the use of optional fields in beacon packets.
Therefore, the use of this mechanism requires vendor support.
%For example, Cisco APs support these features.

A simpler approach, which has been adopted as the comparison baseline by some ChA mechanisms, is to simply sample the RSSI of nearby APs on all channels and choose the channel with the minimum RSSI level.
The literature refers to this approach as \textbf{minRSSI}.

%\todo[inline,color=cyan]{$\checkmark$ how each ap monitors all packets of neighboring aps?}
%\todo[inline,color=cyan]{$\checkmark$ when an ap switches to another channel, then that cannot be involved in tx/rx}  
%\todo[inline,color=cyan]{$\checkmark$ what do you mean it does not consider traffic demand? -clarify} 

We overview ChA mechanisms, categorized into two groups: (i) \textit{traffic-agnostic}, and (ii) \textit{traffic-aware}, as follows.

%Traffic-agnostic techniques do not consider the traffic load of APs and the traffic demands of clients. 
%In contrast, traffic-aware techniques make channel assignment considering the traffic information of APs/clients. 

\subsection{{Traffic-Agnostic Channel Assignment}} 
\label{ChA_traffic_ag}
These mechanisms do not include the traffic load of APs and the traffic demand of clients in their channel assignment process.
In fact, these mechanisms mainly rely on the interference relationship to formulate their graph coloring problems.

\begin{figure}[!t]
	\centering
	\includegraphics[width=0.6\linewidth]{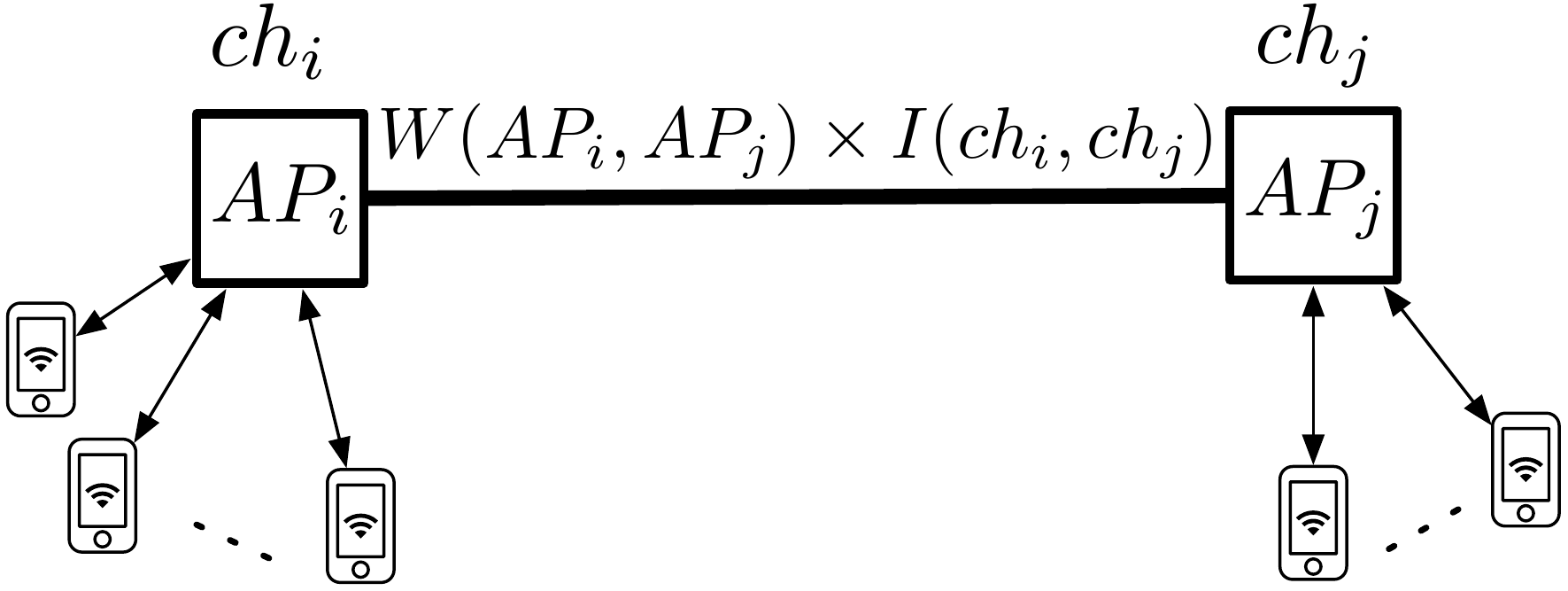}
	\caption{Weighted coloring channel assignment (WCCA) \cite{Wcolor-2005} formulates a weighted graph coloring problem. This figure represents the I-value of the conflicting edge between $AP_{i}$ and $AP_{j}$.}
	\label{fig_coloring}
\end{figure}

\textbf{Weighted coloring channel assignment (WCCA).} 
\label{WCCA}
This seminal work \cite{Wcolor-2005} formulates channel assignment through the weighted graph coloring problem and improves channel re-use by assigning partially-overlapping channels to APs. 
%The proposed model is aware of client distribution by trying to assign less-overlapping channels to the APs with more associated clients. 
Vertices are APs, and edges represent the potential interference between corresponding APs (see Figure \ref{fig_coloring}). 
The colors of the vertices represent the assigned channel number to APs. 
The number of clients associated with two neighboring APs and the interference level between them are used to calculate the weight of conflicting edges. 
\textit{Site-Report} is the proposed method to construct the overlap graph and capture network dynamics. %and provide required information to accurately estimate $W(AP_{i},AP_{j})$ between all AP pairs. 
This method enables APs to request their clients perform channel scanning. 
A list of active APs is generated for each channel.
This list includes the APs in direct communication range and the APs whose associated clients are in direct communication with the client performing Site-Report scan. 
Based on the results of Site-Report, $W(AP_{i},AP_{j})$ is computed as follows,
\begin{equation}
W(AP_{i},AP_{j})= \frac{N_{AP_{i},AP_{j}} + N_{AP_{j},AP_{i}}}{N_{AP_{i}}+N_{AP_{j}}},
\end{equation}
where $N_{AP_{j}}$ is the number of site reports performed by the clients of $AP_{j}$, and $N_{AP_{i}, AP_{j}}$ is the number of site reports of $AP_i$'s clients that reported interference with $AP_j$ or its associated clients. 

The I-factor, $I(ch_i,ch_j)$, is the normalized interference factor between two channels $ch_i$ and $ch_j$.
%------
The value of $I(ch_i,ch_j)$ is computed as follows: 
(i) for two non-overlapping channels (e.g., channel 1 and 6 in the 2.4GHz band) the value is 0; 
(ii) if $ch_i=ch_j$, the value is 1; 
(iii) for two partially-overlapping channels $ch_i$ and $ch_j$, the value is between 0 and 1, depending on the distance between their central frequency.
%-------

The weight of the conflicting edge between $AP_{i}$ and $AP_{j}$ is defined by $Ivalue(AP_{i},AP_{j}) = W(AP_{i},AP_{j})\times I(ch_i,ch_j)$.  
Having $K$ colors (channels), the problem is determining how to color the graph with minimum number of colors in order to optimize the objective function. 
Three different objective functions, $O_{max}$, $O_{sum}$ and $O_{num}$, were used. 
$O_{max}$ represents the maximum I-value amongst all the links.
$O_{sum}$ is the sum of I-values of all conflicting edges. 
The total number of conflicting edges is represented by $O_{num}$. 
In order to minimize the objective functions, two distributed heuristic algorithms are proposed: (i) \textit{Hminmax} aims to minimize $O_{max}$ without AP coordination, and (ii) \textit{Hsum} aims to jointly minimize $O_{max}$ and $O_{sum}$, which requires coordination between APs in order to minimize the interference level of all conflicting edges. %The authors rename \textbf{Hsum} and \textbf{Hminmax} algorithms by \textbf{ADJ-sum} and \textbf{ADJ-minmax} if partially overlapping channels 
Although the proposed algorithms were run distributively, it would be easier to provide them with their required information using a SDWLAN architecture.

The proposed algorithms have been evaluated using NS2 \cite{NS2} simulator and testbed. %on a network consisting 20 APs. 
Two different scenarios are used in simulations: (i) three non-overlapping channels and (ii) partially overlapping channels (11 channels in the 2.4GHz band). 
For the first scenario, the algorithms showed a 45.5$\%$ and 56$\%$ improvement in interference reduction for sparse and dense networks, respectively. 
For the second scenario, there is a 40$\%$ reduction in interference, compared to LCCS. 
%\todo[inline, color=cyan]{what is the network topology?-sparse or dense \\$\checkmark$ \textit{20 APs}} 

%%%%%%%

\begin{figure}[!t]
	\centering
	\includegraphics[width=0.65\linewidth]{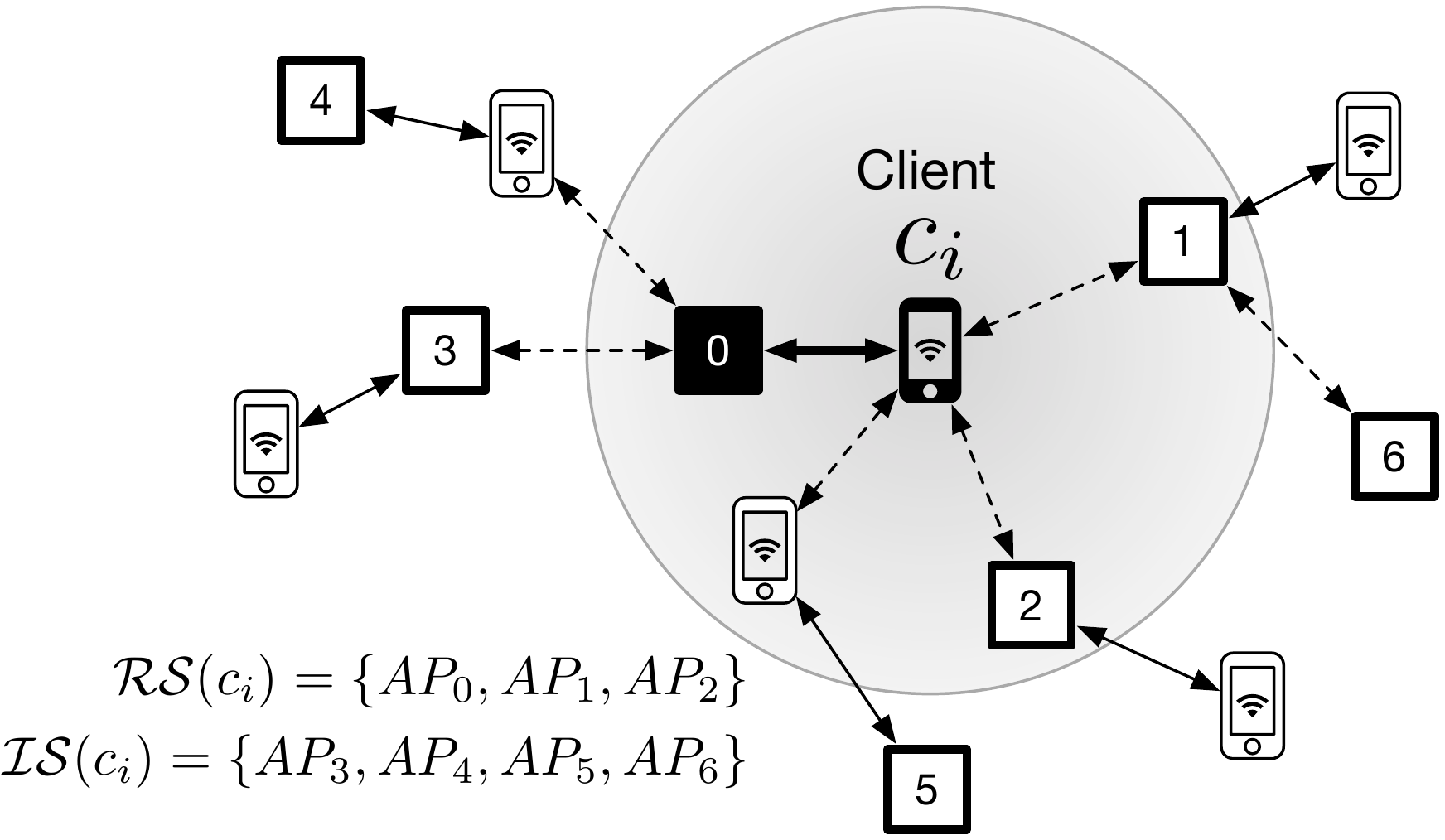}
	\caption{Client-driven channel assignment (CDCA) \cite{J-DCA-LB-2006} assigns two sets per client: \textit{range set} (denoted by $\mathcal{RS}(c_{i})$) and \textit{interference set} (denoted by $\mathcal{IS}(c_{i})$). APs are specified by numbered squares. %Note that the interference set of the client contains all APs that can cause interference on the client directly or via their associated clients.
	}
	\label{fig-conflict-set-coloring}
\end{figure}

\label{CDCA}
\textbf{Client-driven channel assignment (CDCA).}
\cite{J-DCA-LB-2006} formulates a \textit{conflict set coloring} problem and proposes a heuristic algorithm that increases the number of conflict-free clients at each iteration. 
Here, \textit{conflict} refers to the condition in which two nodes (APs or clients) that belong to different BSSs use the same channel.
To measure the interference level experienced by each client, measurement points are selected, and the signal level received from each AP is measured at those points. 
Based on this information, there are two sets assigned to each client: (i) \textit{range set}: the APs that can communicate with the client directly, and (ii) \textit{interference set}: the APs that cannot communicate with the client, but can cause interference. 
Figure \ref{fig-conflict-set-coloring} shows the range set and interference set for a given client.
%
% REVISION - CAPTION REDUCED

This work defines two objective functions: (i) one that maximizes the number of conflict-free clients, and (ii) one that minimizes the total conflict in the network.
To achieve the first objective, if a channel $ch$ is assigned to an AP, then no AP in the range set and interference set of the clients associated with that AP should have channel $ch$ assigned to them.
A client satisfying this condition is referred to as a "conflict-free client".
%Note that due to association, there is only one AP that belongs to the range set of the client and uses channel $ch$. 
Since it may not be possible to satisfy this requirement for all the clients, the algorithm minimizes the overall network conflict.
%\todo[inline,color=cyan]{$\checkmark$ make sure the word "try" is correct here} 

%The authors propose the \textit{conflict-free assignment with randomized compaction} (CFAssign-RaC) algorithm. 
The main step of the channel assignment algorithm randomly selects an AP and assigns a channel that results in the maximum number of conflict-free clients. 
This step is repeated for all APs, and the order of APs is determined by a random permutation. 
%After channel assignment to APs, the number of conflict-free clients is re-calculated. 
This process is repeated as long as it increases the number of conflict-free clients. 
%Since CFAssign-RaC operates based on a randomized order of APs, it is run for several times (with different random permutations) in order to find the best solution among the outputs of different runs.

%After applying CFAssign-RaC, there might be clients left with high level of interference.
After the first step, there might be clients that are left with a high level of interference.
To remedy this condition, the conflict level of all non-conflict-free clients is balanced.
This mechanism implicitly results in traffic load balancing.
%To handle mobility, CFAssign-RaC is either run periodically or it is triggered when the objective function is higher than a threshold.
%The superiority of the proposed algorithms has been demonstrated through simulation and testbed, in terms of the number of MAC collisions, throughput and per-packet delay, compared to LCCS \cite{LCCS}.

%TABLE
%  (1) approach: CA and Load balancing
%  (2) interference model: RSSI-based
%%% (3) problem formulation: Conflict set coloring 
% (4) objective function: number of conflict-free client, Total conflict of all WLAN clients
% (5) channels: non-overlapping (802.11b)
% (6) Client/AP-centric: Client-centric
% (7) Client-aware: Yes.
%%% (8) Perfomance evaluation: simulation, real implementation
%%% (9) Dynamic/Static: Dynamic
%%% (10) Optimality: Sub-optimal
% (11) Measurement: 40 usage-points, compute range and interference sets, based on test clients
%    --> Time and computational complexity is not discussed in paper
%%% (12) Requried collecting information: scan all channels => range set and interference set

\label{FPLN}
\textbf{Frequency planning in large-scale networks (FPLN).}
\cite{CAPWAP-based-CA-11} models the channel assignment problem as a weighted graph coloring problem taking into account the external interference generated by non-controllable APs. 
The interference level caused by $AP_{j}$ on $AP_{i}$, which is the weight of edge $AP_{j} \rightarrow AP_{i}$, is defined as follows,
\begin{equation}
W(AP_{i},AP_{j})= \mathcal{C}^{active}_{AP_{i}}  \times I(ch_i,ch_j) \times P(AP_{i},AP_{j}),
\end{equation}
where $\mathcal{C}^{active}_{AP_{i}} $ is the number of active associated clients with $AP_{i}$. 
An associated client is called \textit{active} in a time interval if it is sending/receiving data packets.
The interference factor between two channels assigned to $AP_{i}$ and $AP_{j}$ is given by $I(ch_i,ch_j)$. 
The power received at $AP_{j}$ from $AP_{i}$ is represented by $P(AP_{i},AP_{j})$. 
%This model neglects transmission power asymmetry (i.e., it assumes $P(i,j)=P(j,i)$). 
The interference factor of $AP_{i}$ is defined as $\sum_{\forall AP_{j} \in \mathcal{AP}}W(AP_{i},AP_{j})$.
The objective function is defined as the sum of all APs' interference factor, including controllable and non-controllable APs. 
The ChA mechanism aims to minimize the objective function when the channels of non-controllable APs are fixed.

The authors propose a two-phase heuristic algorithm to solve the above optimization problem. 
In the first phase, the controllable APs are categorized into separate clusters based on neighborhood relationships, and the local optimization problem is solved for each cluster by finding an appropriate channel per AP to minimize its interference level while the channels of other APs (in the cluster) are fixed. 
This process is started from the AP with largest interference factor. 
In the second phase, a pruning-based exhaustive search is run on each cluster, starting from the AP with highest interference level. 
The algorithm updates the channels of APs to reduce the total interference level of the cluster. 
The pruning strategy deletes the failed channel assignments (i.e., the solutions that increase the interference level obtained in the first phase) from the search space.
FPLN uses CAPWAP (see Section \ref{CUWNarch}) as its south-bound protocol. 
Simulation and testbed experiments show about a 1.2x lower interference level, compared to LCCS.

\textbf{CloudMAC.}
\label{CloudMAC_CM}
The CloudMAC \cite{CloudMAC,CloudMAC3} architecture allows multiple NICs of a physical AP to be mapped to a VAP, where NICs may operate on different channels. 
These NICs  periodically monitor and report channel utilization to the controller. 
Unfortunately, it is not clear how channel measurement is performed.
% controller is OpenFlow \cite{OpenFlow} controller
If a client is operating on a high-interference channel, the controller sends a 802.11h channel switch announcement message to the client (mandatory for 802.11a/n standards), instructing it to switch to another channel with lower interference. 
Since a client is associated with a VAP, the client does not need to re-associate and it can continue its communication through the same VAP using another NIC of the same physical AP.
Additionally, this process does not require any client modification.
Unfortunately, the performance evaluation of this mechanism has not been reported.

%%%%%%%%%%
\label{Odin-CM}
\textbf{Odin.}
\cite{Odin2} proposes a simple channel assignment strategy. 
For each AP, the channel assignment application (running on Odin controller) samples the RSSI value of all channels during various operational hours.
The heuristic algorithm chooses the channel with the smallest maximum and average RSSI for each AP.
Unfortunately, the performance of this mechanism has not been evaluated.

\textbf{Primary channel allocation in 802.11ac networks (PCA).}
%1
\label{PCA}
The \textit{hidden channel} (HC) problem is addressed in \cite{802.11ac-PCA} for 802.11ac networks. 
The HC problem occurs when an AP interferes with the bandwidth of another AP.
This problem occurs due to the heterogeneity of channel bandwidths, different CCA thresholds for primary and secondary channels, and bandwidth-ignorant fixed transmission power. 

\begin{figure}[!t]
	\centering
	\includegraphics[width=0.7\linewidth]{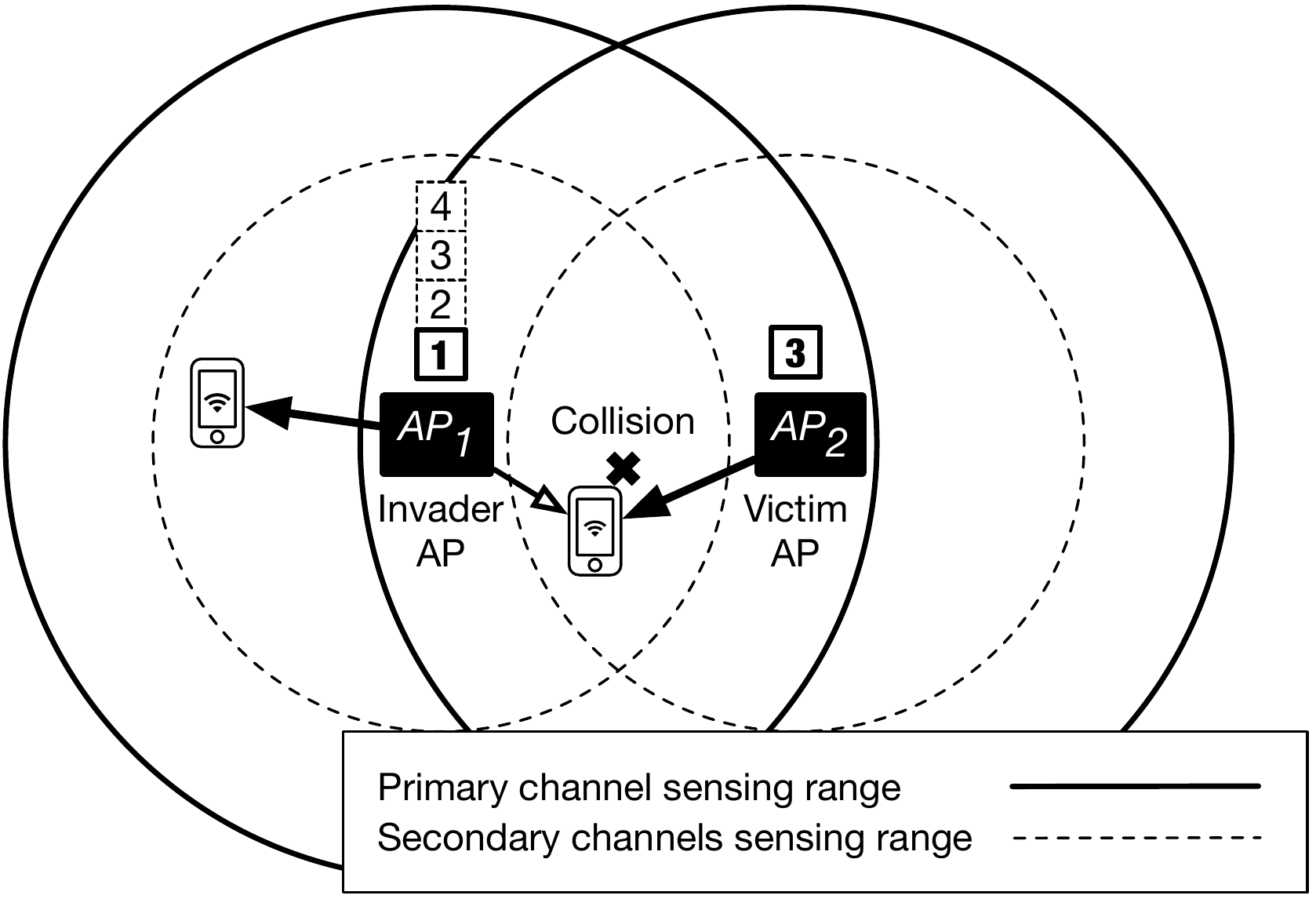}
	\caption{An example of the hidden-channel (HC) problem in 802.11ac \cite{802.11ac-PCA}.}
	\label{invading}
\end{figure}

Figure \ref{invading} illustrates an example of the HC problem. 
The bandwidth of $AP_{1}$ is 80MHz, including four 20-MHz channels 1, 2, 3 and 4, where channel 1 is its primary channel. 
The primary channel sensing range is greater than the secondary channel sensing range due to the difference in CCA thresholds for primary and secondary channels in 802.11ac. 
$AP_{2}$ uses a 20MHz bandwidth operating on channel 3. 
Suppose $AP_{2}$ is communicating with its clients on channel 3. 
When $AP_{1}$ performs carrier sensing on its channels, it cannot detect the presence of $AP_{2}$ on channel 3 because $AP_{2}$ is not in the secondary sensing range of $AP_{1}$. 
Therefore, $AP_{1}$ starts to send data on channels 1, 2, 3 and 4, and invades $AP_{2}$. 
%In this scenario, $AP_{1}$ invades the bandwidth of $AP_{2}$.
%
%
%
%2

In order to study the impact of the HC problem, the authors proposed a Markov chain-based analysis to show the impact of packet length and MAC contention parameters on the performance of APs. 
Furthermore, the effect of various HC scenarios on packet error rate (PER) has been investigated. 
%4
The authors used the graph coloring problem, where APs are vertices and channels are colors, to model the \textit{primary channel allocation} (PCA) problem. 
The edges represent the invasion relationship between APs. 
It is assumed that a controller collects information about the bandwidth of APs and their interference relationship.
The objective is to color the vertices in order to: (i) minimize the interference/invasion relationship between APs, and (ii) maximize channel utilization. 
The problem is formulated as an integer programming optimization problem, which is NP-hard.
%5
A heuristic primary channel allocation algorithm is introduced to solve the problem.
%6
Simulation results show the higher performance of PCA compared to minRSSI and random channel assignment.
Furthermore, PCA is a close-to-optimal solution when compared to the optimal exhaustive search algorithm.

\textbf{EmPOWER2.}
\label{EMPOWER2ChA}
As mentioned in Section \ref{EMPOWERarch}, EmPOWER2 \cite{Primitives} establishes \textit{channel quality and interference map} abstraction at the controller.
The proposed ChA mechanism uses the APIs provided to traverse the map and detect uplink/downlink conflicts between LVAP pairs.
There is an edge between two nodes in the interference map if they are in the communication range of each other.
In addition, the weight of each link corresponds to the channel quality between nodes.
The graph coloring algorithm proposed in \cite{san2012new} is used to assign channels based on the conflict relationships.

\label{Wi5CM}
\textbf{Wi-5 channel assignment (Wi5CA).} 
\cite{DCA-2} proposes a ChA mechanism as part of the Wi-5 project \cite{Wi-5}.
A binary integer linear programming problem is formulated with objective function 
$\textbf{U} = \textbf{G}\times \textbf{A}^{T} . \textbf{I}$, 
where $'\times'$ and $'.'$ are matrix multiplication and element-wise multiplication operators, respectively. $\textbf{G}\in\{0,1\}^{|\mathcal{AP}| \times |\mathcal{AP}|}$ represents the network topology. 
If the average power strength of $AP_i$ on $AP_j$ exceeds a specific threshold, the value of $G_{AP_{i}, AP_{j}}$ will be 1, otherwise it will be 0. $\textbf{A}\in\{0,1\}^{|\mathcal{CH}|\times |\mathcal{AP}|}$ is the current channel assignment for APs.
If channel $ch_i$ is assigned to $AP_j$, the value of $A_{{ch_{i},AP_{j}}}$ is 1, otherwise it is 0. 
$\textbf{I}\in \mathbb{R} ^{|\mathcal{AP}| \times |\mathcal{CH}|}$ represents the predicted interference matrix, where $I_{AP_{i}, ch_{i}}$ is the interference level predicted for $AP_i$ if channel $ch_j$ is assigned to it.
Unfortunately, the interference prediction method is unclear.  
The channel assignment algorithm minimizes 
\begin{equation}
\sum_{\forall AP_{i} \in \mathcal{AP}}\;\sum_{\forall ch_{j} \in \mathcal{CH}} U_{AP_{i}, ch_{j}}, 
\end{equation}
which is the network-wide interference level.
%The authors did not report details about the calculation/measurement of predicted interference matrix $\textbf{I}$. 
%\todo[inline,color=cyan]{$\checkmark$ what do you mean by the "predicted" level of interference?} 
A controller executes the ChA mechanism when the interference level is higher than a threshold.
%However, the controller updates its global knowledge about all APs in the network using OpenFlow. 
%The proposed algorithm is evaluated through MATLAB in terms of three performance metrics: interference level, SINR and spectral efficiency. 
MATLAB simulations show 2dB and 3dB reduction in the average interference level, compared to LCCS and uncoordinated channel assignment, respectively.

%%%%

%  (1) approach: CA
%  (2) interference model: RSSI-based
%%% (3) problem formulation: weighted graph vertex coloring 
% (4) objective function: Total interference of WLAN: Sum of interference level between AP-AP pairs
% (5) channels: (not mentioned: the modeling is general and both partially overlappig and non-overlap channels can be applied.)
% (6) Client/AP-centric: AP-centric
% (7) Client-aware: Yes, the count of active clients associated to APs
%%% (8) Perfomance evaluation: simulation, real implementation
%%% (9) Dynamic/Static: not mentioned (I think, it is static,  But, it can be run in real-time.) --> because it is assumed that the relationship between APs is known.
%%% (10) Optimality: Sub-optimal
% (11) Measurement: Statistical --> non-realtime (column 9)
%%% (12) Requried collecting information: RSSI of APs 

\subsection{Traffic-Aware Channel Assignment} 
\label{ChA_traffic_aware}
Traffic-aware channel assignment mechanisms include the traffic of APs, clients or both in their decision process.
As measuring the effect of interference on performance is complicated, these mechanisms explicitly include metrics such as throughput, delay, and fairness, in their problem formulations.

\label{APCA}
\textbf{AP placement and channel assignment (APCA).} 
\cite{J-AP-DCA-2006} addresses AP placement and channel assignment in order to improve the total network throughput and the fairness established among clients. 
The throughput of clients is estimated as a function of MAC-layer timing parameters, network topology, and data rate of clients. 
The data rate of each client (11, 5.5, 2 or 1Mbps) depends on the link's RSSI. 
The fairness among clients is defined as the throughput deviations of clients based on the Jain's fairness index \cite{Jain's-fairness} as follows,
\begin{equation}
\label{eq-fairness}
J  = \frac{(\sum_{\forall c_{i} \in \mathcal{C}}Th_{c_{i}})^2}{|\mathcal{C}|\sum_{\forall c_{i} \in \mathcal{C}}{(Th_{c_{i}}})^2},
\end{equation}
where $Th_{c_{i}}$ is the throughput of client $c_i$.
Maximizing the objective function, defined as $J\times \sum_{\forall c_{i} \in \mathcal{C}} Th_{c_{i}}$, leads to optimizing both total throughput and fairness among clients. 
%Since exhaustive search is not feasible to find the optimal solution of AP placement and channel assignment, 
The authors propose a heuristic local search method, named the \textit{patching} algorithm. 
In each iteration of this algorithm, an AP is placed on one of the predefined locations, and a non-overlapping channel is assigned to the AP to maximize the objective function.
This process is repeated until a predefined number of APs are placed. 
Simulation and testbed results show that the proposed patching algorithm provides close-to-optimal solutions in terms of total throughput and fairness. 
APCA is used only for the initial network configuration phase.

%\todo[inline,color=cyan]{$\checkmark$ you mostly discuss about the goals- there is not enough discussion about the approach} 
%\todo[inline,color=cyan]{$\checkmark$ it is not clear how ap placement works} 
%  (1) approach: CA, Fairness, AP placement
%  (2) interference model: RSSI-based ==> Client throughput
%%% (3) problem formulation: Throughput per client
% (4) objective function: Throughput and fairness
% (5) channels: non-overlapping (802.11b)
% (6) Client/AP-centric: Client-centric
% (7) Client-aware: Yes. Client throughput and fairness among clients' throughput
%%% (8) Perfomance evaluation: simulation
%%% (9) Dynamic/Static: Static
%%% (10) Optimality: Sub-optimal
% (11) Measurement: passive 
%    --> Time and computational complexity is not discussed in paper
%%% (12) Requried collecting information: RSSI of clients
%\subsubsection{\textbf{Client-agnostic and traffic demand-aware techniques}} In this section, we survey the techniques which consider only the traffic demand of APs to make the channel assignment decisions.

\label{TACA}
\textbf{Traffic-aware channel assignment (TACA).}
% solution 
\cite{Traffic-aware-CA-2007} assigns weights to APs and clients proportional to their traffic demands. 
The \textit{weighted channel separation} (WCS) metric is
\begin{equation}
\label{interference-metric}
WCS = \sum_{\substack{i,j  \in \mathcal{AP} \cup \mathcal{C}, \\ BSS(i)\neq BSS(j) }} W(i,j)\times Separation(i,j),
\end{equation}
where $BSS(k)$ is the BSS including AP or client $k$, $Separation(i,j)$ is the distance between operating channels of node $i$ and node $j$, and $W(i,j)$ is a function of the traffic demands of $i$ and $j$.
$Separation(i,j)=\min{(|ch_i-ch_j|,5)}$, therefore, the maximum separation value is 5, which is the distance between orthogonal channels (e.g., channel 1 and 6).
%The traffic-aware weight between two nodes $i$ and $j$ is defined by $W(i,j)=\beta^{snd}_{i} \beta^{snd}_{j} + \beta^{snd}_{i}\beta^{rcv}_{j} + \beta^{snd}_{j}\beta^{rcv}_{i}$ where $\beta^{snd}_{i}$ and $\beta^{rcv}_{i}$ are sending and receiving demands of node $i$, respectively. 
%In particular, the following parameters are available per $AP_{i}$ every $\Delta t$ second: the number of packets/bytes sent ($Out_{AP_{i}}(t)$), the number of packets/bytes received ($In_{AP_{i}}(t)$), and the number of currently associated clients ($\mathcal{C}_{AP_{i}}$). 
%The sending demand of $AP_{i}$ at a given time $t$ is calculated as $\beta^{snd}_{AP_{i}}(t) = \frac{Out(t)-Out(t-\Delta t)}{\Delta t}$. 
%The sending demand for each client is approximated as $S_{c}(t) = \frac{In(t)-In(t-\Delta t)}{N\Delta t}$. 
Since all nodes are included in the calculation of the WCS metric, all AP-AP, AP-client and client-client interferences are taken into account. 

%
%\todo[inline, color=cyan]{$\checkmark$ how are the sending and receiving demands modeled?} 
%
%
%

\begin{figure}[!t]
	\centering
	\includegraphics[width=0.8\linewidth]{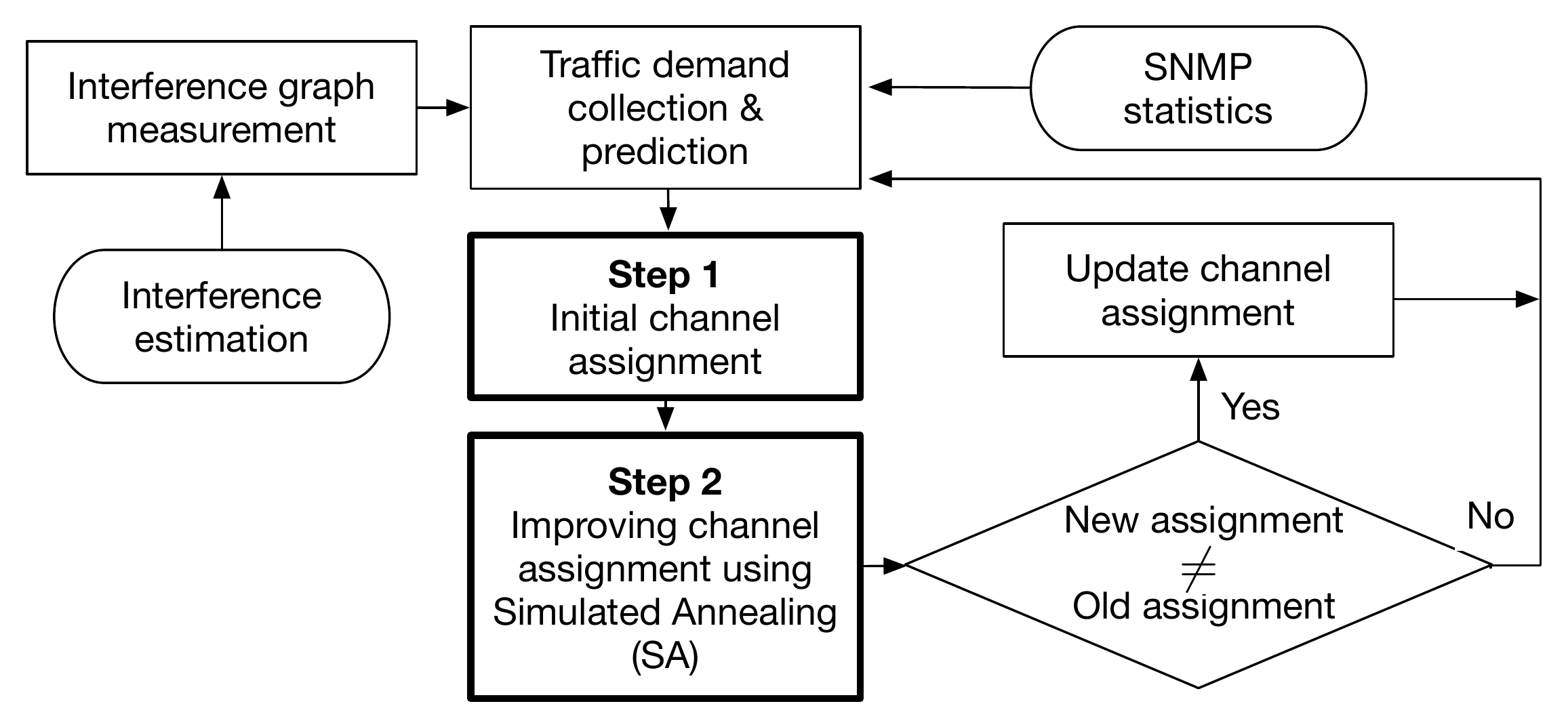}
	\caption{Traffic-aware channel assignment (TACA) \cite{Traffic-aware-CA-2007}.}
	\label{fig_traffic-aware}
\end{figure}

A heuristic algorithm has been proposed to maximize WCS. 
Figure \ref{fig_traffic-aware} shows the flowchart of this algorithm.
The interference graph measurement is performed a few times per day. % when the traffic load is light. 
The traffic demands of APs and clients are collected and predicted using the SNMP statistics collected from APs.
In "Step 1", an initial channel assignment is performed to maximize the WCS metric considering APs' traffic demands only. 
In "Step 2", a simulated annealing (SA) approach is used to update channel assignment, which maximizes WCS by taking into account the traffic demands of AP and clients. 
This step randomly selects an AP and its clients, and assigns a new channel at each iteration in order to maximize the WCS metric. 
%The algorithm is repeated 1000 times to obtain a close-to-optimal channel assignment. 
%\todo[inline,color=cyan]{$\checkmark$ what is the interference metric, and what do you mean by maximizing that?} 
%Performance evaluation is performed using NS2 simulator and a 25-node testbed to measure total network throughput. %for both TCP and UDP flows. 
%Three types of scenarios are considered: (i) client-aware/traffic-agnostic, (ii) client-agnostic/traffic-aware, and (iii) client-aware/traffic-aware. 
%The first scenario ignores traffic demands. The second scenario takes into account the traffic demands of APs, and the third scenario includes the traffic demands of APs and clients. 
Empirical results (using a 25-node testbed) show 2.6 times increase in TCP throughput, compared to a downgraded algorithm in which traffic demands are not included.
%The higher throughput of this traffic-aware channel assignment is presented, especially when the traffic demands are highly variable. 
The authors also showed that one channel switching per 5 minutes does not result in throughput degradation. 

\label{VDTCA}
\textbf{Virtual delayed time channel assignment (VDTCA).}
\cite{Measurement-CA-WCNC-10} introduces a new interference prediction model based on the signal strength and traffic demands of APs and clients. 
The extra transmission delay caused by interference between two APs is represented as $VDT = T_{int} - T_{n}$, where $VDT$ is \textit{virtual delayed time}. 
The term "virtual" refers to the calculation of this metric through interference prediction without actually changing channels.  
$T_{int}$ and $T_{n}$ are the time required to transmit a given amount of traffic in the presence and absence of interference, respectively.

%\todo[inline,color=cyan]{$\checkmark$ what do you mean by virtual? \\ \textit{- there is no directly justification in the paper on why they name it virtual !} \\ \textit{By the way, I added a sentence to justify it.} }  
The transmission time is a function of physical layer data rate, which depends on the signal SINR computed through measuring RSSI at receiver side.
%\todo[inline,color=cyan]{$\checkmark$ I modified the above sentence, is it correct?-- these sen} 
On the other hand, the traffic demands of APs and clients have a great impact on the transmission time of neighboring APs. 
For instance, when there is no traffic on an AP, there will be no transmission delay overhead on the neighboring APs and clients. 
Therefore, $T_{int}$ is calculated based on RSSI values and traffic rates of all neighboring APs and their associated clients.
%\todo[inline,color=cyan]{$\checkmark$ traffic rate of whom?}  
Using the proposed VDT model, it is possible to measure the VDT of each AP pair.
Channel assignment is modeled as a graph vertex coloring problem, which uses $VDT(BSS_i, BSS_j)$ as the edge weight between two interfering BSSs, where $BSS_i$ and $BSS_j$ are two APs and their associated clients.
Non-overlapping channels are the colors of vertices. 
The objective is to color the graph with a minimum number of colors while minimizing the sum of edge weights to achieve interference minimization. 
The semi-definite programming (SDP) \cite{SDP} relaxation technique is used to solve this problem.
The authors show that SDP  can solve the channel assignment problem in the order of seconds for a medium-size network. 
For instance, it takes 27.5 seconds to find the solution for a network with 50 APs and 9 non-overlapping channels. 
Testbed results show $30\%$ throughput improvement, compared to LCCS \cite{LCCS}. 

\textbf{Cisco unified wireless network (CUWN).}
\label{CUWN_CM}
In the CUWN architecture \cite{Cisco} (see Section \ref{CUWNarch}), based on the different RF groups established and the cost metric calculated for each AP, each controller updates the assigned channels periodically\footnote{The period can be adjusted by network administrators.}. 
The controller calculates a cost metric for each AP to represent the interference level of the AP, and prepares a list called \textit{channel plan change initiator} (CPCI), which includes the sorted cost metric values of all APs. 
The leader selects the AP with highest cost metric and assigns a channel to the selected AP and its one-hop neighboring APs in order to decrease their cost metrics. 
These APs are removed from the CPCI list, and this process is repeated for all remaining APs in the list.
A sub-optimal channel assignment is calculated through a heuristic algorithm\footnote{The details of ChA mechanism used by CUWN are not available.} that aims to maximize the frequency distance of selected channels. 
Note that a longer frequency distance results in a lower interference between APs. 
CUWN also enables network administrator to set channels manually.

\label{CAFA}
\textbf{Channel assignment with fairness approach (CAFA).}
\cite{CA-F-WCNC-11} formulates ChA as a weighted graph coloring problem where weights are assigned to the set of edges and vertices. 
The distance of overlapping channels is determined by the normalized interference factor, i.e., I-factor, introduced in \cite{Wcolor-2005} (see WCCA explained in Section \ref{WCCA}). 
The normalized throughput of each client is estimated using the approximation introduced in \cite{Thr-estimation-2005}. 
The weight of $AP_i$ is defined as the total \textit{normalized throughput reduction} of $AP_i$, which is calculated based on clients' normalized throughput and I-factor. 
The weight of the edge between two interfering APs represents the throughput reduction caused by their mutual interference.
Furthermore, the authors formulate the fairness among normalized throughputs of APs using Jain's fairness index \cite{Jain's-fairness}. 
The objective function is defined as the joint minimization of clients' throughput reduction and maximization of client's fairness index. 

%\todo[inline,color=cyan]{$\checkmark$ anything special about the heuristic?}  
An iterative heuristic algorithm is introduced to solve the optimization problem. 
The algorithm sorts the APs based on their weights (i.e., normalized throughput reduction).  
At each iteration, the AP with highest weight is selected for channel assignment. 
The channel of the selected AP is assigned so that the throughput reduction of the neighboring APs is minimized.
This process is repeated until channel assignment is performed for all APs. 
Simulation results show $15\%$ and $6\%$ improvement in total throughput for 4-AP and 8-AP scenarios, respectively, compared to WCCA \cite{Wcolor-2005}. 

%  (1) approach: CA and fairness
%  (2) interference model: Throughput, overlapping channel interference 
%%% (3) problem formulation: weighted graph vertex coloring 
% (4) objective function: normalized throughput reduction and fairness
% (5) channels: overlapping (802.11b)
% (6) Client/AP-centric: AP-centric
% (7) Client-aware: Yes, client throughput 
%%% (8) Perfomance evaluation: simulation
%%% (9) Dynamic/Static: Dynamic (no information about time complexity and information gathering complexity)
%%% (10) Optimality: Sub-optimal
% (11) Measurement: no information
%%% (12) Requried collecting information: clients throughput, [not mentioned in detail]

\label{TFACM}
\textbf{Throughput and fairness-aware channel assignment (TFACA).}
\cite{CA-VTC-14} uses spectrum monitoring information and the traffic demand of APs.
This mechanism defines the utility function of $AP_i$ when operating on channel $ch_{j}$ as,
\begin{equation}
\label{eqTFACM}
U(AP_{i}, ch_{j})=\frac{\min\left\{F(AP_{i},ch_{j}), T(AP_{i}) \right\}}{T(AP_{i})} ,
\end{equation}
where $F(AP_{i},ch_{j})$ is the free airtime of channel $ch_{j}$ on $AP_{i}$, and $T(AP_{i})$ is the total time required to send data of $AP_i$ in a time unit. 
Note that Equation \ref {eqTFACM} represents the percentage of $AP_i$'s data that can be sent over channel $ch_{j}$ in a given time unit.
Two objective functions are defined:
\begin{equation}
 \max\sum_{\forall AP_{i}\in\mathcal{AP}, \forall ch_{j} \in \mathcal{CH}}{U(AP_{i}, ch_{j})},
\end{equation}
and,
\begin{equation}
\max(\min_{\forall AP_{i}\in\mathcal{AP}, \forall ch_{j} \in \mathcal{CH}}{U(AP_{i}, ch_{j})}).
\end{equation}
A heuristic algorithm has been proposed to solve the above NP-hard problem in the following two steps:
(i) temporary channel assignment, which assigns a channel to each AP independent of other APs, to maximize each AP's utility function, and (ii) channel reassignment of the APs with low utility function (starting with the lowest one) without decreasing the sum of all utility function values. 
The second step is repeated for a given number of iterations or until $U(AP_{i}, ch_{j})=1$ for all APs. 
A fairness index similar to Equation \ref{eq-fairness} has been defined to include the utility function of APs. 

The 802.11g protocol with non-overlapping channels are used for performance evaluation through simulation and testbed. 
An $8$ to $15\%$ improvement in the fairness index and an $8$ to $21\%$ increase in the mean utility function are achieved, compared to minRSSI. 
%Furthermore, the authors report the performance of proposed technique in the presence of non-controllable APs (which are working using the minRSSI technique). 

%It is worth noting that the channel and data traffic information required to calculate utility function in the central controller are not collected in a real-time manner.
%Rather, this information is collected passively and based on statistical values. 

%  (1) approach: CA and fairness
%  (2) interference model: available throughput, percentage of operating on the assigned channel
%%% (3) problem formulation: calculus-based 
% (4) objective function: total throughput 
% (5) channels: non-overlapping (802.11g)
% (6) Client/AP-centric: AP-centric
% (7) Client-aware: Yes, traffic demand and the number of clients associated to each AP
%%% (8) Perfomance evaluation: simulation, real implementation
%%% (9) Dynamic/Static: Static
%%% (10) Optimality: Sub-optimal
% (11) Measurement: Statistical --> non-realtime (column 9)
%%% (12) Requried collecting information: traffic demand in AP, the acheivable throughput in a time unit (on a given channel)  

\label{FBWA}
\textbf{Frequency and bandwidth assignment for 802.11n/ac (FBWA)}.
%The authors of \cite{CA-BW-VTC-14} propose a channel and bandwidth allocation technique for 802.11a/n/ac-based WLANs (the proposed approach is similar to \cite{CA-VTC-14}). 
\cite{CA-BW-VTC-14} addresses channel and bandwidth allocation to benefit from the channel bonding feature of 802.11n/ac. 
The utility function of $AP_i$ that works on primary channel $ch_{j}$ and bandwidth $\beta$ is defined as
\begin{equation}
U(AP_{i}, ch_{j}, \beta)=\frac{Th(AP_{i}, ch_{j}, \beta)} {\min \left\{ Th^{\mathrm{max}}(AP_{i}, ch_{j}, \beta),\; L_{AP_{i}}^{\mathrm{mean}}   \right\} } 
\end{equation}
where $Th(AP_{i}, ch_{j}, \beta)$ is the expected throughput of $AP_i$ operating on primary channel $ch_{j}$ and channel bandwidth $\beta$,
 $Th^{\mathrm{max}}(AP_{i}, ch_{j}, \beta)$ is the maximum achievable throughput of $AP_i$, and $L_{AP_{i}}^{\mathrm{mean}}$ is the mean traffic load on $AP_i$. 
Through a heuristic algorithm similar to TFACA \cite{CA-VTC-14}, the appropriate primary channels and bandwidths are determined for all APs to maximize the sum of the APs' utility functions.
%\todo[inline,color=cyan]{$\checkmark$ and a jain based ... ??} 

Testbed results show a $65$ to $89\%$ reduction in the number of overlapping BSSs, and a mean throughput that is 3 times higher, compared to random channel assignment and minRSSI. 
%Similar to \cite{CA-VTC-14}, the process of collecting information from APs is not performed in a real-time manner. 

%
\label{ATCM}
\textbf{Normalized-airtime channel assignment (NATCA).}
%In \cite{DCA-residential-2015}, a WiFi Union (WU) framework is proposed to manage AP channel assignment.
\cite{DCA-residential-2015} requires each AP to measure its busy time (referred to as \textit{normalized airtime}) and report to a controller periodically. 
The busy time of an AP is the percentage of time that a channel assigned to that AP is occupied by its neighbors. 
An AP measures this value when sending data towards its clients. 
%The downlink traffic from an AP to its clients is used as the dominant traffic (compared to uplink) to measure APs' busy times. 
A threshold value is defined to categorize APs into two groups based on their busy time: \textit{heavily congested} and \textit{lightly congested}. 
An optimization problem has been proposed to minimize the sum of the normalized airtime of all APs without modifying the status of lightly congested APs. 
A Tabu search algorithm is proposed to solve the optimization problem. 
NS3 simulations show a 1.5x throughput improvement, compared to LCCS \cite{LCCS}. 
The measurement of normalized airtime for all APs is performed passively using the NS3 tracing system, which provides packet-level trace files for all APs.  

\begin{table*}
	\centering
	\scriptsize
	\caption{Comparison of Channel Assignment (ChA) Mechanisms}
	\label{CA-table} 
	\def\arraystretch{1}
	\begin{tabular}{|c|c|c|c|c|c|c|c|}
		\Xhline{3\arrayrulewidth}
		%	\textbf{Ref.} & \multicolumn{2}{|c|}{OneTwoThree} & \multicolumn{2}{|c|}{OneTwoThree}&\textbf{Channels} & \textbf{Dynamic/Static} & \textbf{\multicolumn{2}{|c|}{OneTwoThree}}\\ \Xhline{3\arrayrulewidth}
		%		
		\multirow{2}{*}{\quad \quad \textbf{Mechanism} \quad\quad}& \multicolumn{2}{c|}{\quad\quad  Traffic-Aware \quad\quad}& \multicolumn{2}{c|}{Standards $\&$ Channels} & \multirow{2}{*}{\textbf{\quad Dynamic \quad}}&  \multicolumn{2}{c|}{\quad Performance Evaluation \quad}\\ \cline{2-5}\cline{7-8}
		
		&\textbf{\quad Downlink \quad}& \textbf{\quad Uplink \quad}& \textbf{\quad\quad Standard \quad\quad} & \textbf{\quad Partially Overlapping \quad} & & \textbf{Simulation} & \textbf{Testbed}\\ \Xhline{3\arrayrulewidth}
		%1
		WCCA \cite{Wcolor-2005} &$\times$&$\times$&802.11b&$\checkmark$&$\checkmark$&$\checkmark$&$\checkmark$\\\hline
		%2
	    CDCA \cite{J-DCA-LB-2006}&$\times$&$\times$&802.11b&$\times$&$\checkmark$&$\checkmark$&$\checkmark$\\\hline
		%3
		%4
		%
		%		
		FPLN \cite{CAPWAP-based-CA-11}&$\times$&$\times$&802.11b&$\checkmark$&$\times$&$\checkmark$&$\checkmark$\\\hline					
		%		
		%6
		% jCI \cite{DCA-2016-1}&$\checkmark$&$\checkmark$&$\times$&802.11b&$\times$&$\checkmark$&$\checkmark$&$\checkmark$\\\hline
		%
		%cloudmac-2013
		CloudMAC \cite{CloudMAC}&$\times$&$\times$&802.11a/b/g/n&$\checkmark$&$\checkmark$&$\times$&$\checkmark$\\\hline		
		%		
		%Odin 2014
		Odin \cite{Odin2}&$\times$&$\times$&802.11a/b/g/n&$\checkmark$&$\checkmark$&$\times$&$\checkmark$\\\hline				
		%		
		% -PCA 2015
		PCA \cite{802.11ac-PCA}&$\times$&$\times$&802.11ac&$\times$&$\checkmark$&$\checkmark$&$\times$\\\hline		
		EMPOWER2 \cite{Primitives}&$\times$&$\times$&802.11a/b/g/n/ac&$\times$&$\checkmark$&$\times$&$\checkmark$\\\hline		
		%
		% -DCA-2 2016
		Wi5CA \cite{DCA-2}&$\times$&$\times$&802.11b/g&$\times$&$\checkmark$&$\checkmark$&$\times$		 \\\hline
		APCA \cite{J-AP-DCA-2006}&$\times$&$\checkmark$&802.11b&$\times$&$\times$&$\checkmark$&$\times$\\\hline		
		TACA \cite{Traffic-aware-CA-2007}&$\checkmark$&$\checkmark$&802.11b/g&$\checkmark$&$\checkmark$&$\checkmark$&$\checkmark$\\\hline
		VDTCA \cite{Measurement-CA-WCNC-10}&$\checkmark$&$\checkmark$&802.11a/g&$\times$&$\checkmark$&$\times$&$\checkmark$\\\hline					
		CUWN \cite{Cisco}&$\checkmark$&$\checkmark$&802.11a/b/g/n/ac&$\checkmark$&$\checkmark$&$\times$&$\times$\\\hline					
		CAFA \cite{CA-F-WCNC-11}&$\checkmark$&$\checkmark$&802.11b&$\checkmark$&$\times$&$\checkmark$&$\times$ \\\hline
		TFACA \cite{CA-VTC-14}&$\checkmark$&$\times$&802.11g&$\times$&$\times$&$\checkmark$&$\checkmark$\\\hline		
		FBWA \cite{CA-BW-VTC-14}&$\checkmark$&$\times$&802.11a/n/ac&$\times$&$\times$&$\checkmark$&$\times$\\\hline		
		NATCA \cite{DCA-residential-2015}&$\checkmark$&$\times$&802.11g&$\times$&$\checkmark$&$\checkmark$&$\times$ 
		%\cite{Ch-usage-based-2016}&$\times$&$\times$&$\times$&$\times$&$\times$&$\times$&\\\hline		
		%
		%
		\\\Xhline{3\arrayrulewidth}
	\end{tabular}
\end{table*}

\subsection{Channel Assignment: Learned Lessons, Comparison, and Open Problems}
\label{ChAProblems}
Table \ref{CA-table} presents and compares the features of ChA mechanisms.
In the following, we study these features and identify research directions.

\subsubsection{\textbf{Dynamicity and Traffic-Awareness}}
\label{asc_dyn_tra_awar}
As Table \ref{CA-table} shows, not all the ChA mechanisms support dynamic channel reallocation.
In addition, most of the ChA mechanisms are either traffic-agnostic, or they do not recognize uplink traffic.
%For example, although EmPOWER2 architecture (see Section \ref{empower2_arch})  provides traffic information for network applications, its employed ChA mechanism is very simple and does not benefit from these features.
Furthermore, many of the ChA mechanisms do not actually discuss the data gathering policy employed, and the rest focus on the mean traffic load of APs and the mean throughput demand of clients. 
These limitations are due to two main reasons: First, the higher the rate of central interference map generation, the higher the overhead of control data exchanged by the controller. 
Second, although heuristic algorithms have been proposed to tackle the NP-hardness of the channel assignment problems, the execution duration of these algorithms might not be short enough to respond to network dynamics.
The two approaches earlier proposed in Section \ref{asc_dyn_overhead} (i.e., prediction and hybrid design) can be employed to cope with these challenges.
For example, depending on the time window and accuracy of predictions, a ChA proactively responds to dynamics, which in turn reduces the burden on real-time network mapping and fast algorithm execution.
%Meanwhile, dynamic adjustment of evaluation period and designing monitoring mechanisms that do not consume network resources excessively are open research challenges.
The second solution, i.e., hybrid design, employs multi-level decision making to support fast and low-overhead reaction to network dynamics. 
For the hybrid designs, however, the topology of controllers may be tailored depending on the control mechanism employed.
For example, while a suitable controller topology for a ChA mechanism depends on the interference relationship between APs, a more suitable topology for an AsC mechanism is to connect all the APs of a hallway to a local controller to improve the QoS of clients walking in that area.
Therefore, it is important to design architectures that support flexible communication between controllers.

\subsubsection{\textbf{Joint ChA and AsC}}
\label{asc_joint_des}
As channel switching may lead to re-association (depending on the architecture used), clients may experience communication interruption and violation of their QoS requirements \cite{SDWLAN2,Lv2013,jin2011fast}.
Therefore, it is necessary to measure and reduce the channel switching overhead of dynamic ChA mechanisms. 
In this regard, joint design of ChA and AsC is essential for uninterrupted performance guarantee.
Such a joint mechanism, for example, monitors and predicts clients' traffic pattern, performs channel assignment to maximize spatial reuse, and moves VAPs between APs to support seamless handoff as the traffic changes.
In addition, mobility prediction \cite{manweiler2013predicting,dong2012evaluation} may be employed to minimize the delay of AsC and ChA by enabling the use of proactive control mechanisms instead of using reactive mechanisms that are triggered by performance drop.
Note that mobility prediction is specifically useful in SDWLANs as these architectures enable the central collection and analysis of metrics such as RSSI to apply localization and mobility prediction algorithms.
Unfortunately, the existing joint ChA-AsC mechanisms (e.g., \cite{xu_channel_2011,Zheng2016}) operate distributively and do not provide any performance guarantee, which is required for  mission-critical applications such as medical monitoring or industrial control \cite{REWIMO}.

%most of those are implemented in a distributed way and do not address delay constraints of applications.

\subsubsection{\textbf{Partially Overlapping Channels and AP Density Management}}
\label{cha_part_overlap}
Most ChA mechanisms designed for 2.4GHz networks utilize only non-overlapping channels to decrease the interference level (as shown in Table \ref{CA-table}).
However, due to the dense deployment of APs, this approach limits the trade-off between interference reduction and channel reuse \cite{zhao_dapa:_2016}.
To address this concern, research studies have improved the performance of channel assignment by dynamically assigning partially-overlapping channels \cite{CA-F-WCNC-11}. 
These mechanisms, for example, rely on the interference relationship between APs to increase the distance between assigned channels as the pairwise interference increases.
However, these mechanisms require the underlying SDWLAN to provide APIs for efficient collection of network statistics.

In addition to utilizing partially-overlapping channels, ChA mechanisms can further improve network capacity through AP topology management, which is achieved by three main strategies:
(i) \textit{AP placement}: the site survey is performed to determine the location of APs during the installation phase \cite{zvanovec2003wireless};
(ii) \textit{AP power control}: a SDWLAN controller dynamically adjusts the transmission power level of APs \cite{li2011achieving};
(iii) \textit{AP mode control}: a SDWLAN controller dynamically turns on/off APs \cite{EmPOWER}.
Note that the second and third strategy require architectural support.
%For example, a ChA mechanism needs APIs to monitor network and make decisions that aim to reduce interference and maximize spatial reuse through both channel assignment and transmission power control.
It is also important to coordinate the aforementioned strategies with AsC to avoid clients' service interruption.
For example, by relying on the global network view established, clients association may be changed before channel assignment or transmission power control cause client disconnection. 
Although these problems have been addressed in isolation or distributively \cite{bejerano2009cell,wang2014coverage,DenseWLAN1,huang2010distributed}, centralized and integrated solutions are missing.

\subsubsection{\textbf{Integration with Virtualization}}
\label{ch_disc_int_virt}
Our review of ChA and AsC mechanisms shows that these mechanisms are oblivious to virtualization.
Specifically, they do not take into account how the pool of resources is assigned to various network slices.
However, virtualization has serious implications on network control.
For example, mobility management becomes more challenging when network virtualization is employed.
When a client requires a new point of association due to its mobility, in addition to parameters such as fair bandwidth allocation, the available resources of APs should also be taken into account. 
More specifically, for a client belonging to slice $n$, the AsC mechanism should ensure that after the association of this client with a new AP, the QoS provided by slice $n$ and other slices is not violated.
However, as this may require client steering, the re-association cost of other clients should be minimized.
%Specifically, as discussed in Section \ref{archComp}, SDWLANs enable the slicing of network resources such as airtime.
%As another example, an AsC mechanism may dynamically update and maintain a subset of APs that should be used for handoff of the clients that belong to a slice $n$.
As an another example, a ChA mechanism may assign non-overlapping and overlapping channels to a mission-critical slice and a regular slice, respectively.
The problem becomes even more complicated when multiple controllers collaborate to manage network resources.
The integration of network slicing and control mechanisms would pave the way to use SDWLANs in emerging applications, such as mission-critical data transfer from mobile and IoT devices.

In addition to network-based slicing of resources, it is desirable to support personalized mobility services as well.
To this end, AsC mechanisms require architectural support to enable client association based on a variety of factors.
As an example, a client's VAP stores the client's handset features and exposes that information to the AsC mechanism to associate the client with APs that result in minimum energy consumption of the client.
Such interactions between architectural components and control mechanisms are unexplored.

Another important challenge of slicing at low level is the complexity of coordination.
For example, assume that different transmission powers are assigned to flows. 
In this case, simply changing the transmission power of APs working on the same channel would cause significant interference and disrupt the services being offered.
Therefore, offering low-level virtualization requires coordination between the network control mechanisms.

\subsubsection{\textbf{Coexistence of High-Throughput and Legacy Clients}}
Although new WiFi standards (e.g., 802.11n/ac) supporting high throughput are broadly used in various environments, we should not neglect the existence of legacy devices (e.g., 802.11b), especially for applications that require low throughput \cite{Tozlu2012,CYW43907,BCM4343}.
In heterogeneous networks, the throughput of high-rate 802.11ac devices is jeopardized by legacy clients, due to the wider channel widths used by 802.11n/ac through channel bonding \cite{Zeng2014,han_fair_2016}. 
Specifically, the channel bonding feature of 802.11n/ac brings about new challenges such as \textit{hidden-channel problem}, \textit{partial channel blocking}, and \textit{middle-channel starvation} \cite{802.11ac-PCA,han_fair_2016,zhang2011adaptive}. 
The bandwidth of a 802.11n/ac device (80/160MHz) is blocked or interrupted by a legacy client (20MHz) due to the partial channel blocking and hidden-channel problem.
In the middle-channel starvation problem, a wideband client is starved because its bandwidth overlaps with the channels of two legacy clients.
Hence, the wideband client can use its entire bandwidth only if both legacy clients are idle. 
Despite the heterogeneity of WLAN environments, our review shows that addressing the challenges of multi-rate networks to satisfy the diverse set of QoS requirements is still premature. 
In fact, only PCA \cite{802.11ac-PCA} proposes a central solution, and its performance has been compared with distributed mechanisms.

%\subsubsection{Multiple spatial streams}
%The emergence of new standards (802.11n/ac/ad) necessitates the use of new mechanisms. For example, when channel bonding feature is available, a DCA mechanism may address both channel assignment and channel width allocation. 
%Although there are a few techniques available for joint frequency and channel width assignment (e.g., \cite{802.11ac-PCA,CA-BW-VTC-14}), there is no technique addressing the effect of MIMO and beamforming on channel assignment.
%For example, a potential approach would be to rely on user location and coverage pattern to achieve a higher level of spatial reuse with channel assignment. 

\subsubsection{\textbf{Security}}
Wireless communication is susceptible to jamming attacks through which malicious signals generated on a frequency band avoid the reception of user traffic on that channel. 
The vulnerability of 802.11 networks to such denial of service (DoS) attacks have been investigated through various empirical measurements \cite{pelechrinis_denial_2011,bayraktaroglu2008performance,pelechrinis2009ares}.
Although the research community proposes various mechanisms (such as frequency hopping \cite{pelechrinis2010efficacy}) to cope with this problem, the existing approaches do not benefit from a central network view.
In particular, by relying on the capabilities of SDWLANs, clients and APs can periodically report their operating channel information (such as channel access time, noise level and packet success rate) to the controller, and mechanisms are required to exploit this information to perform anomaly detection.

%Consequently, we have four types of channel assignment techniques as follows
%\begin{enumerate}[(I)]
%	\item Client-agnostic, traffic-agnostic
%	\item Client-aware, traffic-agnostic
%	\item Client-agnostic, traffic-aware
%	\item Client-aware, traffic-aware.
%\end{enumerate}

%Furthermore, the performance of channel assignment techniques are evaluated through either \textit{simulations} or \textit{empirical experiments}. Some works consider real implementation through small-scale testbeds to reveal the superiority of proposed algorithms such as \cite{Wcolor-2005,J-DCA-LB-2006,Traffic-aware-CA-2007,DCA-2016-1,CA-VTC-14,CAPWAP-based-CA-11,Measurement-CA-WCNC-10}. However, other techniques perform the performance evaluation using simulations. 

%Flashback [11] proposes a control channel technique for
%WiFi networks, by allowing stations to send short control
%messages concurrently with data transmissions, without
%affecting throughput. This ensures a low overhead con-trol plane for WiFi networks that is decoupled from the
%data plane.

%
\section{Related Surveys}
\label{exis_surveys}
In this section we mention the existing surveys relevant to SDN and WLAN.

A comprehensive study of software-defined wireless cellular, sensor, mesh, and home networks is presented in \cite{wSDN1}. 
This work classifies the existing contributions based on their objective and architecture.
A survey of data plane and control plane programmability approaches is given in \cite{macedo2015programmable}, where the authors review the components of programmable wired and wireless networks, as well as control plane architectures and virtualization tools.
\cite{richart2016resource} presents a survey of virtualization and its challenges in wireless networks. 
%This paper reviews resource allocation and isolation as the two main challenges of virtualization.
%
A survey of OpenFlow concepts, abstractions and its potential applications in wired and wireless networks is given in \cite{hu2014survey}.
\cite{NextGen-WiFi-Survey-2016} and \cite{NGwlans-survey-2016} review the 802.11 standards for high-throughput applications, TV white spaces and machine-to-machine communications. 
\cite{S1-energy-MAC} surveys energy-efficient MAC protocols for WLANs.
A survey of cellular to WLAN offloading mechanisms is given in \cite{Offloading-survey-2016}.

Channel assignment strategies are reviewed in \cite{Channel-assignment-survey-2010}, and the dynamics of channel assignment in dense WLANS are investigated in \cite{S9-Channel}. 
QoS support in dense WLANs through layer-1 and layer-2 protocols are surveyed in \cite{S2-QoS,S3-QoS,S4-QoS}. 
Client-based association mechanisms and AP placement strategies are surveyed in \cite{Association-DenseWLANs-2014}.
Distributed load-balancing mechanisms are reviewed in \cite{S5-LB}.
\cite{S6-Intf} provides a survey on spectrum occupancy measurement techniques in order to generate interference maps in WLANs. 
In \cite{S7-ax}, the future demands of WLANs and the potential new features of ongoing 802.11ax standard are investigated and reviewed. 
A review of wireless virtualization mechanisms is given in \cite{liang2015wireless}.
This work also presents the business models, enabling technologies, and performance metrics of wireless virtual networks.
%
%\cite{claise2012overview} presents a review of network management standards, including CAPWAP, SNMP and Netconf.
%

Our review of existing surveys shows that there is no study of WLANs from a SDN point of view.
Therefore, the main motivation for conducting this study was the lack of a comprehensive survey on SDWLAN architectures and essential centralized network control mechanisms. 

%To the best of our knowledge, this is the first effort on reviewing, investigating and qualitative comparison of existing SDWLAN architectures and centralized network control mechanisms.
%

\section{Conclusion}
\label{Conclusion}
%Conventional wisdom prefers to manage WLANs in a distributed way to provide more scalable approaches. However, industry and academia have switched to centrally manage of these networks due to the dense deployment of APs in enterprise WLANs. 
SDWLANs enable the implementation of network control mechanisms as applications running on a network operating system.
Specifically, SDWLAN architectures provide a set of APIs for network monitoring and dissemination of control commands.
Additionally, network applications can benefit from the global network view established in the controller to run network control mechanisms in a more efficient way, compared to distributed mechanisms.
In this paper, we categorized and reviewed the existing SDWLAN architectures based on their main contribution, including \textit{observability and configurability}, \textit{programmability}, \textit{virtualization}, \textit{scalability}, \textit{traffic shaping}, and \textit{home networks}. 
Through comparing the existing architectures as well as identifying the growing and potential applications of SDWLANs, we proposed several research directions, such as establishing trade-off between centralization and scalability, investigating and reducing the overhead of south-bound protocols, exploiting layer-1 and layer-2 programmability for virtualization, and opportunities and challenge of SDWLANs with respect to security provisioning.

After reviewing architectures, we focused on association control (AsC) and channel assignment (ChA), which are the two widely-employed network control mechanisms proposed to benefit from the features of SDWLAN architectures.
We reviewed centralized AsC and ChA mechanisms in terms of their metrics used, problem formulation and solving techniques, and the results achieved compared to distributed mechanisms.
%We discussed that due to the diversity of clients and emerging new applications with different QoS requirements and traffic behavior, it is more effective and reasonable to directly include the application-level demands and behavior of clients in formulating association optimization problems. 
%Our study showed that the AM mechanisms tend to be client-oriented with the aim of satisfying QoS requirements in terms of throughput, delay, packet loss probability and re-association costs. 

At a high level, we categorized AsC mechanisms into two groups: \textit{seamless handoff}, used to reduce the overhead of client handoff, and \textit{client steering}, to improve the performance of clients.
Furthermore, we specified several research directions towards improving AsC mechanisms, such as the design and measurement of metrics to include both uplink and downlink traffic into the decision making process, establishing proportional demand-aware fairness among clients, and including the application-level demand and behavior of clients in formulating AsC problems.

We reviewed central ChA mechanisms, and we focused on the approaches employed by these mechanisms to model the interference relationship among APs and clients.
Based on the inclusion of traffic in the decision making process, we categorized ChA mechanisms into \textit{traffic-aware} and \textit{traffic-agnostic}.
Our comparisons revealed that, given the heterogeneity of clients, the new ChA mechanisms should address the coexistence of low- and high-throughput devices.
%In particular, mitigating interference in densely deployed SDWLANs with heterogeneous clients requires optimal use of partially-overlapping channels in the 2.4GHz band and variable channel widths in the 5GHz band, which is a challenging problem.
We also discussed the challenges and potential solutions for the integration of virtualization with AsC and ChA, as well as the importance and potential benefits of AsC and ChA co-design.

\section{Acknowledgment}
This research is partially supported by a grant from the Cypress Semiconductor Corporation (Grant No. CYP001).

\bibliographystyle{IEEEtran}
\bibliography{IEEEabrv,references}

\end{document}